**Effect of superplastic deformation on the intergranular corrosion resistance of austenitic steel 321 with different types of UFG structure: A case of homogeneous and heterogeneous microstructure obtained by ECAP**

A.V. Nokhrin[(*)], V.I. Kopylov, M.K. Chegurov, S.V. Shotin, C.V. Likhnitskii, N.V. Melekhin, D.A. Zotov, K.A. Rubtsova, O.E. Pirozhnikova, N.N. Berendeev, D.D. Kulagin

*Lobachevsky State University of Nizhny Novgorod, 603022 Nizhny Novgorod, Gagarin Avenue 23, Russia*

E-mail: nokhrin@nifti.unn.ru

**Abstract**

The effects of superplasticity (SP) on the microstructure, microhardness, and intergranular corrosion (IGC) resistance of ultrafine-grained (UFG) 321L steel are reported for the first time. By increasing the number of the Equal Channel Angular Pressing (ECAP) cycles (*N*) and varying the ECAP temperature, and annealing temperature, the degree of the microstructural heterogeneity was changed. Regardless of the microstructural homogeneity degree, UFG steels demonstrated the possibility of realizing the SP effect at temperatures below the onset temperature of dynamic recrystallization. The dependence of ductility of the UFG steel on the test temperature is non-monotonic with a minimum at 450-600ºC. The maximum elongation to failure (~250%) was achieved at a SP temperature of 800°C. The SP deformation of the UFG steel is controlled by the power law creep. It is shown that the hardness of UFG steel samples after SP is equal than the hardness of UFG steels annealed at the same temperature. It is also shown what SP leads to a decrease in the IGC resistance of the UFG steel. Heterostructured steel have reduced ductility, as well as lower resistance to IGC after tension tests at room temperature (RT). The minimum IGC resistance is ensured at $N = 2$ ECAP cycles, and the maximum – at $N = 4$. The IGC resistance of homogeneous UFG samples after tension testing exceeds the IGC resistance of coarse-grained steel. Tension test at a RT has the greatest negative effect on the IGC resistance of heterostructured steel, and the least negative effect on the IGC resistance of homogeneous steel. Increasing the number of ECAP cycles enhances the ductility of the UFG steel at 750–900°C. Increasing the ECAP temperature results in a simultaneous decrease in ductility and flow stress. Heterostructured steel exhibits minimal values of the activation energy of SP deformation. Increasing the number of ECAP cycles and increasing the ECAP temperature lead to an increase in the activation energy of SP deformation.

---

[(*)] Corresponding author: Aleksey Nokhrin, Dr.Sc., Head of Laboratory (e-mail: nokhrin@nifti.unn.ru; ORCID: 0000-0002-0328-1923)

**Abbreviations:** CG – coarse grained (steel); ECAP – Equal Channel Angular Pressing; EDS – Energy Dispersion (analysis); El – Elongation to failure; GBS – Grain Boundary Sliding; DLEPR – Double Loop Electrochemical Potentiodynamic Reactivation (method); OMDs – Orientation Misfit Dislocations; RT – Room Temperature; SEM – Scanning Electron Microscopy; SP – Superplasticity; SPD – Severe Plastic Deformation; TEM – Transmission Electron Microscopy; UTS – Ultimate tensile strength; UFG (steel) – Ultrafine-Grained (steel).

## 1. Introduction

Austenitic stainless steels (304, 316, 321, etc.) are applied extensively in the petrochemical industry, nuclear power engineering, and mechanical engineering. The low strength of coarse-grained (CG) austenitic steels is one of their drawbacks, which prevents the achievement of the necessary level of reliability in products. The ductility of the CG steels is low even at elevated temperatures, and the elongation to failure (El) in the tensile tests seldom exceeds 50–100% [1]. At lower deformation temperatures of CG metastable steel, strain aging effects are frequently observed. This further restricts the temperature range in which CG steel can be effectively deformed. The insufficient ductility of CG austenitic steel during warm deformation leads to the necessity to increase the stamping temperature of products to temperatures exceeding 900–1100°C. At these temperatures, the process of dynamic recrystallization commences, which enhances the ductility of CG austenitic steel.

Formation of ultrafine-grained (UFG) microstructure in austenitic steels is a conventional method of improving their strength. Deformation of austenitic steels at room temperature (RT) or moderate temperatures followed by annealing is the most common method of the UFG microstructure formation [1–3]. Cold Severe Plastic Deformation (SPD) of metastable austenitic steel leads to strain-induced martensite formation, and annealing initiates the reverse transformation of martensite into austenite [2–5]. It should be noted that martensite reduces the corrosion resistance of austenitic steel; therefore, to minimize this negative effect, either SPD is performed at elevated temperatures, or the resulting dual-phase UFG steel is subjected to annealing, which leads to the reversible transformation of martensite back into austenite [6–8]. UFG austenitic steels obtained by warm SPD have high strength and hardness, as well as satisfactory ductility [9–13], and good corrosion resistance [14–17]. The forming of the UFG microstructure allow increasing the ductility of steels at elevated temperatures and small strain rates [1, 18–21]. Earlier, it was shown that the UFG materials can demonstrate low-temperature and high-strain-rate superplasticity (SP) effect [22–25]. Reducing the temperature and strain time while simultaneously increasing ductility reduces energy consumption of SP stamping process, extends the service life of die tooling for hot stamping, enables the production of more complex-shaped components with minimal defects, and decreases steel oxidation.

Note that in spite of the great interest in the SP of ferritic and duplex steels [26], there are very few studies on the SP of the austenitic steels. At present, the major attention is attracted to the SP of steels, which exhibit a fine-grained microstructure formed by the reverse transformation of martensite into austenite during heating [27].

A major challenge in the development of corrosion-resistant, superplastic, UFG steels is the ambiguous effect of martensite on ductility. Studies indicate that, under SP deformation conditions, UFG austenitic steels containing martensite exhibit the best plasticity [2, 5–7]. However, the presence of martensite simultaneously leads to a reduction in the steel's corrosion resistance.

The low-temperature SP of the UFG steel 304 obtained by cold rolling followed by annealing was described in [28]. At RT, the σ(ε) curves had a plastic instability region, which was typical for the Lüders deformation. At 600–650°C, the σ(ε) curves had a shape typical for the SP flow. The maximum El of 268–296% was achieved at 600°C and a strain rate of $\dot{\varepsilon} = 2.5 \cdot 10^{-4}\ s^{-1}$. After pre-annealing at 700°C for 5 min, the severely deformed steel had a completely austenite UFG microstructure with the lower ductility. Since the austenite grains remained equiaxial after testing, the authors of [28] concluded the grain boundary sliding (GBS).

The authors of [28, 29] suggested residual martensite to suppress the dynamic austenite grain growth at the SP of UFG steel. Note the work [29] where the elongation to failure of UFG steel 18Cr–9Ni was shown to depend on the α′-martensite content strongly. The maximum El in steel 18Cr–9Ni was achieved with the α′-martensite content of ~10%. At this composition, the El was 270% at a temperature of 650°C and a strain rate $10^{-3}\ s^{-1}$ [29].

A comparative analysis of the SP mechanisms for CG and UFG steel 18Cr–8Ni was carried out in [30]. The El for the CG steel was 113% at 800°C. In the temperature range of 25–400°C, the ductility of the UFG steel was smaller than the one of the CG steel whereas at 600–800°C it was close to the ductility of the CG steel or slightly higher. The σ(ε) curve for the UFG steel at 800°C had an elongated stage of stable plastic flow. It is interesting to note that higher martensite content was observed in the UFG steel samples in the tensile tests compared to the UFG steel before the start of the test. The authors of [30] related the presence of a large number of small recrystallized grains at the grain boundaries after SP at 600°C and 800°C to GBS and the onset of dynamic recrystallization.

The achievement of very high ductility for UFG steel 304 was reported in [31]. The grain size of the UFG steel was ~0.2 μm and the martensite content was small (< 0.5%). At 600°C and the strain rate $< 3 \cdot 10^{-4}\ s^{-1}$, the elongation to failure was > 500%. The authors of [31] referring to their previous works, also reported the possibility to achieve the El in steel 304 > 600% at $\sim 5.5 \cdot 10^{-5}\ s^{-1}$ and 600°C.

In [32], unexpected results of SP investigations of fine-grained steels 321L were presented. The ductility of CG steels is known to depend non-monotonically on the test temperature [28, 33].

The maximum El ($\delta_{max}$ = 43%) of the fine-grained steel 321L was achieved at 500°C. This value exceeds the ductility of the CG steel at this temperature (~6%). Among other interesting results presented in [32], it should be mentioned the absence of the El reduction effect in the temperature range of 400–500°C in the UFG steel 321.

Summarizing the results of the brief analysis, one can conclude that there are some contradictions in the question of the effect of various factors on the strain character in the UFG austenitic steels in the SP conditions at present. It should be noted that fine-grained steels with an optimal initial proportion of martensite are often used to ensure good SP properties [28-31]. Although most studies note the positive effect of martensite on the ductility of UFG austenitic steels, there are currently no studies describing how SP deformation conditions affect the corrosion resistance of such steels.

Note that SPD is an effective method to obtain heterostructured metallic materials [34]. Equal Channel Angular Pressing (ECAP) results in heterogeneity in austenitic steel, characterized by the formation of shear macro- and micro-bands of localized deformation [35]. This leads to an increase in martensite in the shear-bands. A heterogeneous microstructure significantly influences the strength, ductility, creep resistance, and other performance characteristics of alloys [34, 36, 37]. At present, the SP deformation mechanisms in heterostructured steels remain unexplored.

As noted above, SP deformation can provide significant economic advantages in the manufacture of complex-shaped products in engineering. However, to achieve this, steel products of complex shape produced by the SP stamping method must possess good corrosion resistance. To ensure high corrosion resistance, steels after SP stamping must have an austenitic structure in which there is no precipitation of corrosion-prone particles (such as carbides, σ-phase, martensite, etc.). It is generally assumed that in CG steels, martensite negatively affects the corrosion resistance of austenitic steel [1, 38]. In nanostructured and UFG steels, the situation is more complex – there is evidence of both increased [14-16] and decreased corrosion resistance in steels containing martensite.

It should be noted that intergranular corrosion (IGC) and SP deformation can be closely related in UFG materials. The primary mechanism of SP deformation in UFG materials is grain boundary sliding (GBS). Particles precipitated at grain boundaries can impede grain boundary shear and reduce the GBS rate. This counteracts their positive effect on the stability of UFG structure at elevated temperatures. Particles precipitated at grain boundaries also trigger IGC in steels [1, 35, 38]. Therefore, studying IGC in UFG steels provides a convenient method for indirectly assessing the purity of grain boundaries and their susceptibility to GBS.

The primary objective of this study is to examine the effect of SP deformation temperature on the IGC resistance of UFG austenitic steel 321L. The SP deformation was conducted at temperatures not exceeding 900°C. This temperature corresponds to the onset of dynamic recrystallization at conventional strain rates [6, 40, 41]. Testing at temperatures above 600°C prevents the formation of martensite during SP and eliminates the influence of this factor. For convenience, the temperature range of 600-900°C (0.48-0.65$T_m$, where $T_m$ = 1810 K is the melting point of steel) will be referred to as warm superplastic deformation. Test temperature ranges below 600°C ($T < 0.48T_m$) and above 900°C ($T > 0.65T_m$) will be referred to as cold deformation and hot deformation, respectively.

To date, the SP deformation of this steel in the heterogeneous UFG state hasn't been studied. Therefore, the study of the SP deformation mechanism of heterostructured UFG steel 321L was a also objective of this research. To achieve this, the degree of microstructural homogeneity in the metastable austenitic steel was varied by altering the number of cycles and the temperature of ECAP, as well as the temperature of subsequent annealing.

The selection of steel 321L as the object of investigation was motivated by its wide application in the oil chemical industry, mechanical engineering, and nuclear power engineering. The heat exchangers, muffles, pipes, shut-off valve parts, and other parts operated under conditions of elevated temperatures and corrosion-aggressive media are fabricated from Ti-containing steel 321L by warm and hot deformation. This work continues a series of studies [35, 42] devoted to investigating the corrosion resistance and deformation behavior of UFG steel 321L produced by the ECAP method.

## 2. Materials and methods

The commercial metastable austenitic steel AISI 321L (composition, in % wt.: Fe–0.08%C–17.9%Cr–10.6%Ni–0.5%Si–0.1%Ti, Russian trademark 08Cr18Ni10Ti) was the object of investigations. The UFG microstructure formation in the steel was performed by ECAP. Prior to ECAP, the rods were annealed at 1050°C for 30 min followed by quenching in water. The angle of crossing the working channel and the output one was 90° (π/2). The workpieces were processed in *A* ECAP route. The ECAP temperatures ($T_{ECAP}$) were 150°C and 450°C, the number of ECAP cycles (*N*) varied from one to four. The features of the ECAP regimes are detailed in [35]. 30-minute annealing of the samples was carried out in a PlavkaPro air furnace, in the temperature range of 600-900°C. The samples were cooled in air. The designations of the samples are presented in Table 1.

Table 1. Designation of steel samples

| $T_{ECAP}$, °C | *N* | Series |
|---|---|---|
| 150 | 1 | #150-1 |
| | 2 | #150-2 |
| | 3 | #150-3 |
| | 4 | #150-4 |
| 450 | 1 | #450-1 |
| | 2 | #450-2 |
| | 3 | #450-3 |
| | 4 | #450-4 |

The investigations of the microstructure were carried out using Jeol JSM-6490 and Tescan Vega Scanning Electron Microscopes (SEMs), Jeol JEM-2100F Transmission Electron Microscope (TEM) and Leica IM DRM optical microscope (OM). Samples for the microstructure study were cut out along the workpiece axis. The average grain size (*d*) was determined by the chord method using GoodGrains 2.0 software. The local chemical analysis was performed using Oxford Instruments INCA 350 EDS spectrometer. X-ray diffraction (XRD) phase analysis of the steels was carried out using Shimadzu XRD-7000 X-ray diffractometer (Cu$K_\alpha$ emission, scanning step – 0.02°, exposure time – 3 s). The unit cell parameters and mass fraction of austenite and martensite were calculated by Rietveld method using Topas software. The phase identification was performed using PDF and ISCD databases.

The microhardness (*HV*) of the steel was measured with HVS1000 tester. The measurements of *HV* were performed at the load of 2 kg. The uncertainty of the *HV* measurements was ± 50 MPa. For the mechanical tests, was performed using flat dog-bone shaped specimens (Fig. 1a). The sizes of the working part of the dog-bone shaped specimens were 2×2×3 mm. The tensile tests were carried out using Tinius Olsen H25K-S machine with the strain rates $\dot{\varepsilon}$ = 3.3·10$^{-4}$ s$^{-1}$ (the tension rate was 10$^{-3}$ mm·s$^{-1}$) and 3.3·10$^{-3}$ s$^{-1}$ (the tension rate was 10$^{-2}$ mm·s$^{-1}$). The tensile tests were performed at RT and in the temperature range of 450–900°C. The specimens were heated up to the test temperatures in 5 min. The samples were kept at the test temperatures for 10 min to establish the thermal equilibrium. From the engineering stress (σ) – engineering strain (ε) curves, the magnitudes of the ultimate tensile strength (UTS, $\sigma_b$) and the El ($\delta_{max}$) were determined. The flow stress (FS, $\sigma_s$) and maximum uniform strain ($\varepsilon_u$) were determined from the true stress–true strain curves.

The fractographic analysis of the fractures after the tensile tests was carried out using Jeol JSM-6490 SEM. The IGC studying, microstructure investigations and the *HV* measurements were performed in the fracture zones and in the non-deformed areas (Fig. 1b, c). The study results of the fracture region (Zone II) were averaged from two data series obtained from both parts of the fractured sample. In contrast, the study results of the Zone I were averaged over six series (three series for each half of the fractured sample, as shown in Fig. 1b, c). During the analysis of the experimental data, it was assumed that the strain was uniformly distributed in Zone II (Appendix A).

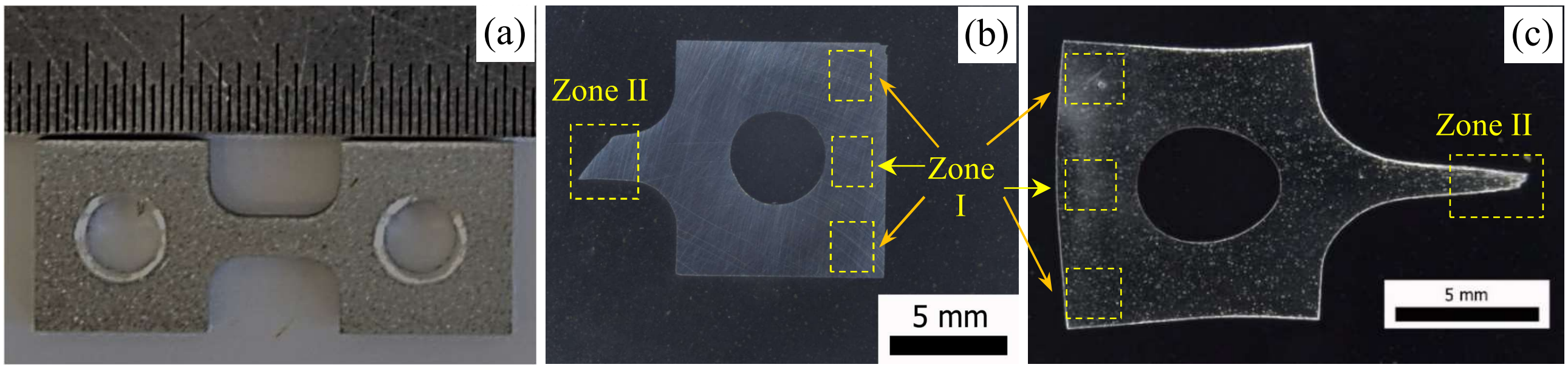

**Fig. 1** General appearance of the sample from the #150-3 series before (a) and after testing at RT (b) and at 800°C (c). In Fig. 1b, c the areas of the study structure and the *HV* measurements are marked

The IGC resistance of the steels was investigated using an R-8 potentiostat-galvanostat, employing the double loop electrochemical potentiodynamic reactivation (DLEPR) method [43-45]. The tests were conducted at RT in an aqueous solution of 10% $H_2SO_4$ + 0.0025 g/l KSCN according to the Russian National Standard GOST 9.914-91. An auxiliary electrode was made from a Pt grid, while the reference electrode was made from silver chlorine (AgCl). The sample under study served as the working electrode. The sample under study was polarized by a cathode at a potential φ = –550 mV for 2 min. The voltage-current density curves were recorded in the potential range from –550 mV to +1200 mV at a rate of 3 mV/s. Prior to testing, the surfaces of all samples were mechanically polished and ground to achieve a roughness level not exceeding 1 μm. A protective, corrosion-resistant varnish was applied to the surfaces of the samples, covering the entire area except for the fracture zone (Zone II in Fig. 1b, c). The area of the active corrosion zone on the sample surfaces was measured using a Leica IM DRM optical microscopy, employing GoodGrains 2.0 software.

The tendency of the steel to IGC was determined by the DLEPR method from the ratio of the areas under the reactivation curve ($S_r$) and the activation curve ($S_a$): $K = S_r/S_a$. In this method, the sample is initially passivated by varying the potential at a specified rate, transitioning from the corrosion potential to the passivation potential. The potential is then altered in the opposite direction. During the test, reactivation of the sample occurs, accompanied by the breakdown of the passive film. It is generally assumed that the degradation of the passive film begins at the surface of the austenitic steel grain boundaries, which may be depleted in chromium or contain of the σ-phase particles, etc. According to the Russian National Standard GOST 9.914-91, the increasing of the coefficient of $K$ to critical value $K_{IGC}$ = 0.11 means that the austenite steel 321L demonstrates an increased tendency to IGC. Hereinafter, the coefficient $K$ for Zone I will be denoted as $K_{\mathrm{I}}$, and for Zone II – as $K_{\mathrm{II}}$.

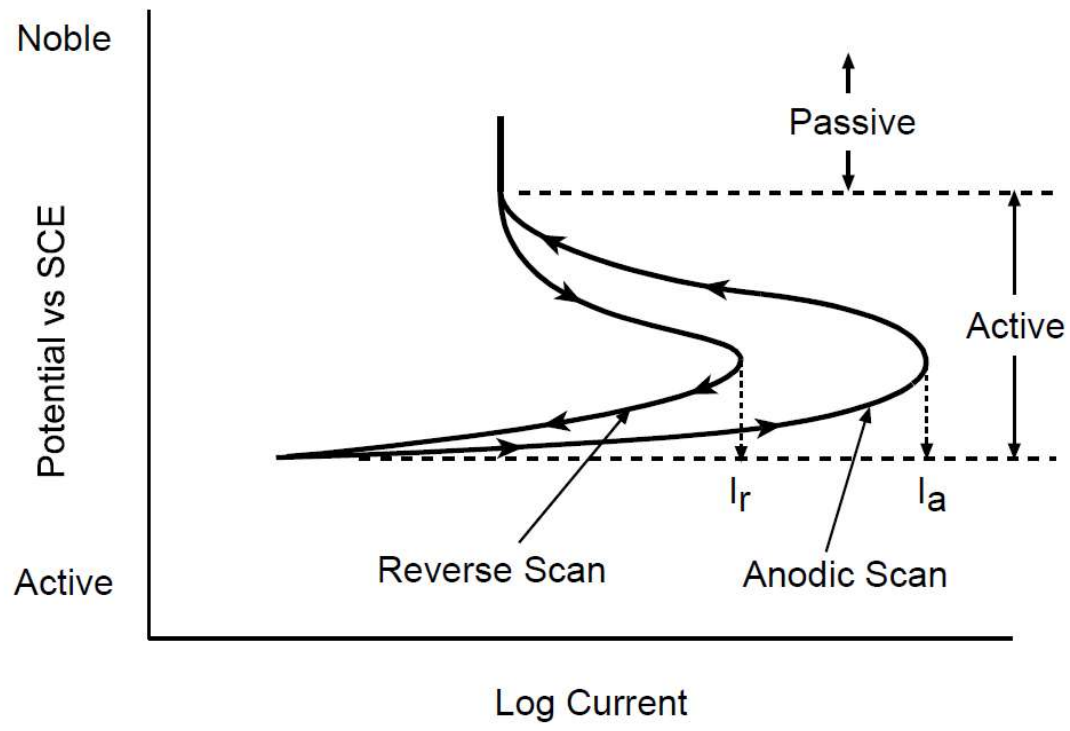

**Fig. 2** Methodology for determining the IGC susceptibility of austenitic steel by DLERP method [46]

In terms of its meaning, the coefficient $K$ is analogous to the Degree of Sensitization (DOS), which is defined as the ratio of the reactivation current ($I_r$) to the activation current ($I_a$) (Fig. 2): DOS = $I_r/I_a$ [47]. In GOST 9.914-91, the coefficient $K$ is considered a more precise parameter for determining the susceptibility of 321L austenitic steel to IGC. It should be emphasized that GOST 9.914-91 applies only to the Russian steel grades 08Cr18Ni10Ti and 12Cr18Ni10Ti, which are equivalent to AISI 321L steel.

The results of corrosion tests on samples after SP testing were compared with those of steel 321 samples subjected to ECAP and annealing. These findings were previously reported in [35].

Additional test for IGC resistance were conducted in accordance with the Russian National Standard GOST 6032-2003 (ISO 3651-1:1998, ISO 3651-2:1998). IGC tests on samples after SP tests were performed using a boiling sulfuric acid (250 cm$^3$ $H_2SO_4$) and copper sulfate (50 g $CuSO_4 \cdot 5H_2O$) in an aqueous solution (1000 cm$^3$ $H_2O$), with the addition of metallic copper. The tests were carried out for 8 h in the boiling solution. Corrosion products were removed from the surface of the samples using a stream of warm water. Persistent corrosion products were removed by etching in a 20% nitric acid solution at RT. The nature of the corrosion fracture of the surface of the samples was studied using a Leica IM DRM optical microscope. The fracture of grain boundaries to a maximum depth of 30 μm is considered a sign of resistance to IGC, while in samples with a thickness of less than 1.5 mm – to a depth of no more than 10 μm.

## 3. Results

### 3.1 Microstructure investigations

Steel in the initial state had a uniform CG austenite microstructure (Fig. 2a), which thin δ-ferrite particles were present in. The average austenite grain size is 20–25 μm. The thicknesses of the δ-ferrite particles were ~3–5 μm (Fig. 2a), the lengths of the ones exceeded 500 μm. The δ-ferrite particles intersect the austenite grains. The OM study and XRD phase analysis results revealed the mass fraction of δ-ferrite in the CG steel to be ~2%. There were some non-metallic particles in the steel microstructure, which may form the agglomerates (Fig. 2a) or be located separately. EDS analysis indicates that Ti is present in the composition of these particles. This suggests that these particles are Ti(C,N) compounds. In Fig. 2a, the largest separate particles are marked by dark lines.

The ECAP process led to the formation of the UFG austenite microstructure and formation of α′-martensite. It has been demonstrated in [35] that steel 321L exhibits a heterogeneous macrostructure following the first cycle of ECAP, characterized by macro-shear bands. The shear bands are aligned with the primary deformation axes during ECAP. The microstructure of UFG steel after $N = 4$ cycles of ECAP is a set of macro shear-bands (Fig. 3b) with an increased content of α′-martensite. Increasing the $N$ from 1 to 4 and increasing the $T_{ECAP}$ from 150°C to 450°C leads to an increase in the degree of the macrostructure homogeneity. However, it is important to note that in the microstructure of steel #450-4, macro-shear bands remain distinctly visible (see Fig. 4b in [35]).

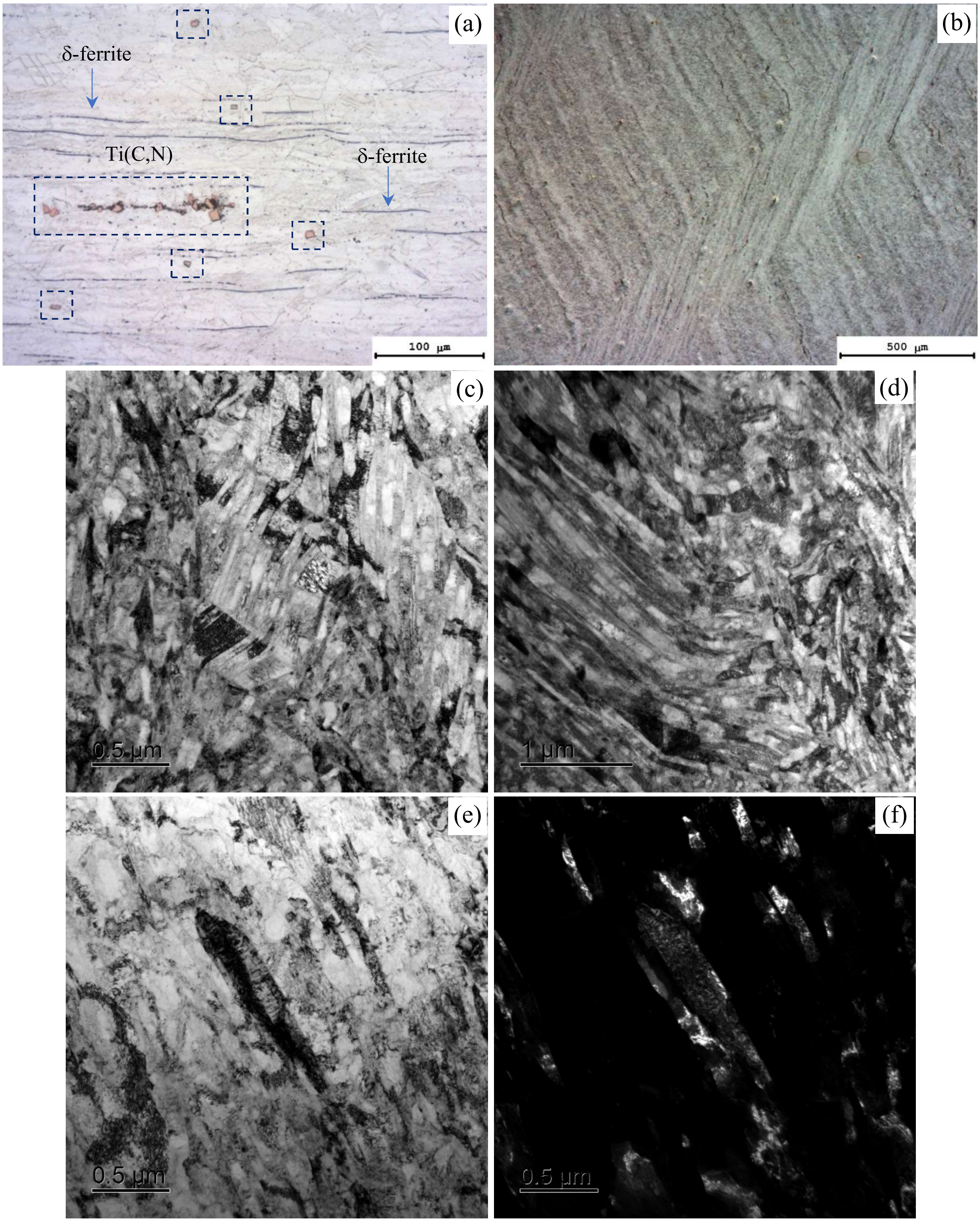

**Fig. 3** Structure of the CG steel (a), steels #150-4 (b, c, d) and #450-4 (e, f). OM (a, b) and TEM (c-f)

As can be seen from Fig. 3c, d, austenitic steel after ECAP at 150°C had a strongly fragmented austenite microstructure. The average sizes of the austenite fragments were 0.3–0.5 μm. The deformation twins are seen inside the grains. The grains in the microstructure of the UFG steel after ECAP at 450°C had more equiaxial shapes (Fig. 3e, f). The average austenite grain sizes of the UFG steels #150-4 and #450-4 were 0.3–0.4 μm and 0.5–0.7 μm, respectively.

ECAP leads to the refinement of the δ-ferrite bands. In the UFG steel structure, the δ-ferrite bands are preserved (Fig. 4a). However, at high magnifications, it is observed that each band consists of numerous individual particles (Fig. 4b). It is interesting to note that metallographic studies reveal large fragments in the structure of the UFG steels (Fig. 4c). The size of these fragments (~20-40 μm) corresponds to the size of austenite grains in the CG steel. It should be emphasized that TEM studies indicate the formation of a highly deformed microstructure with a fragment size of ~0.5 μm (Fig. 3c, d, e, f). Comparison of Fig. 3c, d, e and Fig. 4a indicates that ECAP primarily results in the formation of a highly fragmented structure within the original large austenite grains. At the same time, ECAP does not lead to the transformation of the boundaries of the original austenite grains themselves. Consequently, the steel microstructure after ECAP can be described as large grains of initial austenite, within which a highly fragmented structure is formed. This conclusion aligns well with the results [40].

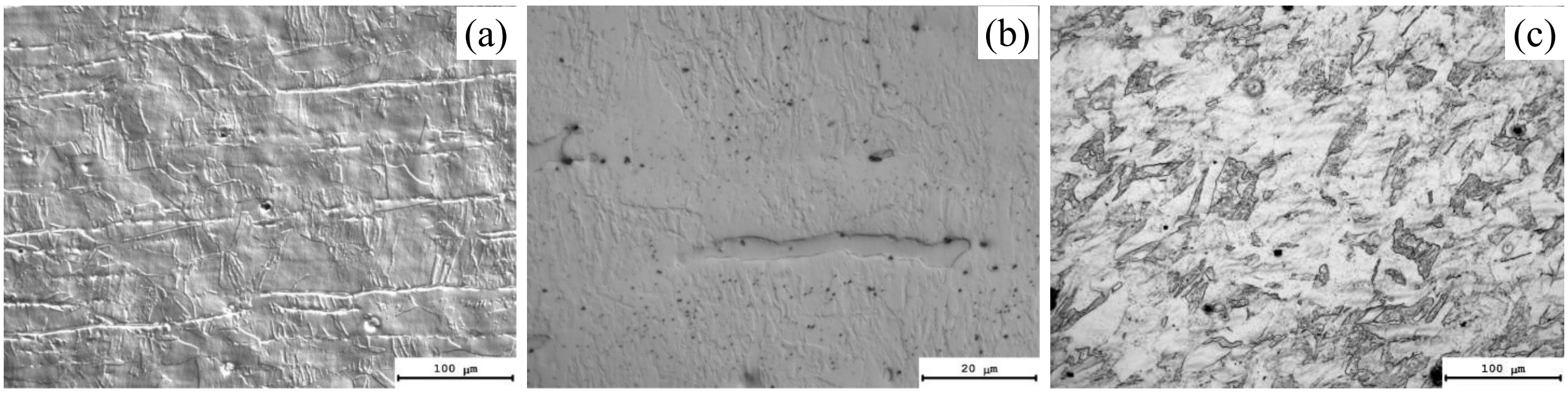

**Fig. 4** δ-ferrite (a, b) in the steel #450-4 and macrostructure of steel #450-2 after ECAP (c): (a, c) low magnification; (b) high magnification. SEM (a) and OM (b, c)

The sizes and the character of spatial distribution of the large non-metallic particles in the UFG steel samples were close to respective parameters of the CG steel.

Note that even after ECAP at 450°C, the steel retained a two-phase austenite-martensite microstructure. XRD patterns of the samples exhibit maxima corresponding to α-Fe (PDF #00-006-0696, ICSD #52258) and γ-Fe (PDF #01-071-4649). As can be seen from Fig. 5, the (110) α-Fe peaks are present in the XRD curves of the CG and UFG steel samples. In the coarse-grained steel, the (110) α-Fe peak is attributed to the presence of the δ-ferrite particles. The mass fractions of α′-martensite after four ECAP cycles at 150°C and 450°C were ~5–15%. The α′-martensite content didn't depend on the $T_{ECAP}$ within the measurement uncertainty that corresponds to the results of [41]. It is not possible to determine the martensite content with greater precision due to the high texture of the UFG samples. No peaks corresponding to ε-martensite (PDF #00-034-0529, ICSD #631723) were observed in the XRD curves of the UFG steel samples.

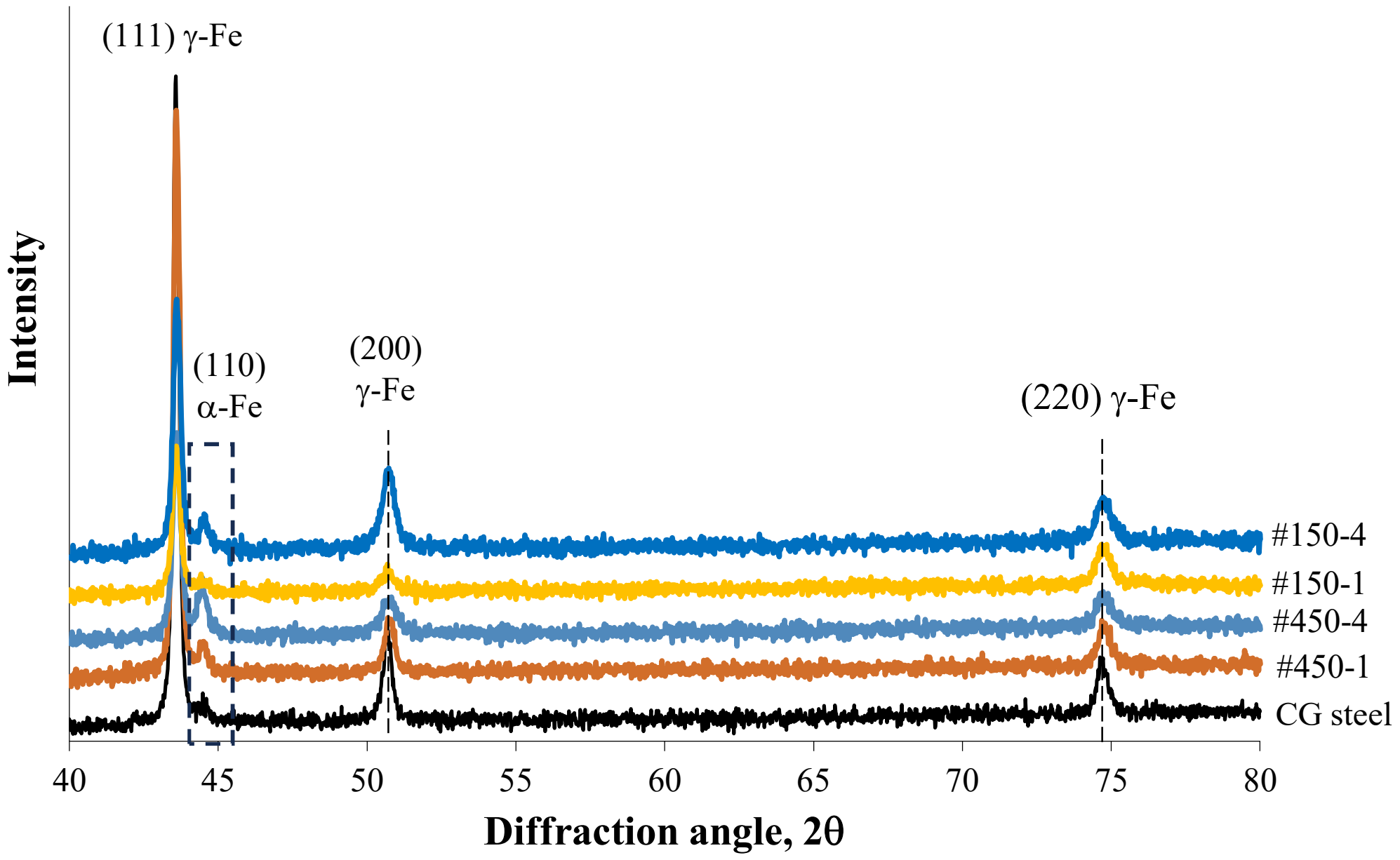

**Fig. 5** Results of XRD analysis of the CG and UFG samples after ECAP at 150°C and 450°C [35]

The heating of the samples up to the temperatures over 650-700°C led to the beginning of the nucleation of the σ-phase particles. Fig. 6 presents the results of *in-situ* investigation of microstructure evolution of the steel #450-4. As can be seen from Fig. 6, the UFG structure remained stable up to 700°C. High level of thermal stability of the UFG structure is provided by the nucleation of the σ-phase particles of ~10–20 nm in sizes. The σ-phase nanoparticles are light in color and equiaxed in shape (Fig. 6). The results of EDS analysis evidenced the nucleated particles to consist of Fe–Cr intermetallic compound. When the heating temperature was increased to 800°C (the holding time 1 h), the sizes of the σ-phase particles increased to 30–50 nm. As can be seen from Fig. 6, the σ-particles were distributed uniformly over the section. Note, that the nucleation of the σ-phase particles during annealing of UFG steel 321L was reported earlier [48, 49]. It is important to note that the most intense formation of σ-phase particles occurs in the shear-bands of localized deformation.

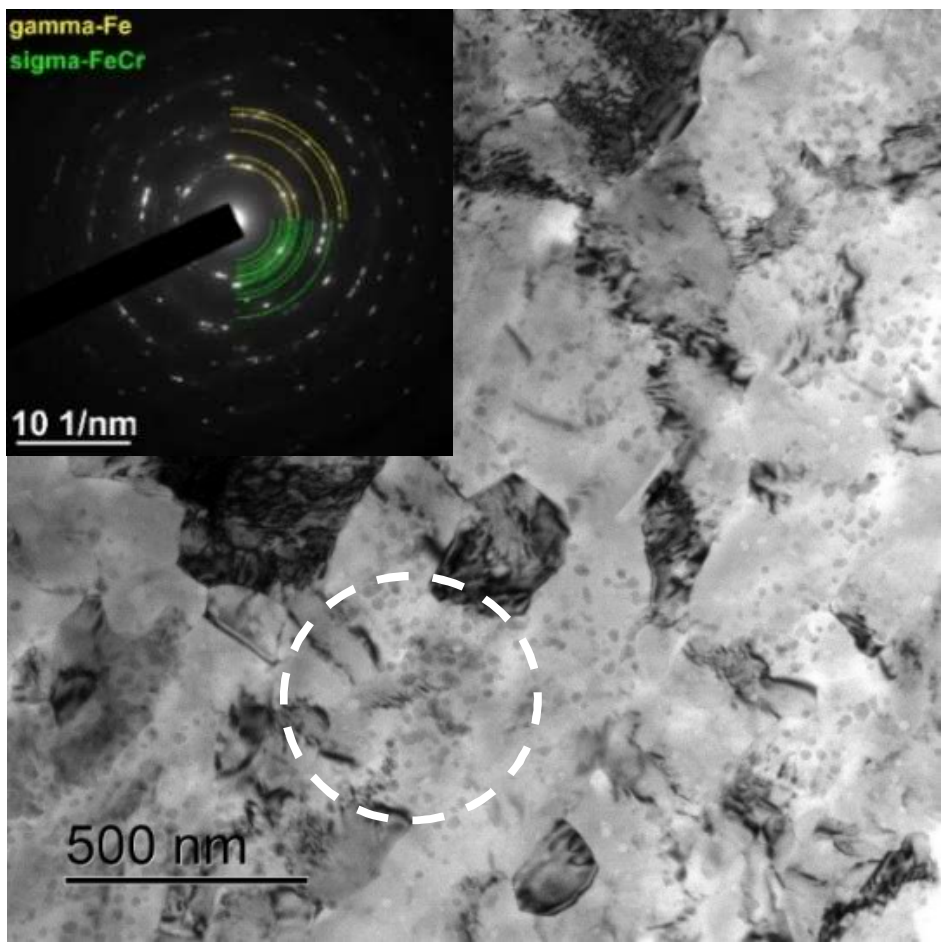

**Fig. 6** Nucleation of the σ-phase particles during heating of the steel #450-4 after heating to 700°C and holding for 1 h. TEM [35]

Grain growth in the UFG steel begins after heating to 750°C (Fig. 7). After 30-minute annealing of the steel #450-4 at 750°C, a uniform recrystallized structure with an average grain size of 2.5 μm is formed, and the maximum size of the recrystallized grains does not exceed 6 μm. The volume fraction of unrecrystallized material after annealing at 750°C for 30 min does not exceed 3% vol. (Fig. 6c). The heating to 900°C leads to an increase in the average austenite grain size to 4 μm and the maximum austenite grain size to 12 μm (Fig. 7d). It is important to note that in the UFG steel produced at a lower ECAP temperature, a heterogeneous microstructure is formed during annealing. This microstructure consists of areas with a fully recrystallized fine-grained structure (Fig. 7a) alternating with areas of partially recrystallized structure (Fig. 7b).

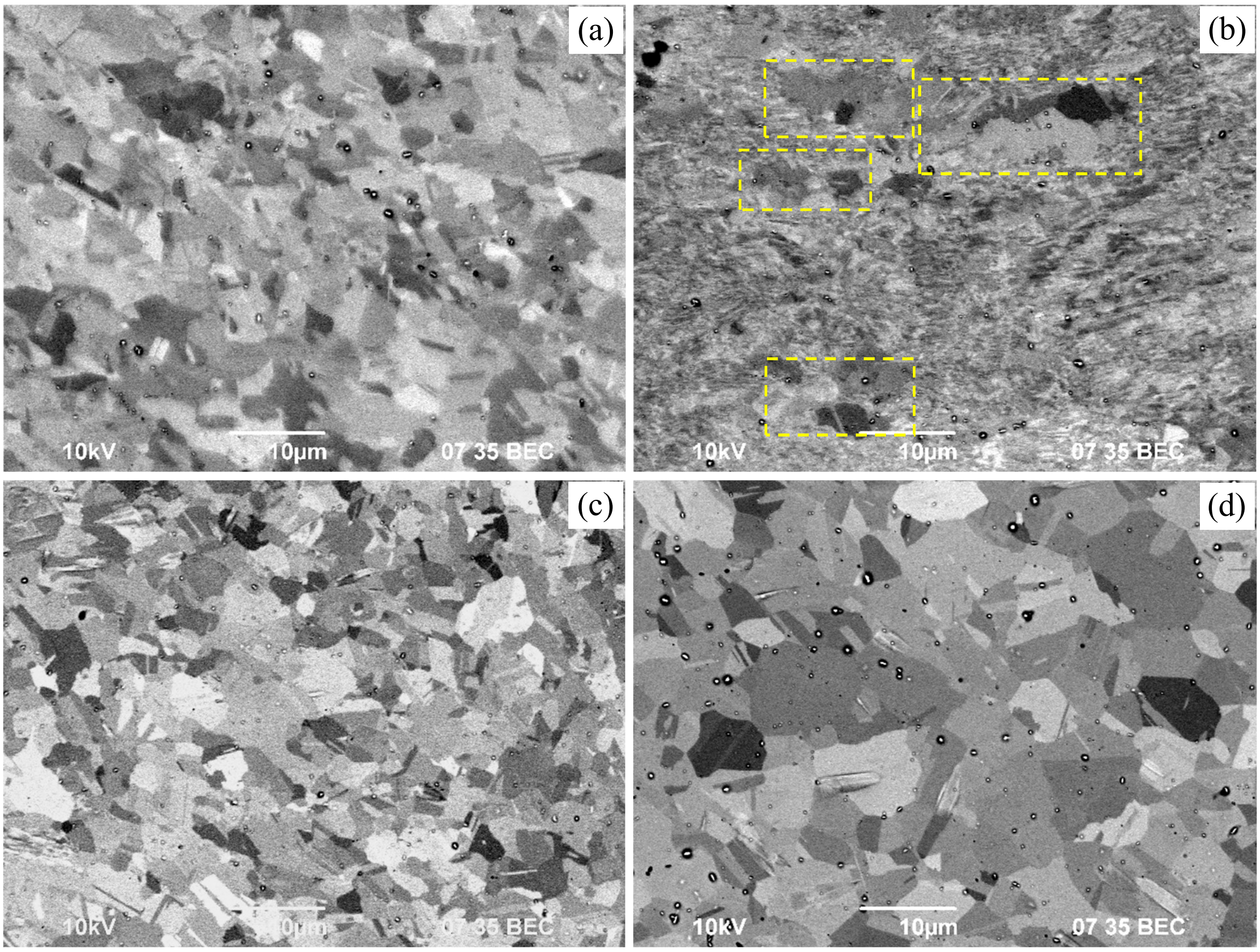

**Fig. 7** SEM-images of the structure of the UFG steel after 30-min annealing at 750°C (a, b, c) and 900°C (d). (a, b) series #450-3; (c, d) series #450-4. The areas of recrystallized structure are marked with a dashed line in Fig. 7b

3.2 Mechanical testing at RT

The curves σ(ε) of the CG steel specimens and the UFG ones at RT are presented in Fig. 8. The σ(ε) curve of the CG steel had a conventional form, with a long strain hardening stage. The duration of the uniform plastic deformation stage in the CG steel is relatively long ($\delta_u$ ~ 60–65%).

The values of the UTS of the investigated CG steel samples (UTS = 720 MPa) exceeded the usual ones for commercial steel AISI 321L (UTS = 490–520 MPa) essentially. In our opinion, high strength of steel 321 originates from the presence of the δ-ferrite particles (Fig. 3a), which played the role of stoppers of the deformation propagation inside the sample during the tensile testing (see also [35]).

**Table 2.** Mechanical properties and IGC resistance of steels after tension testing at RT

| Steel | | Calculation of characteristics by engineering curves σ(ε) | | | Microhardness *HV*, GPa | | IGC resistance *K* | |
|---|---|---|---|---|---|---|---|---|
| | | UTS, MPa | $\delta_u$, % | El, % | Zone I | Zone II | Zone I | Zone II |
| CG state | | 690–720 | 80–90 | 90–125 | 2.0–2.1 | 3.65–3.7 | $9.0 \cdot 10^{-3}$ | $56 \cdot 10^{-3}$ |
| UFG state | #150-1 | 850–900 | 20–25 | 55–70 | 3.05-3.35 | 4.0–4.10 | $9.0 \cdot 10^{-3}$ | $69 \cdot 10^{-3}$ |
| | #150-2 | 955–970 | 15–20 | 50–65 | 3.55–3.7 | 4.25–4.4 | $2.8 \cdot 10^{-3}$ | $87 \cdot 10^{-3}$ |
| | #150-3 | 1080–1100 | 20–25 | 35–40 | 3.85–3.9 | 4.45 | $3.6 \cdot 10^{-3}$ | $47 \cdot 10^{-3}$ |
| | #150-4 | 1090–1100 | 25 | 40–45 | 3.95–4.0 | 4.5–4.55 | $3.3 \cdot 10^{-3}$ | $38 \cdot 10^{-3}$ |
| | #450-1 | 825–860 | 30–35 | 65–75 | 3.0–3.1 | 3.95–4.10 | – | – |
| | #450-2 | 930–950 | 25–30 | 60–70 | 3.35–3.55 | 4.2–4.35 | $3.5 \cdot 10^{-3}$ | $92 \cdot 10^{-3}$ |
| | #450-3 | 950–960 | 25–30 | 55–65 | 3.5–3.6 | 4.35–4.4 | $3.8 \cdot 10^{-3}$ | – |
| | #450-4 | 1010–1020 | 20–25 | 50–60 | 3.55–3.6 | 4.4–4.45 | $3.5 \cdot 10^{-3}$ | $35 \cdot 10^{-3}$ |

Near the fracture region of the CG samples, multiple elongated pores oriented along the austenite grain boundaries are observed (Fig. 9a). In contrast, away from the fracture region, no pores or cracks along the grain boundaries were detected. The surface of the deformed area of CG steel samples after tensile testing is characterized by the presence of roughness in the largest grains (Fig. 9b). This indicates the formation of strain twins and martensite in the CG steel following tensile tests conducted at RT (Fig. 9b). The formation of martensite is likely responsible for the extended strain hardening stage observed on the σ(ε) curve for the CG steel.

An indirect indicator of martensite formation is a significant increase in the steel hardness in the fracture region after tensile testing at RT. The CG steel microhardness in Zone II is 3.3 GPa. This value is ~30% higher than the CG steel hardness (2.0-2.1 GPa) and is comparable to the UFG steel microhardness (~3.35 to 4.0 GPa, see Table 2). Since the contribution of dislocations to the strengthening of austenite is small [50], the observed increase in microhardness by $\Delta HV_{CG}$ = 1.55–1.70 GPa is primarily associated with the formation of martensite during the tensile testing of the samples.

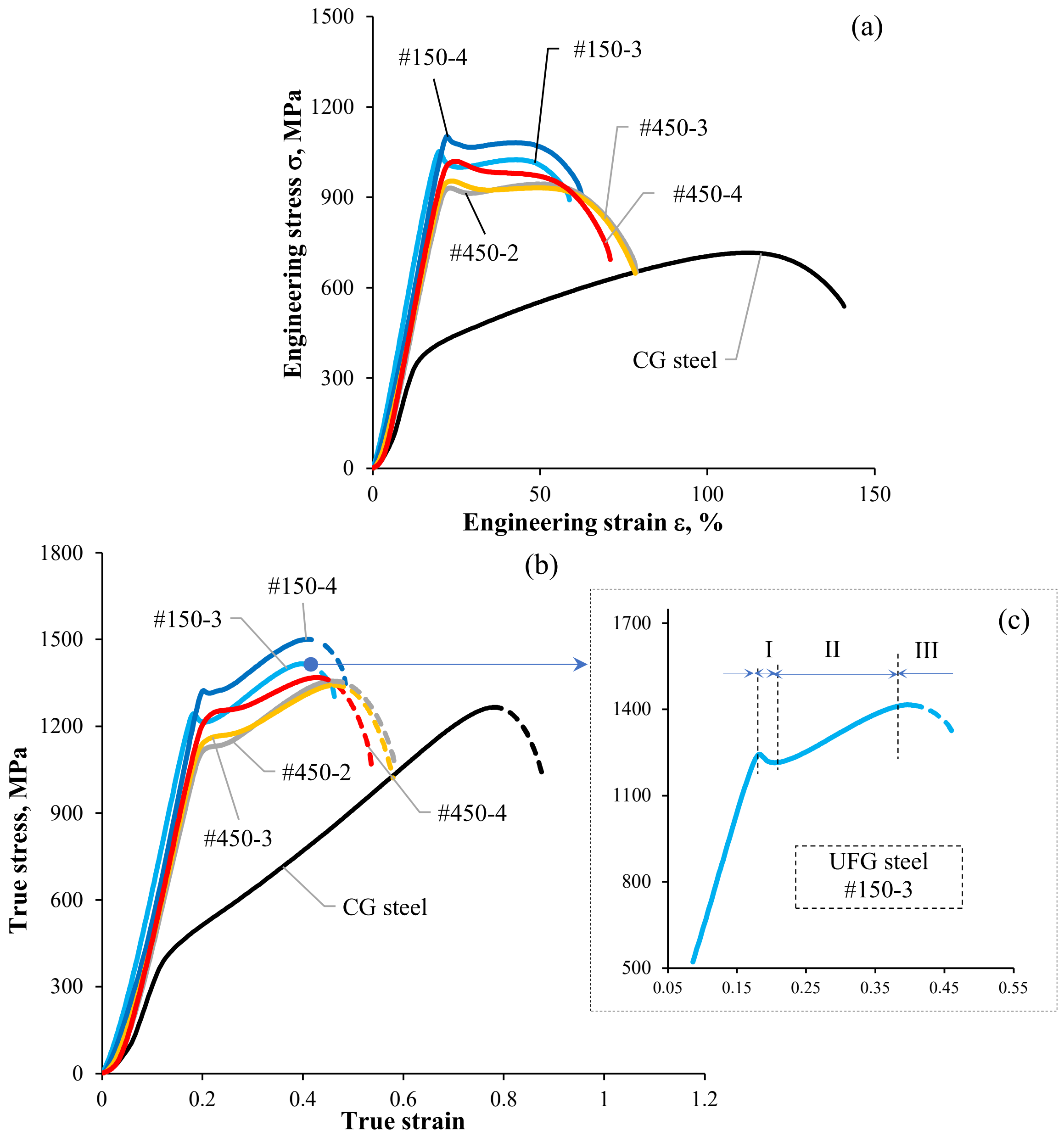

**Fig. 8** Stress–strain curves for the steel samples at RT: (a) engineering stress – engineering strain curves; (b) true stress – trues strain curves; (c) curve σ(ε) fragment

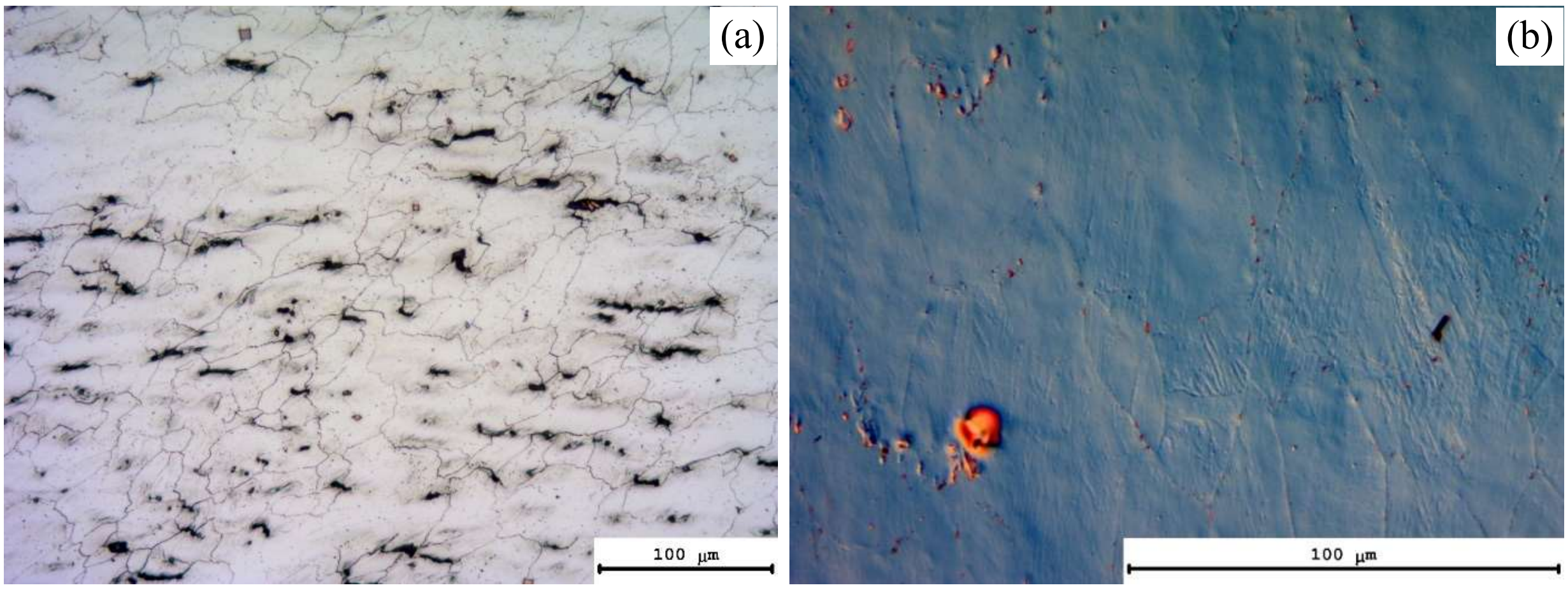

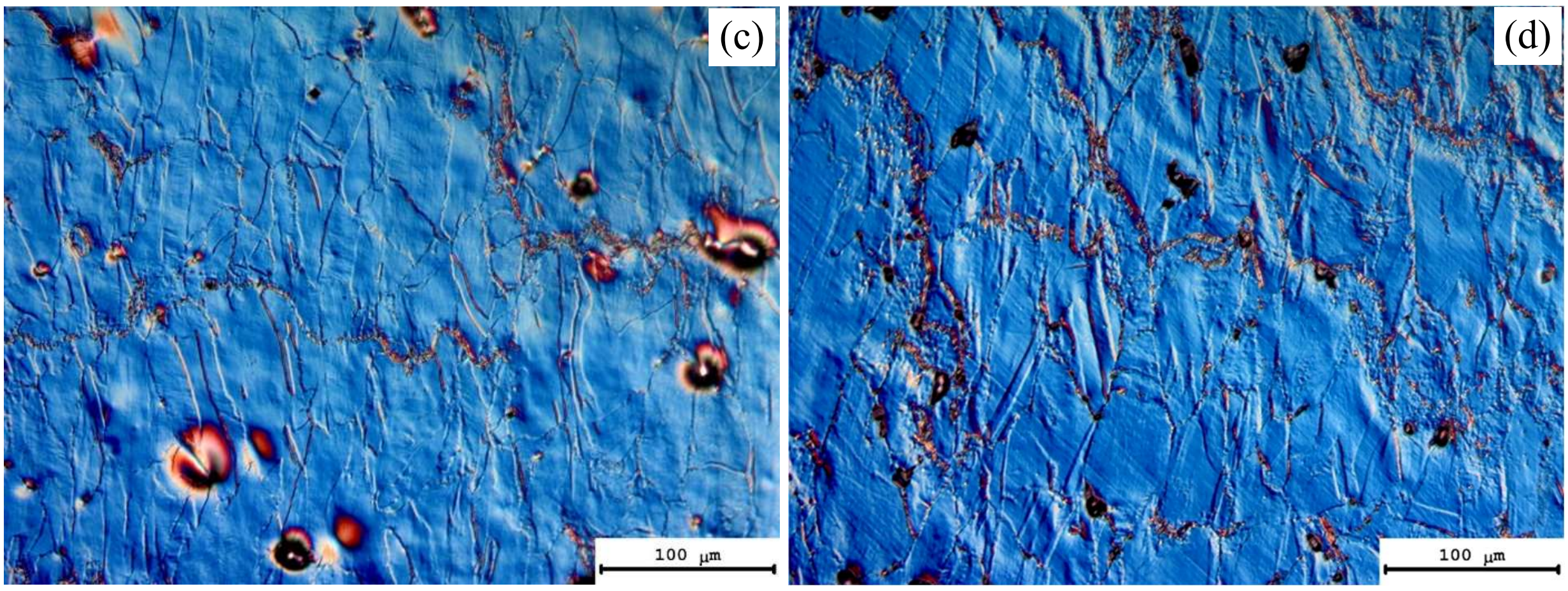

**Fig. 9** Microstructure of the CG steel sample in the vicinity of the fracture area (a) and deformed area (b, c, d) after testing at RT (a, b), 750ºC (c), and 800ºC (d). OM

For all UFG samples a non-monotonous character of the σ(ε) curves was observed at the initial stages of the unstable plastic flow (Fig. 9a, b). The stage of rapid stress decreasing in 30–40 MPa (Stage I) is changed by a strain hardening stage (Stage II) of the UFG material (Fig. 8c). In the stress–strain curves σ(ε) of the UFG steels, there are a short stage of strain hardening (Stage II), which transforms into a stage of localization of the strain (Stage III). The uniform strain degree in the UFG steel ($\delta_u$ ~ 25–30%) is less than in the CG steels. Increasing the ECAP temperature from 150 to 450°C resulted in an increase in the duration of the strain hardening stage (Fig. 9a, b).

The formation of the UFG structure by ECAP ($N$ = 4, 150°C) to result in a decrease in the El of the steel 321L from 90–125% to 40–45% and an increase in the UTS from 690–720 MPa to 1080–1100 MPa (Table 2). ECAP at 450°C resulted in a decrease in the UTS to 1010–1020 MPa and an increase in the elongation to failure to $\delta_{max}$ ~ 50–60%.

The increased spread of characteristics is observed in the heterostructured steel obtained after a low number of the ECAP cycles ($N$ = 2). This spread is primarily due to the formation of a heterogeneous structure during ECAP, which is characterized by the presence of large shear bands (Fig. 2b, [35]). The structure heterogeneity decreases, and the spread of characteristics becomes smaller with an increase in the number of ECAP cycles and ECAP temperature.

Intense fragmentation of δ-ferrite is observed on the surface of the UFG steel samples after testing at RT (Fig. 10). However, there is no significant formation of pores and/or cracks at the δ/γ interfacial boundaries or on the Ti(C,N) particles (Fig. 10b). Some bands of δ-ferrite fragment into individual particles that are oriented at an angle of ~45° to the axis of tension. These particles are indicated by dotted lines in Fig. 10b. Metallographic studies using interference contrast indicate that the deformation process of the UFG steel proceeds heterogeneously. On the grinding surface, areas of highly deformed metal are visible, which alternate with areas of slightly deformed metal (Fig. 10c).

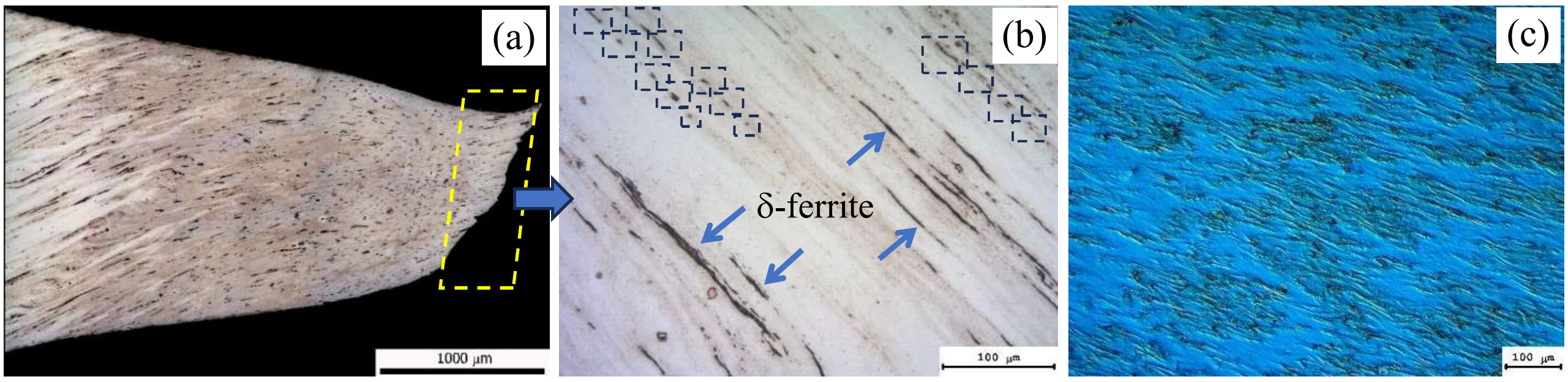

**Fig. 10** OM-images of the fracture region of the sample #450-3 (a, b) and #150-3 (c) after testing at RT

Anticipating the results of studies on the fractures of samples, it is important to note that the fracture of the UFG steel samples occurs at an angle of 45° to the direction of stretching (Fig. 1b). In contrast, the fracture of the CG samples occurs in a conventional manner, characterized by a uniform reduction in the cross-sectional area of the working part of the sample.

The fractographic analysis revealed three characteristic zones on the fractures of both the CG steel specimens and the UFG ones after the tensile tests (Fig. 11). These are: the dimples zone (Area #1 in Fig. 11a), the radial one (Area #2 in Fig. 11a), and the break zone (Area #3 in Fig. 11a). It is worth noting that the break zone in the CG steel occupied ~50% of the whole fracture area. In the UFG steel after ECAP ($N$ = 4), the break zone occupied ~70%. So far, the formation of the UFG microstructure resulted in an increase in the break zone fraction of the fracture area and, hence, in a decrease in the ductile component of the fracture.

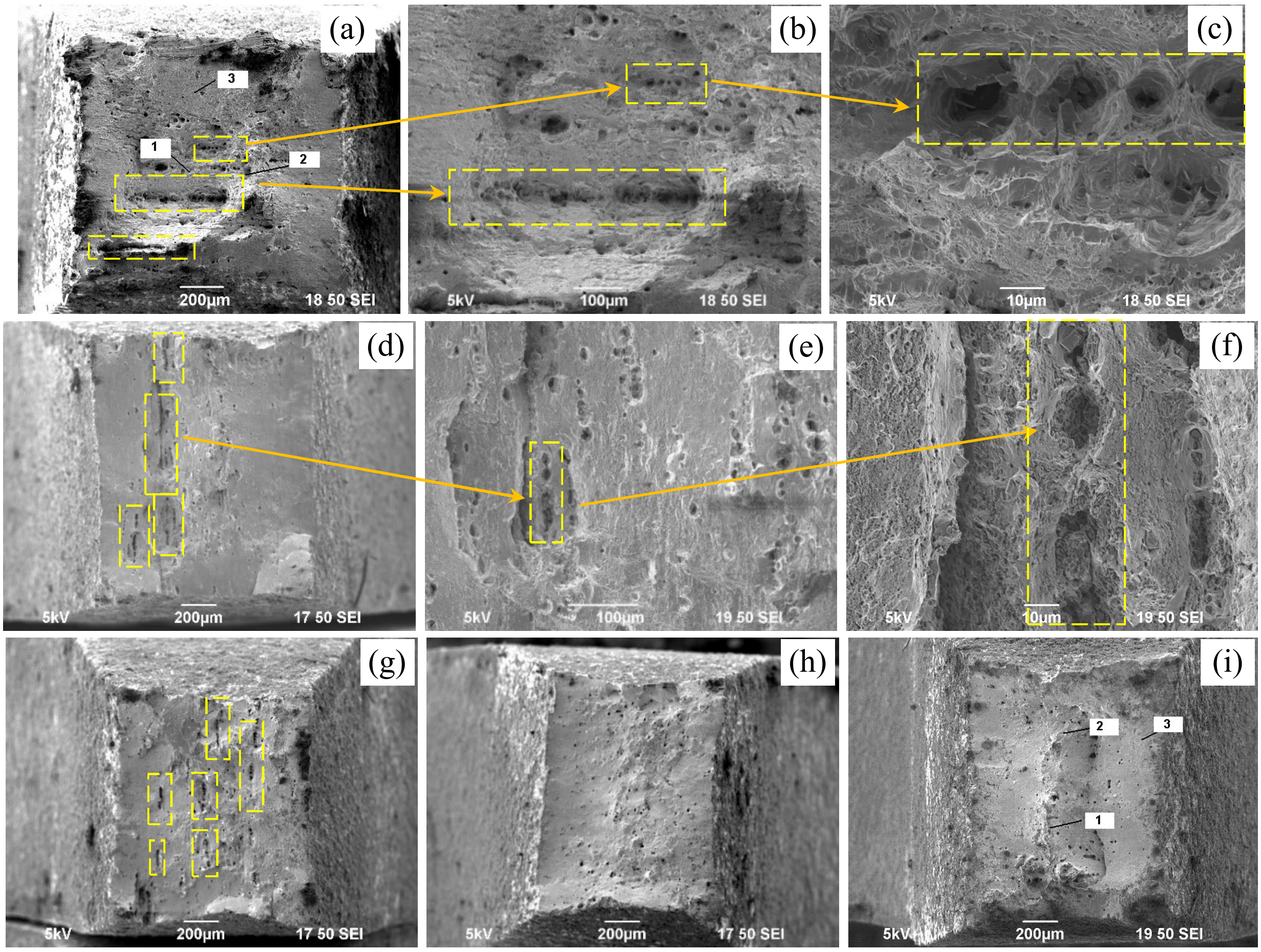

**Fig. 11** SEM-images of the fractures of the CG (a, b, c) and UFG (d, e, f, g, h, i) samples after the tensile tests at RT. Fig. 11d, e, f present the fractures of the sample #150-4, Fig. 11g, h, i – the fractures of the sample #450-2 (g), #450-3 (h), #450-4 (i)

The elongated voids up to 0.5 mm long on the fractures of the CG samples are marked by yellow dashed lines in Fig. 11a. These voids formed probably in the areas of δ-ferrite location. Also, some separate voids of 10–20 μm in size are seen clearly on the fractures of the samples. These voids are formed likely due to the non-metallic inclusions. Some separate voids of 5–10 μm in size are seen mainly on the surfaces of the fractures of the UFG samples after ECAP at 450°C (Fig. 11g, h, i). Sometimes, areas containing fragmented δ-ferrite particles are observed after ECAP at 150°C. These areas are marked by white dashed lines in Fig. 11d, e, f. It is noteworthy that as the number of ECAP cycles increases, the quantity of large voids associated with the degradation of δ-ferrite particles decreases (Fig. 11g, h, i). This finding indirectly suggests that a higher strain during ECAP facilitates the refinement of δ-ferrite particles. Consequently, the results of fracture studies on the samples indicate a significant influence of δ-ferrite and Ti(C,N) particles on the fracture of the UFG steel at RT.

It is also worth paying attention to the *HV* values for the UFG samples after tensile testing (Table 2). Table 2 shows that tension at RT leads to an increase in *HV*. However, the magnitude of this change $\Delta HV = HV_{\mathrm{ZoneII}} - HV_{\mathrm{ZoneI}}$ is much smaller in the UFG steels than in the CG steel ($\Delta HV_{\mathrm{CG}}$ = 1.55–1.7 GPa). Increasing the number of ECAP cycles at 150°C leads to a decrease in $\Delta HV_{\mathrm{UFG}}$ from 0.65–1.0 to 0.5–0.6 GPa. In samples processed at 450°C, a decrease in $\Delta HV_{\mathrm{UFG}}$ from 0.85–1.1 to 0.8–0.9 GPa is observed. Thus, the UFG steels exhibit a lower tendency toward martensite formation, which is consistent with the findings in [30]. Increasing the degree of deformation and lowering the temperature of ECAP leads to an increase in microhardness, but reduces the extent of strain hardening in the UFG steel during subsequent tension at RT (Table 2).

### 3.3 Tensile testing at elevated temperatures

Fig. 12 presents the true stress–true strain curves σ(ε) for the tests at elevated temperatures. The σ(ε) curves in engineering stress–engineering strain coordinates are presented in Appendix B. The σ(ε) curves for the CG steel specimens had the form typical for high-ductility materials (Fig. 12a). The duration of the localized strain stage was much smaller than of the strain hardening one. The duration of the strain hardening stage decreases, and the slope of the σ(ε) curve becomes less steep as the test temperature increases. The stage of strain hardening is virtually absent from the σ(ε) curves for the CG steel at test temperatures of 700 and 800°C. The stage of uniform plastic flow is observed immediately after the end of the elastic deformation stage (Fig. 12a).

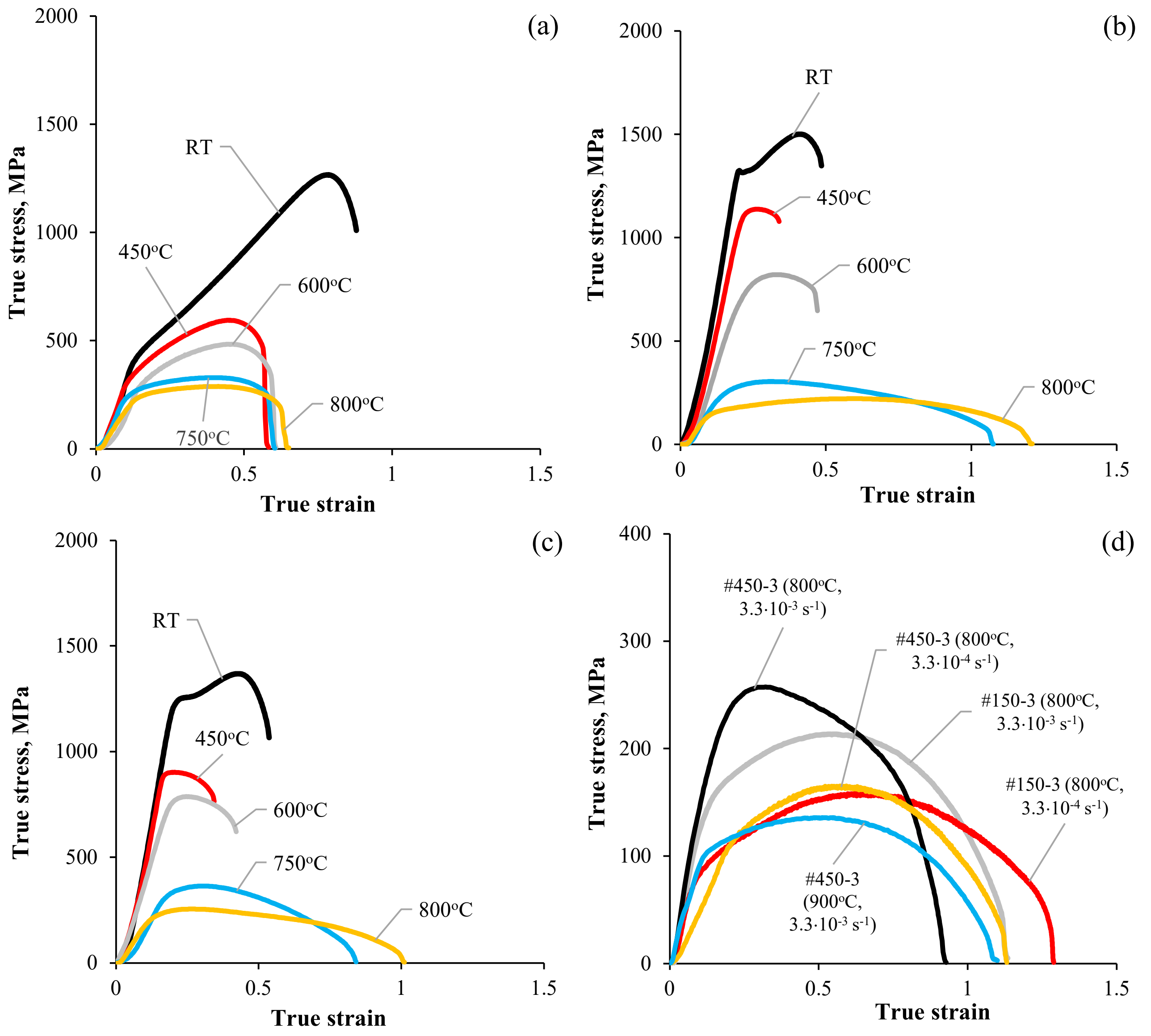

**Fig. 12** True stress–true strain curves for the CG (a) and UFG (b, c, d) steel at elevated temperatures: (b) steel #150-4; (c) steel #450-4; (d) analysis of the effect of strain rate on the ductility of the UFG samples ($N = 3$)

The σ(ε) curves for the UFG steel specimens at the test temperatures of 750°C and 800°C had the form non-typical for highly ductility materials – the stage of an elastic strain transformed into a long stage of unstable localized strain flow (Figs. 12b, c). There is no distinct stage of deformation hardening evident in the σ(ε) curves for the UFG steels.

Analysis of the results of tensile tests at elevated temperatures demonstrated increasing the temperature from RT to 750°C to result in a monotonic decrease in the UTS from 720 MPa to 250 MPa for the CG steel and from 950–1100 MPa to 240–290 MPa for the UFG steel, respectively (Appendix B). No unstable plastic flow regions were observed in the high-temperature σ(ε) curves for the CG and UFG steels (Fig. 12). It is interesting to note that the UTS of the CG steel ($\sigma_b$ = 250 MPa) exceeds the UTS of the UFG steel ($\sigma_b$ = 150–205 MPa) at elevated test temperatures (800, 900°C). It is assumed that the primary reason for this effect is the presence of large δ-ferrite particles in the CG steel. During ECAP, the intense grinding of the δ-ferrite particles occurs, leading to a significant reduction in their contribution to the UTS of austenitic steel.

Fig. 13 shows the dependence of El and UTS on test temperature. Note that increasing the test temperature resulted in a non-monotonic variation of the El for the UFG steel that differs from the same dependencies for the CG steel. The analysis of the data presented in Fig. 13a shows that the El for the CG steel monotonically decreases from 125 to 70% with increasing the test temperature from RT to 750°C. At the test temperature of 800°C, the ductility of the CG steel increased insufficiently ($\delta_{max}$ ~ 90%). As already mentioned in the Introduction, the non-monotonic (with a minimum) character of the $\delta_{max}(T)$ dependence is typical for the metastable CG austenitic steel.

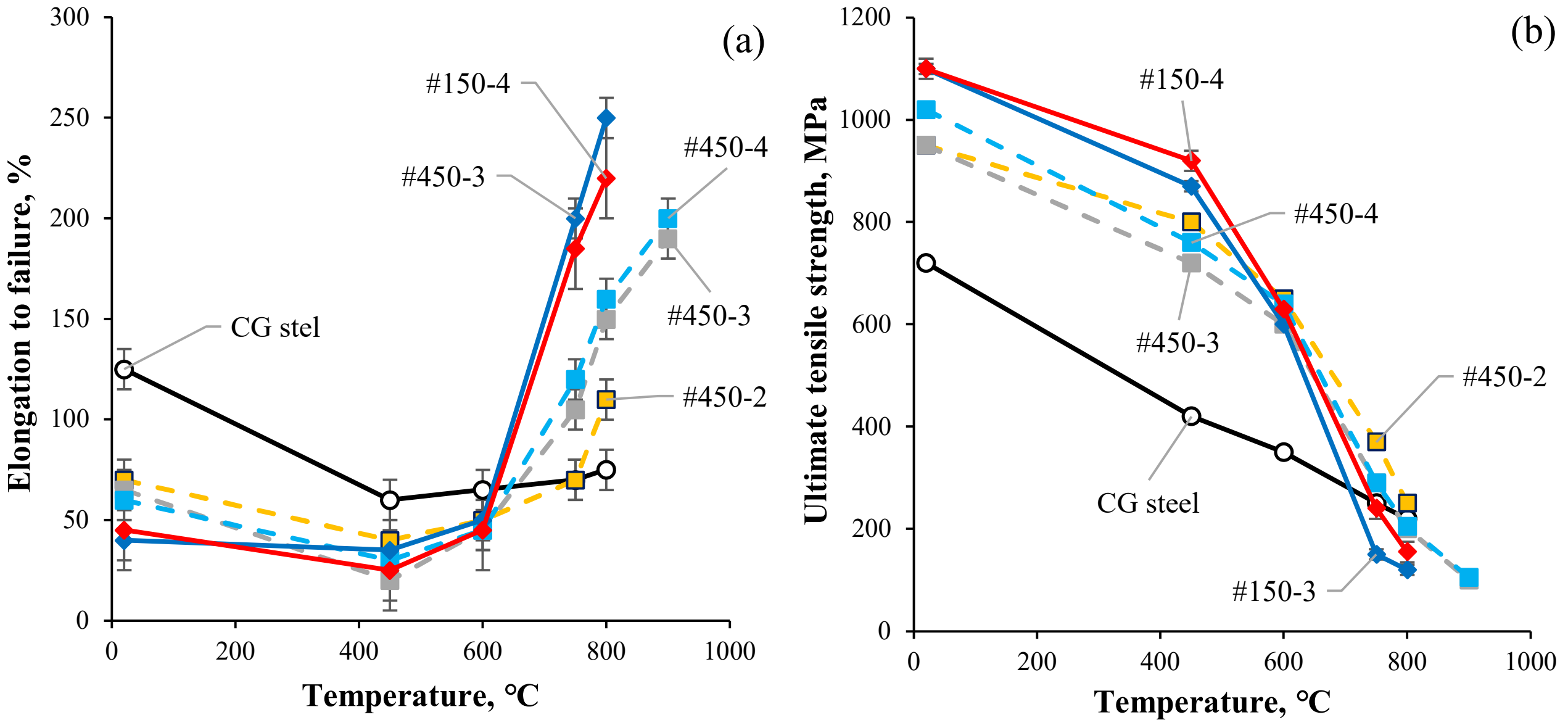

**Fig. 13** Effect of test temperature on El (a) and UTS (b) of the CG and UFG steels. $\dot{\varepsilon}$ = 3.3·10$^{-3}$ s$^{-1}$

The character of the $\delta_{max}(T)$ dependence for the UFG steel was more complex. The El decreased with increasing the test temperature from RT to 450°C (Fig. 13a). The increasing of the number of ECAP cycle resulted in a decrease in the El at 450°C. At further increasing the temperature from 450 to 750–800°C, the El of the UFG steel increased and was several times higher than $\delta_{max}$ of the CG steel. For the UFG steel specimens obtained by ECAP at 150°C, the elongation to failure at the test temperature of 800°C reached 250%.

The analysis of the data presented in Fig. 13 indicates that the UFG steels fabricated at a reduced ECAP temperatures ($T_{ECAP}$ = 150°C) exhibit higher El and lower UTS at 600–900°C. Samples of the UFG steels fabricated using the ECAP method at $T_{ECAP}$ = 450°C exhibit lower elongations to failure and higher UTS values. It is interesting to note that at test temperatures exceeding 750°C, the flow stresses of the UFG steels are lower than those of the CG steel. The only exception is the heterostructured steel, which was produced by two ECAP cycles.

To analyze the impact of the initial grain size on the superplasticity characteristics, UFG steel samples of the #450-4 series were subjected to 30-minute preliminary annealing at temperatures of 750, 800, and 900°C. Due to the annealing temperature, the average grain size of austenite in highly deformed steel varied (Fig. 7). The annealed samples were tested at a temperature of 600°C with a strain rate of $\dot{\varepsilon}$ = 3.3·10$^{-4}$ s$^{-1}$. The annealing temperature of 600°C corresponds to the end of the reverse transformation of martensite into austenite in steel 321. However, it does not exceed the temperature at which grain growth begins in the UFG steel 321. This allows for the exclusion of the effects of strain-induced martensite and grain growth on the ductility of UFG steel. The unannealed sample was maintained at 600°C for 30 minutes in the working chamber of the testing machine. Consequently, the unannealed sample effectively corresponded to an annealing temperature of 600°C. The σ(ε) curves are presented in Fig. 14, and Table 3 summarizes the test results.

**Table 3**. The effect of the annealing temperature on the deformation behavior of the UFG steel at a temperature of 600°C, $\dot{\varepsilon}$ = 3.3·10$^{-4}$ s$^{-1}$

| Annealing temperature, $T_{ann}$, °C | #450-1 | | #450-2 | | #450-3 | | #450-4 | |
|---|---|---|---|---|---|---|---|---|
| | UTS, MPa | El, % | UTS, MPa | El, % | UTS, MPa | El, % | UTS, MPa | El, % |
| Without annealing | 540 | 21 | 545 | 25 | 480 | 21 | 560 | 25 |
| 750°C | 510 | 18 | 470 | 17 | 360 | 26 | 345 | 52 |
| 800°C | 440 | 22 | 305 | 68 | 300 | 58 | 320 | - |
| 900°C | 310 | 58 | 285 | 70 | 310 | 71 | 320 | 64 |

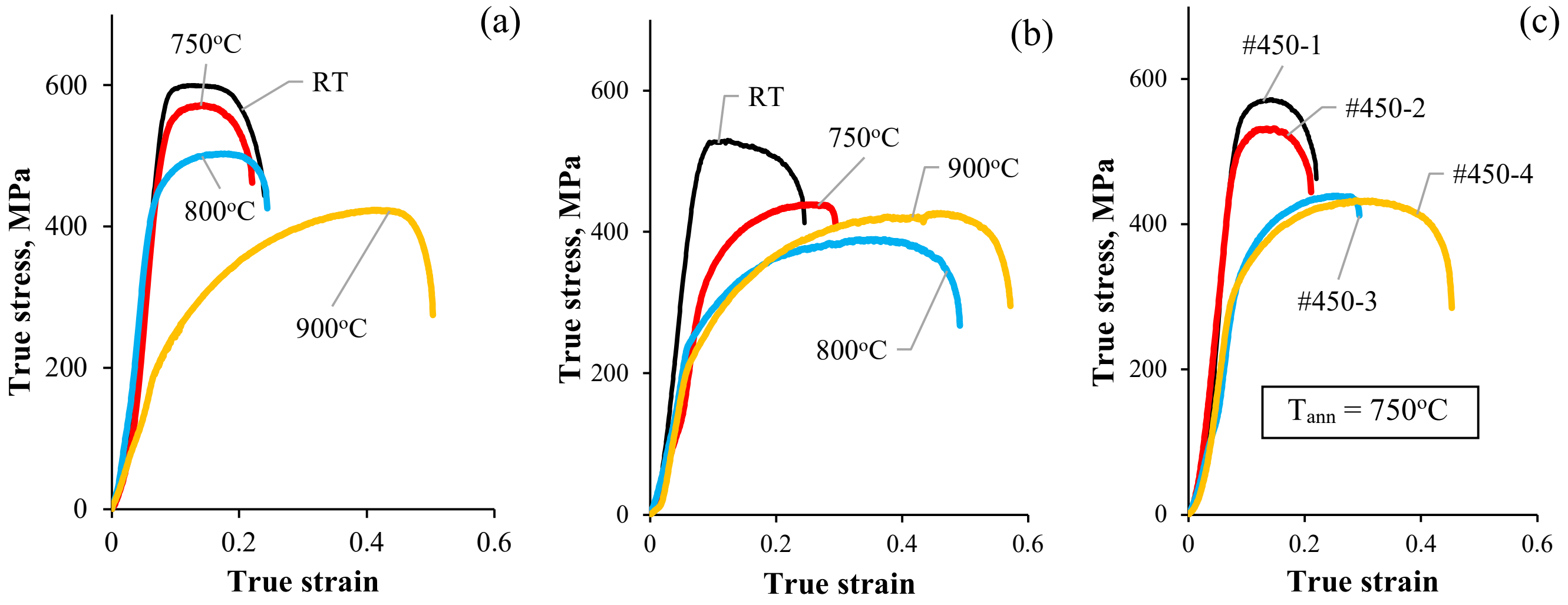

**Fig. 14** Curves of σ(ε) for the UFG steel samples obtained by the preliminary annealed at various temperatures: (a) #450-1; (b) #450-3. Tests at 600°C, $\dot{\varepsilon}$ = 3.3·10$^{-4}$ s$^{-1}$. The temperatures of 30-minute annealing (Fig. 13a, b) are presented in the corresponding figures. Fig. 14c shows the σ(ε)curves for samples subjected to different number of ECAP cycles at 450°C and subsequent annealing at 750°C

Analysis of the data presented in Table 3 shows that preliminary annealing leads to a decrease in the tensile strength of UFG steels, regardless of the deformation treatment regime. Preliminary annealing has a different impact on the deformation behavior of heterostructured steels (series #450-1, #450-2) compared to steels with a more homogeneous UFG structure (series #450-3, #450-4). Annealing at 750°C leads to a decrease in the ductility of heterostructured steels to 17-18%. A further increase in the annealing temperature to 900°C leads to an increase in the ductility of heterostructured steels to 58% (steel #450-1) and 70% (steel #450-2). An increase in the annealing temperature of the steels with a homogeneous UFG microstructure leads to a monotonic increase in the ductility from 21-25% to 71% (steel #450-3) and 64% (steel #450-4) (Table 3).

Secondly, it is important to note that preliminary annealing influences the shape of the curves of σ(ε). As can be seen from Fig. 14a, b, the σ(ε) curves for unannealed samples have an extended stage of localized deformation. The stage of uniform plastic flow is virtually absent from these curves. Preliminary annealing at temperatures above 750-800°C leads to the appearance of a uniform plastic flow stage on the σ(ε) curves. The uniform deformation stage on the σ(ε) curves becomes evident only after annealing at 900°C (Fig. 14a) in heterostructured steels (series #450-1, #450-2). In contrast, the uniform plastic flow stage is observed after annealing at 750°C (Fig. 14b) in steels with an homogeneous UFG microstructure. The influence of the number of ECAP cycles on the shape of the σ(ε)curve can be conveniently illustrated using Fig. 14c, which shows the σ(ε) curves for samples that were preliminary annealed at 800°C. As can be seen from Fig. 14c, an increase in the number of ECAP cycles leads to an increase in the uniform deformation stage.

We will analyze the fracture behavior of the CG and UFG samples during SP testing.

The fractures of the CG steel samples after tensile testing at elevated temperatures had a viscous character (Fig. 15). The central dimples zones of the sample fractures (Area #1) after tests at 450–600°C comprised sets of small voids of 5–20 μm in size (Fig. 15a, b). There were few deep pores in the dimple zones, the shapes of which were close to the spherical ones. Some of these pores are marked with white dotted lines in Fig. 15b. Large elongated voids associated with the fracture of δ-ferrite particles were not observed, but large spherical pores up to 100-150 μm in size are visible on the fracture surface (Fig. 15a). It is important to note that the fractures of the CG steel samples remained viscous even after the tensile tests at 450–600°C, although the ductility of the CG steel at these temperatures was lower than the one at RT. After the tests at 750–800°C, the areas of the radial zones of the fractures (Area #2) increased essentially whereas the areas of the break zones (Area #3) tend to zero. The sizes and depths of the dimples in the central zones of the CG steel fractures increased essentially after the tests at 750–800°C, the coalescence of the dimples into the elongated voids up to 50-150 μm long was observed. Such areas are marked with yellow dotted lines in Fig. 15c.

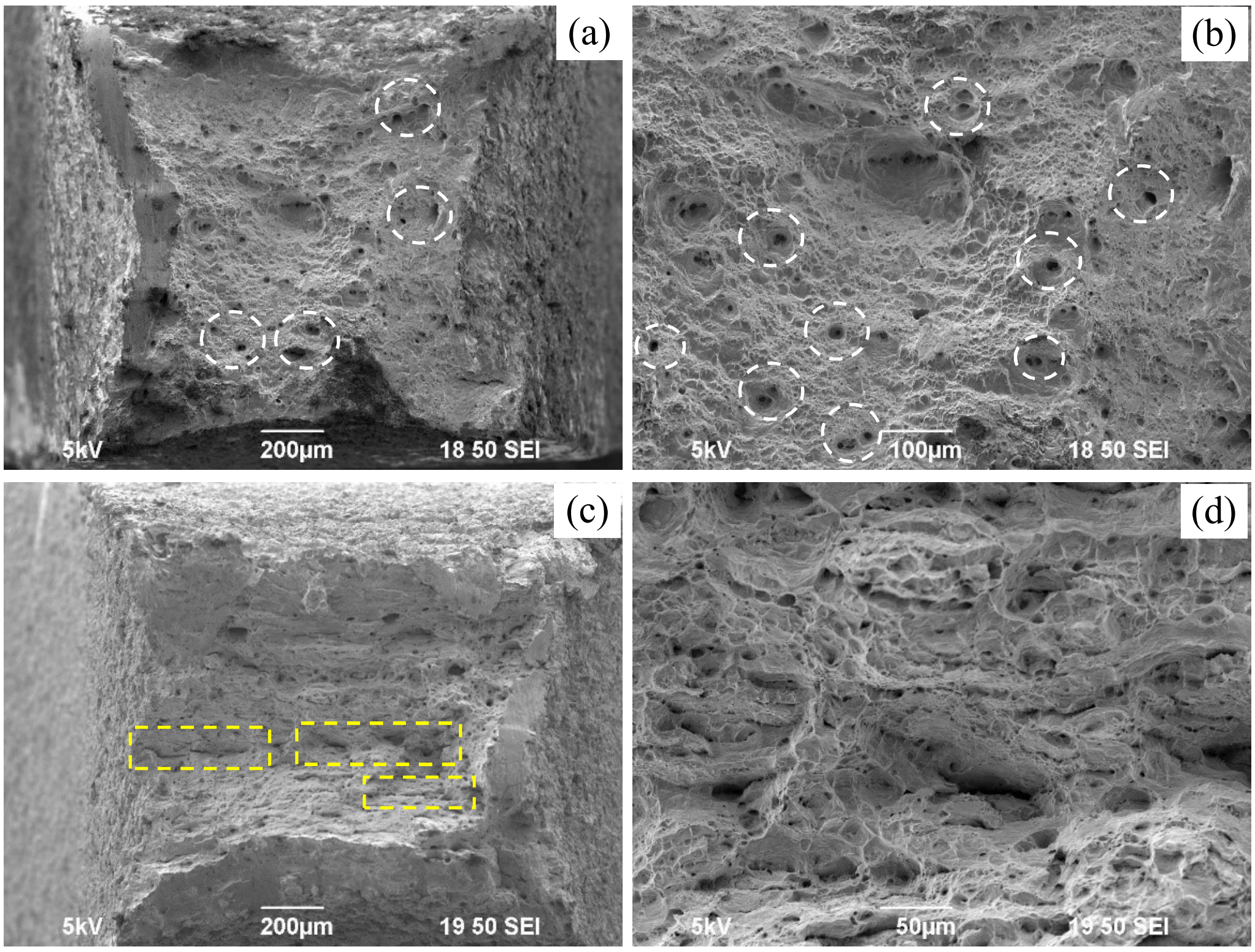

**Fig. 15** SEM-images of the CG steel fractures after the tests at 450°C (a, b) and 750°C (c, d). Fig. 15a, c shows a general view of the fracture, Fig. 15b, d presents an enlarged image of the central part of the fracture. $\dot{\varepsilon} = 3.3{\cdot}10^{-3}$ s$^{-1}$

The investigations of the UFG samples after testing at elevated temperatures evidenced an intensive pore formation in the Zone II (Fig. 16). An increase in the temperature results in an increase in the pore size. The pore size increases proportionally to the strain along the working part of the sample (Appendix A). The lengths of such pores reach ~100–150 µm. Some such defects are marked with solid white lines in Fig. 16a. Large elongated defects, the size of which can reach 100-500 µm, are observed near the fracture of the UFG samples. A fragment of such a defect is marked with a white dotted line in Fig. 16b. In the deformed zone, small micron-sized pores distributed uniformly over the UFG sample surfaces are present also. Such defects are marked with yellow dotted lines in Fig. 16.

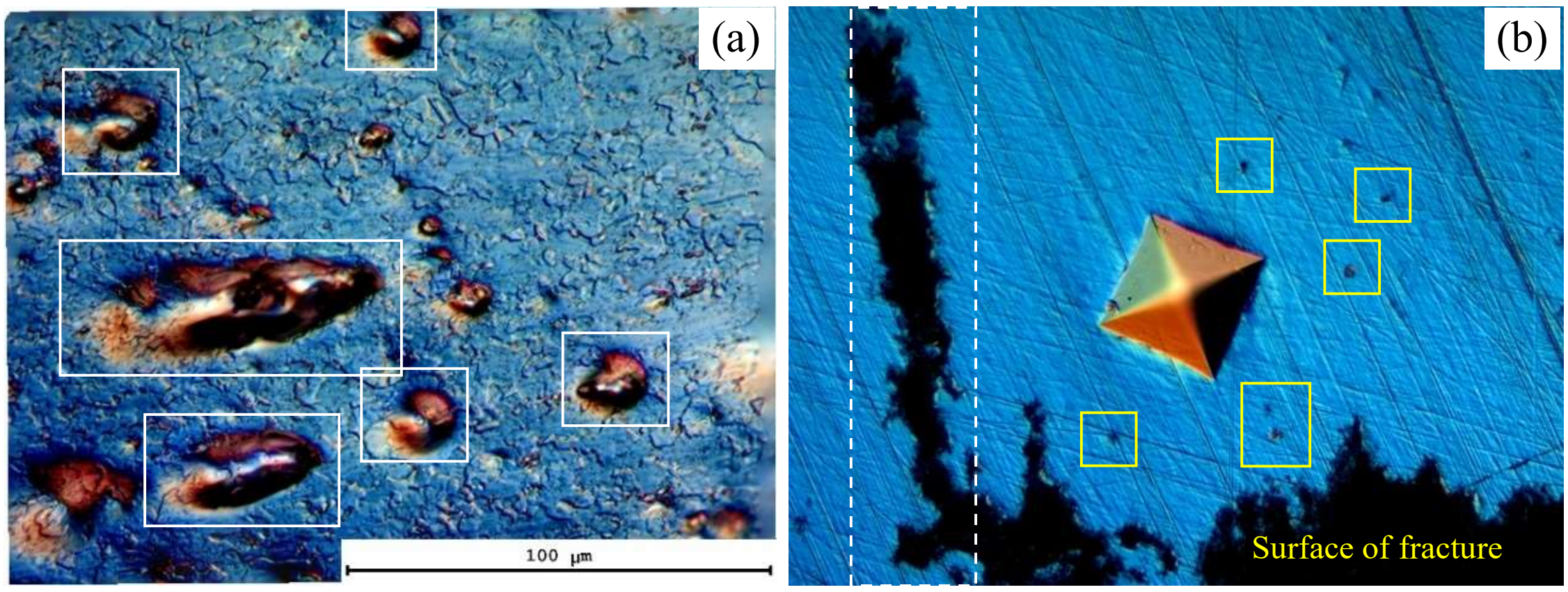

**Fig. 16** OM-images of pores on the surface of steel #450-3 after testing at 800°C. $\dot{\varepsilon} = 3.3{\cdot}10^{-3}$ s$^{-1}$

The fracture region of the UFG steel samples, following tensile testing at elevated temperatures, exhibits a conventional character. The fracture occurs due to the uniform thinning of the working part of the sample (Fig. 1c). The SEM analysis of the fractures of the UFG steel demonstrated the areas of the central zones and the radial ones to increase and the areas of the break zones – to decrease with increasing test temperature. At the test temperature of 600°C, the break zone area didn't exceed 5–10% of the whole fracture area. For the UFG steels ($T_{ECAP}$ = 450°C), the break zones were absent that also evidences an increased ductility of the UFG material as compared to the CG state.

The character of the fractures of the UFG steel samples made by ECAP at 150°C and 450°C is different. After the tensile tests at low temperatures (450–600°C), the elongated voids up to 500 μm long remained on the fracture surfaces of the UFG steel samples (ECAP at 150°C) (Fig. 17a, b). The largest voids are marked by the dashed lines in Fig. 17a, b for clarity. Similar voids being likely the destruction regions of δ-ferrite were observed on the fractures of the CG and UFG steel after the tests at RT. The sampled cross section areas decreased with increasing test temperature; the elongated voids transformed into the equiaxial pores. After testing at 800°C, the UFG steel fracture has a conventional dimple structure (Fig. 17d).

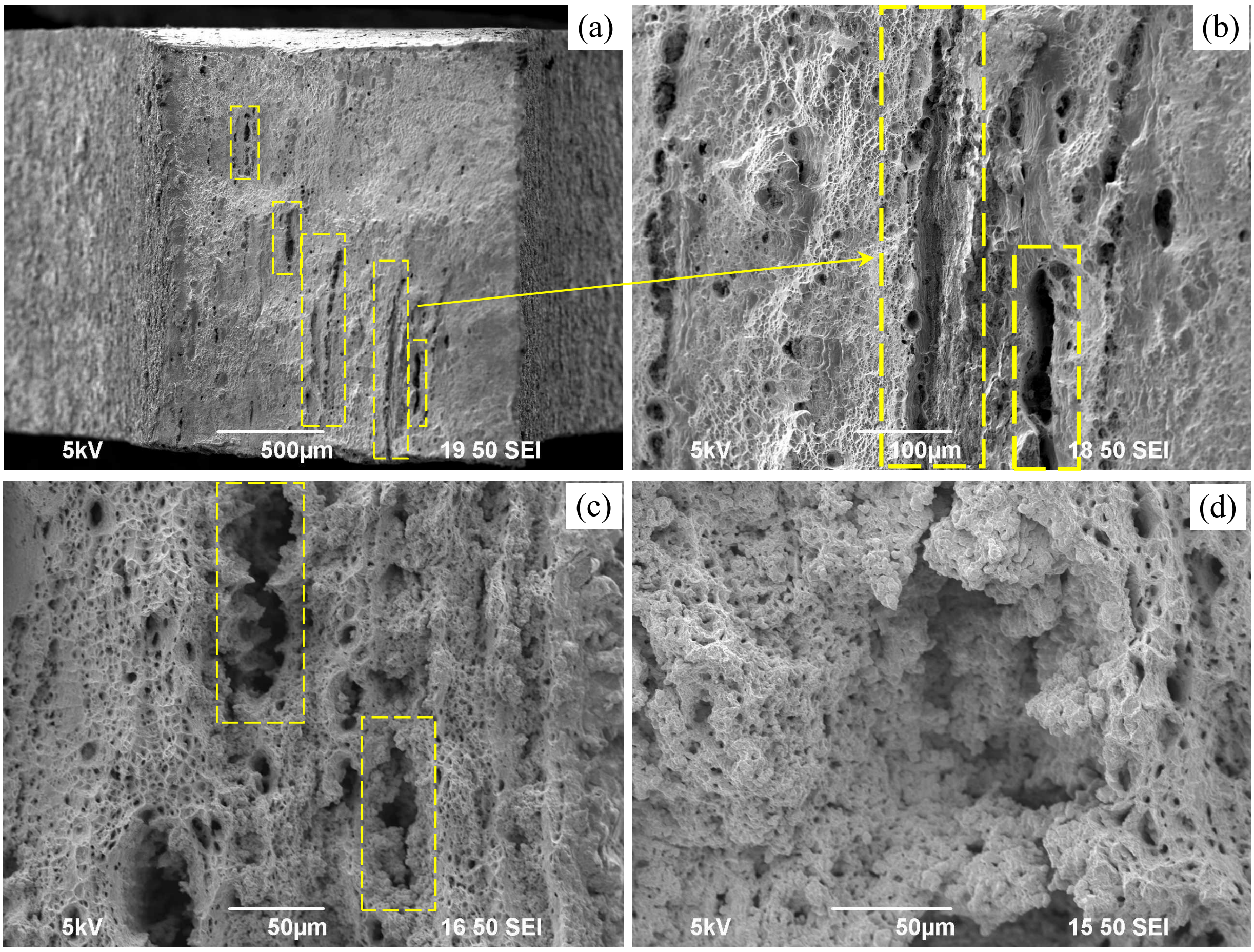

**Fig. 17** SEM-images of the steel #150-4 fractures after the tests at 600°C (a, b), 750°C (c), 800°C (d)

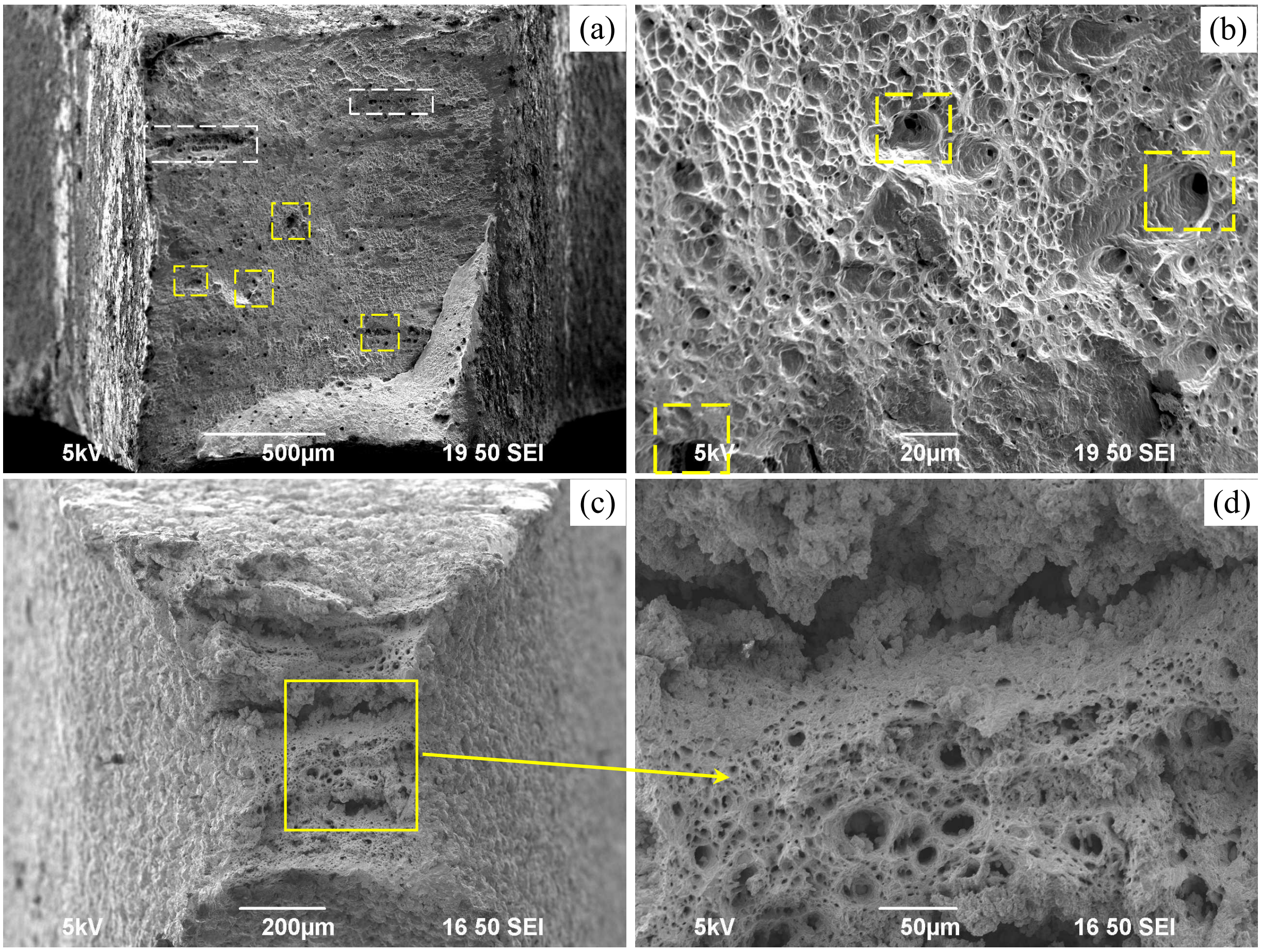

**Fig. 18** SEM-images of the steel #450-4 fractures after the tests at 450°C (a, b) and 800°C (c, d)

Fig. 18 presents the images of the fractures of the UFG samples produced by ECAP at 450°C, following superplasticity testing. Large, elongated voids formed in the samples from series #450-1 и #450-2, but were absent in samples #450-3 and #450-4. There were few large pores and shallow dimples in the central zones of the fracture surfaces of the UFG steel samples (ECAP at 450°C) after the tests at 450–600°C. Large voids in Fig. 18a, b are outlined by dashed yellow lies. The sizes of some voids reach 20 μm. Some voids are combined into large agglomerates of several hundred microns in size, which are indicated by white dotted lines in Fig. 18a. At further increasing the tensile temperature up to 800°C, the cross-section areas of the samples decreased essentially; the fractures had fully viscous character and comprised an ensemble of deep pores and large dimples (Fig. 18c, d). The areas of the radial zones of the fractures decreased with increasing test temperature up to 800°C whereas the central zones of the fractures consisted of dimples and pores of various sizes completely.

The following will provide a description of the microstructural evolution of the CG and UFG steels during superplasticity testing.

In the non-deformed part of the CG samples, a uniform austenite microstructure remained after tensile testing at elevated temperatures (Fig. 19a). The average austenite grain sizes almost didn't depend on the test temperature. Pores of submicron and micron sizes are formed at the austenite grain boundaries. It is important to note that pores along the austenite grain boundaries were observed in both the deformed (Fig. 19a) and non-deformed parts of the samples (Fig. 19b, c, d). Large twins are seen inside the non-deformed austenite grains. In the deformed parts of the CG samples, the elongated austenite grains were observed inside some equiaxial austenite grains (Fig. 18b, c, d). At the same time, a significant refinement of the δ-ferrite bands is observed. At elevated deformation temperatures, thin δ-ferrite bands of 0.5-1 mm in length are fragmented into individual particles several tens of microns in length (Fig. 11c, d, e, f). Intense martensite formation in the deformed part of the CG samples is not observed. This is probably due to the fact that the test temperature significantly exceeds the temperature of the end of the martensitic transformation in austenitic steel 321, consequently, a reverse transformation (from martensite back to austenite) could have occurred during cooling following the high-temperature test.

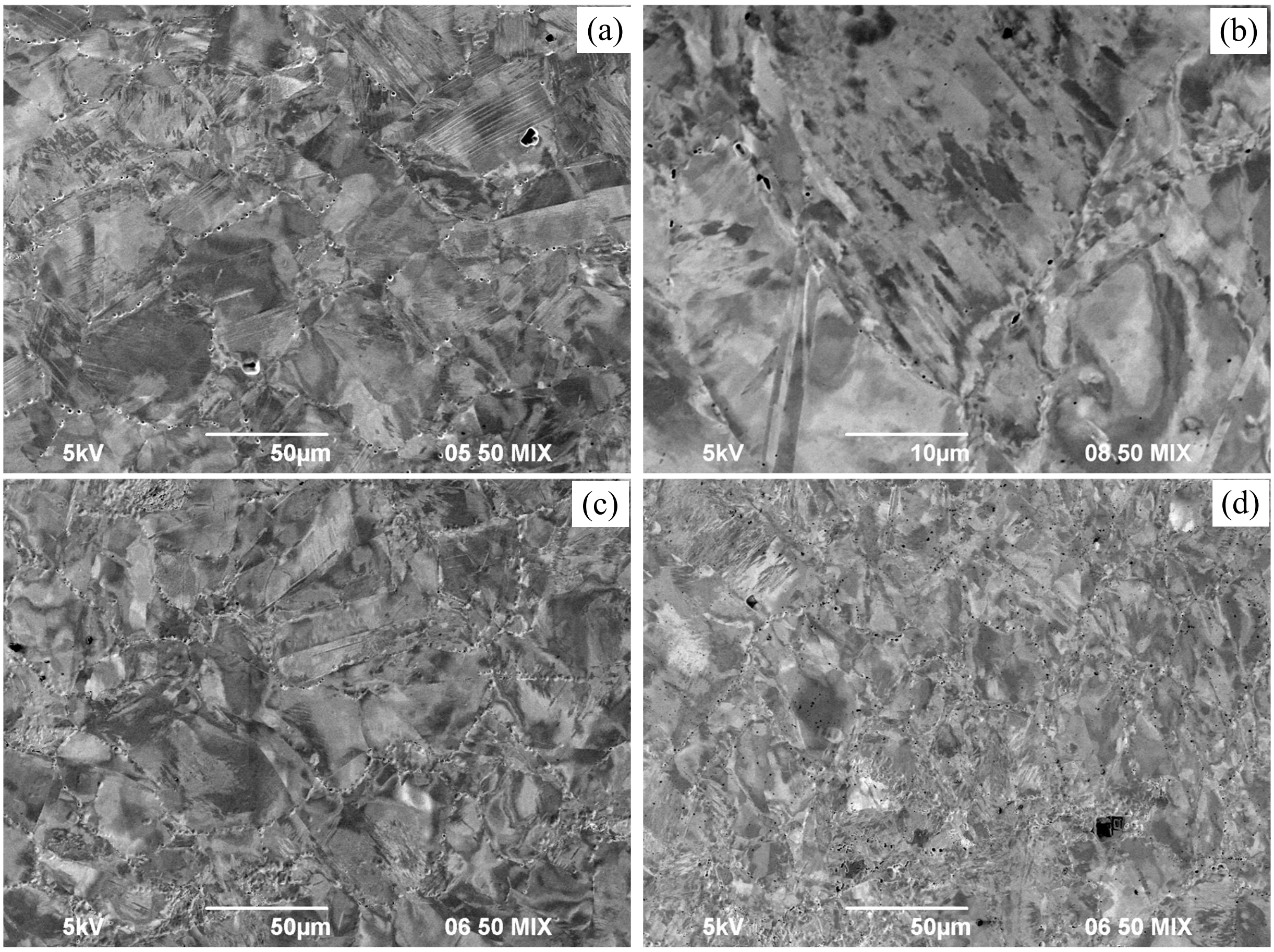

**Fig. 19** SEM-images of the microstructure of the nondeformed areas (a, b) and the deformed ones (b, c, d) of the CG steel specimens after the tests at 450°C (a, b), 750ºC (c) and 800°C (d)

The regularities of the microstructure evolution of the UFG steel during the deformation were more complex. At reduced test temperatures (700–750°C) close to the recrystallization temperature, the average grain sizes in the deformed zones were slightly greater than the ones in the non-deformed zones (Fig. 20a, b). This indicates that the strain-induced growth of austenite grains occurs during deformation. At elevated temperatures (800–900°C), the average grain sizes in the deformed zones were slightly smaller or comparable with the grain sizes in the non-deformed zones (Fig. 20c, d, Table 2). The grains preserved the equiaxial shapes. No elongated austenite grains expressed clearly were found when etching the deformed zones of the samples (Fig. 20b, d). This result allows suggesting the strain-induced grain growth to take place at reduced deformation temperatures whereas at higher deformation temperatures the dynamic recrystallization starts. This result is rather unexpected, as it is generally assumed that dynamic recrystallization in coarse-grained austenitic steel begins at temperatures exceeding 950–1000°C [6, 39]. It is important to note that the intensity of this process at such low temperatures remains relatively low, and the σ(ε) curves do not exhibit a jump-like change in stress, which is characteristic of actively occurring dynamic recrystallization.

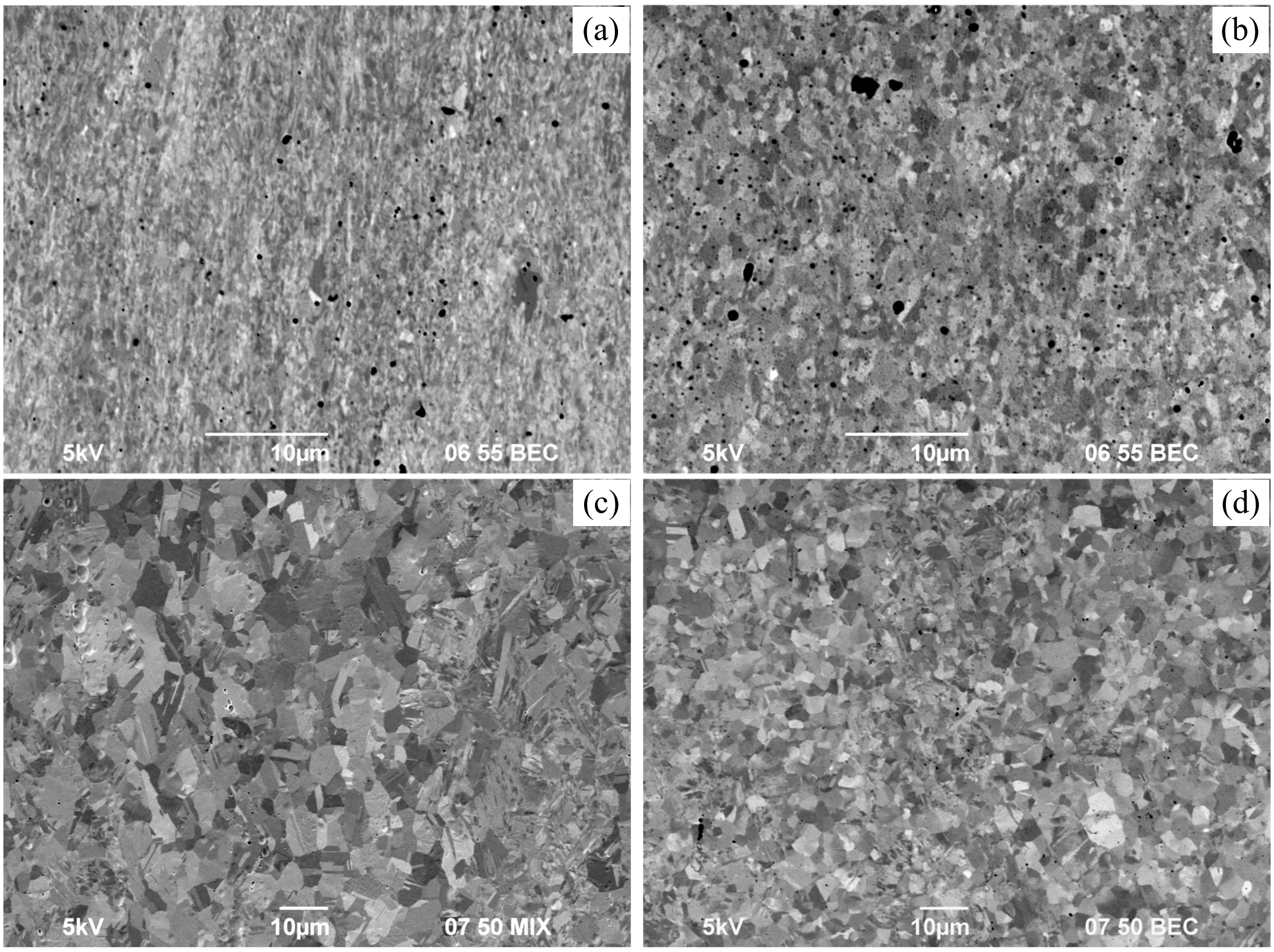

**Fig. 20** SEM-images of the microstructure of the nondeformed areas (a, c) and the deformed ones (b, d) of the steel specimens #450-3 after the tensile tests at 750°C (a, b) and 900°C (c, d)

3.4 The influence of superplastic deformation on the microhardness

The microhardness testing results (Table 4, Fig. 21) evidence the recrystallization at high test temperature of the UFG steel. The microhardness measurements of the specimens after the tensile tests demonstrated the increasing the test temperature from 450 to 800°C to result in a decrease in the microhardness both in the deformed areas and the non-deformed ones. This conclusion is supported by the results of the microstructure investigations in the deformed areas of the UFG steel specimens and the non-deformed ones after the tensile tests (Fig. 20). As can be seen from Fig. 20 and the data presented in Table 4, the testing at 800°C resulted in the formation of a well uniform fine-grained microstructure. No essential grain growth was observed. The average grain sizes in the deformed parts were slightly smaller than in the non-deformed ones.

It is important to note that the microhardness of the deformed areas of the specimens produced by ECAP at 150°C is higher than that of the specimens produced be ECAP at 450°C (Fig. 21). Analysis of the data presented in Table 4 and Fig. 21 shows that the microhardness of the Zone II of the CG steel samples significantly exceeds the microhardness of the Zone I. The reason for this is the intensive strain hardening and the formation of strain-induced martensite under tensile strain. It is noteworthy that the microhardness of the fracture region of UFG steels (Zone II) is close to or slightly higher than the microhardness of the undeformed part of the samples (Zone I). This means that the microhardness of superplastically deformed samples of UFG steels is close or slightly higher than the microhardness of UFG steels annealed at the same temperature. The result indicates that the strain hardening process during SP deformation of UFG steels is counterbalanced by intensive de-strengthening due to strain-induced grain growth.

**Table 4.** Microhardness of the samples after testing at different temperatures and $\dot{\varepsilon} = 3.3\cdot10^{-3}\ s^{-1}$. The average recrystallized grain sizes [$d$, μm] for the samples tested at 800°C and 900°C are given in braces

| $T_{test}$, °C | Microhardness ($HV$, GPa) | | | | | | | | | | | |
|---|---|---|---|---|---|---|---|---|---|---|---|---|
| | | | #450-2 | | #150-3 | | #450-3 | | #150-4 | | #450-4 | |
| | Zone I | Zone II | Zone I | Zone II | Zone I | Zone II | Zone I | Zone II | Zone I | Zone II | Zone I | Zone II |
| 450 | 1.87 | 2.95 | – | – | – | – | 3.64 | 3.55 | 4.07 | 4.13 | 3.76 | 3.75 |
| 600 | 1.88 | 2.68 | 2.94 | 3.38 | 4.05 | 3.63 | 3.56 | 3.49 | 4.25 | 3.75 | 3.77 | 3.60 |
| 750 | 1.76 | 2.38 | – | – | – | – | 2.79 (~0.5) | 2.77 (~1) | 3.33 | 2.52 | 3.38 | 2.71 |
| 800 | 1.84 (33) | 2.26 (41) | 2.40 (21.5) | 2.54 (2.0) | 2.13 (2.6) | 2.39 (2.3) | 2.15 (2.9) | 2.48 (1.6) | 2.96 (2.4) | 3.05 (1.9) | 2.21 (2.6) | 2.29 (1.6) |
| 900 | – | – | – | – | – | – | 1.90 (6.8) | 1.99 (4.4) | – | – | – | – |

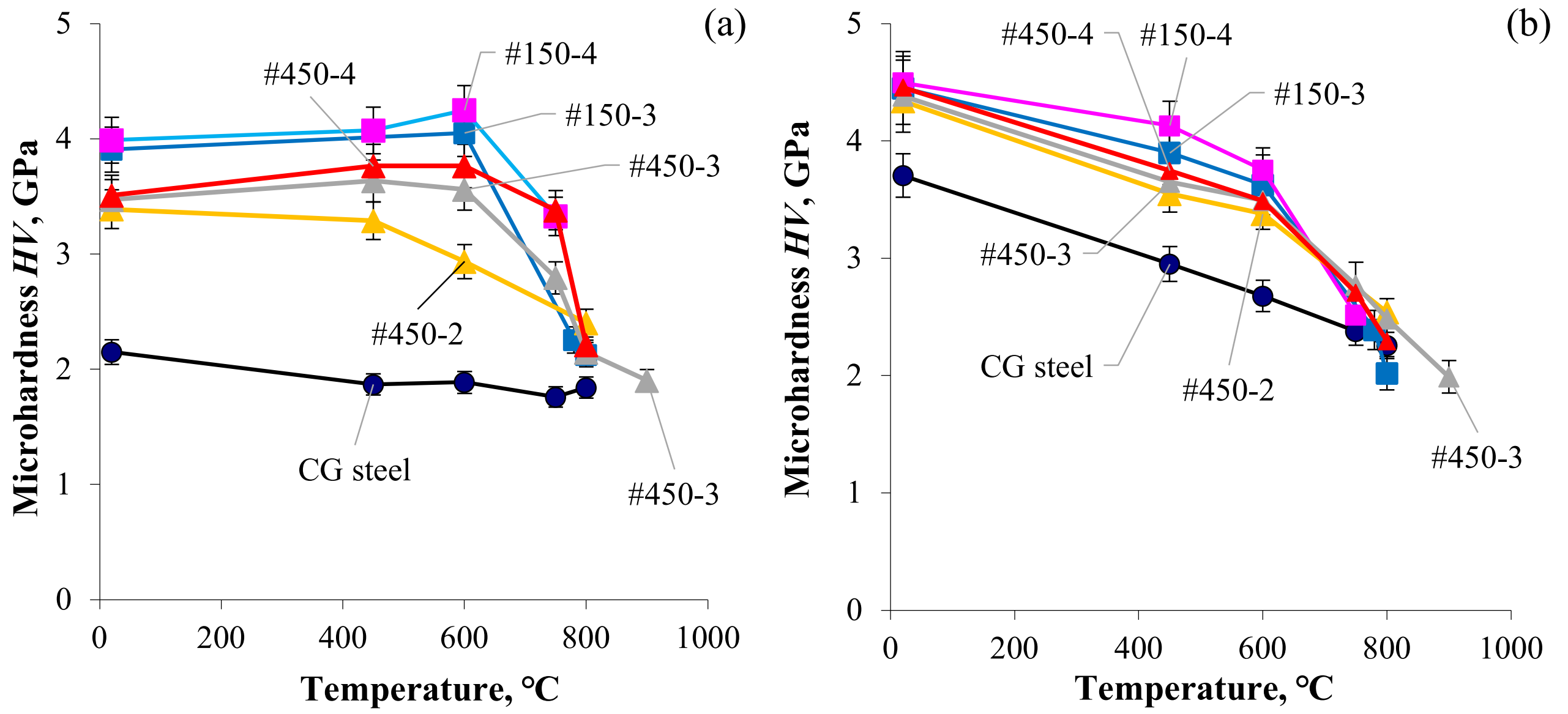

**Fig. 21** The dependence of microhardness on test temperature for Zone I (a) and Zone II (б) areas of the specimens. Strain rate $\dot{\varepsilon} = 3.3 \cdot 10^{-3}$ s$^{-1}$

3.5 The influence of SP deformation on the IGC resistance

The results of IGC tests conducted on UFG steel 321L samples obtained by ECAP were previously described in [35]. CG steel is characterized by a very high IGC resistance ($K_{CG} = 0.9 \cdot 10^{-4}$) [35]. UFG steel #150-4 and #450-4 has $K_{150\text{-}4} = 2.3 \cdot 10^{-4}$ and $K_{450\text{-}4} = 2.2 \cdot 10^{-4}$, respectively [35]. A 1-hour annealing of the UFG steel at 750-800°C results in an increase in the IGC rate ($K_{150\text{-}4} = (0.5–0.8) \cdot 10^{-2}$ and $K_{450\text{-}4} = (1.1–1.3) \cdot 10^{-2}$) [35]. At higher annealing temperatures, a slight decrease in the *K* coefficient was observed. It should be noted that the value of the *K* coefficient remained very small and did not exceed the critical value of $K_{IGC} = 0.11$.

Fig. 22 presents the results of electrochemical studies on the resistance to IGC of steel samples in both the CG state and the UFG state after ECAP. The analysis shows that the DLEPR curves of current density *i* – potential *E* for the deformed and undeformed sections of the samples after superplasticity testing are similar. For all samples, the area of the reactivation region ($S_r$) in the DLEPR curves is significantly smaller than that of the activation region ($S_a$).

Comparison of the DLEPR curves presented in Fig. 22a, b shows that the material in the fracture zone (Zone II) is characterized by a smaller $S_a$ and a larger $S_r$ than the material in Zone I. Thus, the material in Zone II has lower IGC resistance than the material in Zone I (Fig. 22c). It is noteworthy that two maxima are observed in the activation curves for the Zone II in the UFG samples (Fig. 22a). The primary maximum in the activation curve, with a high corrosion current density (up to 40-70 mA/cm$^2$), is observed at a potential value close to zero, while a second, smaller maximum is observed at a potential value of ~200 mV (Fig. 22a). In contrast, the activation curves for the undeformed part of the UFG samples have a classical profile, with a single maximum at 150-260 mV (Fig. 22b). This indirectly indicates the presence of two structural factors that provoke IGC (see [51]).

The DLEPR curves of UFG samples are characterized by a larger $S_r$ than those of CG steel samples. This conclusion is valid for both Zone I and Zone II.

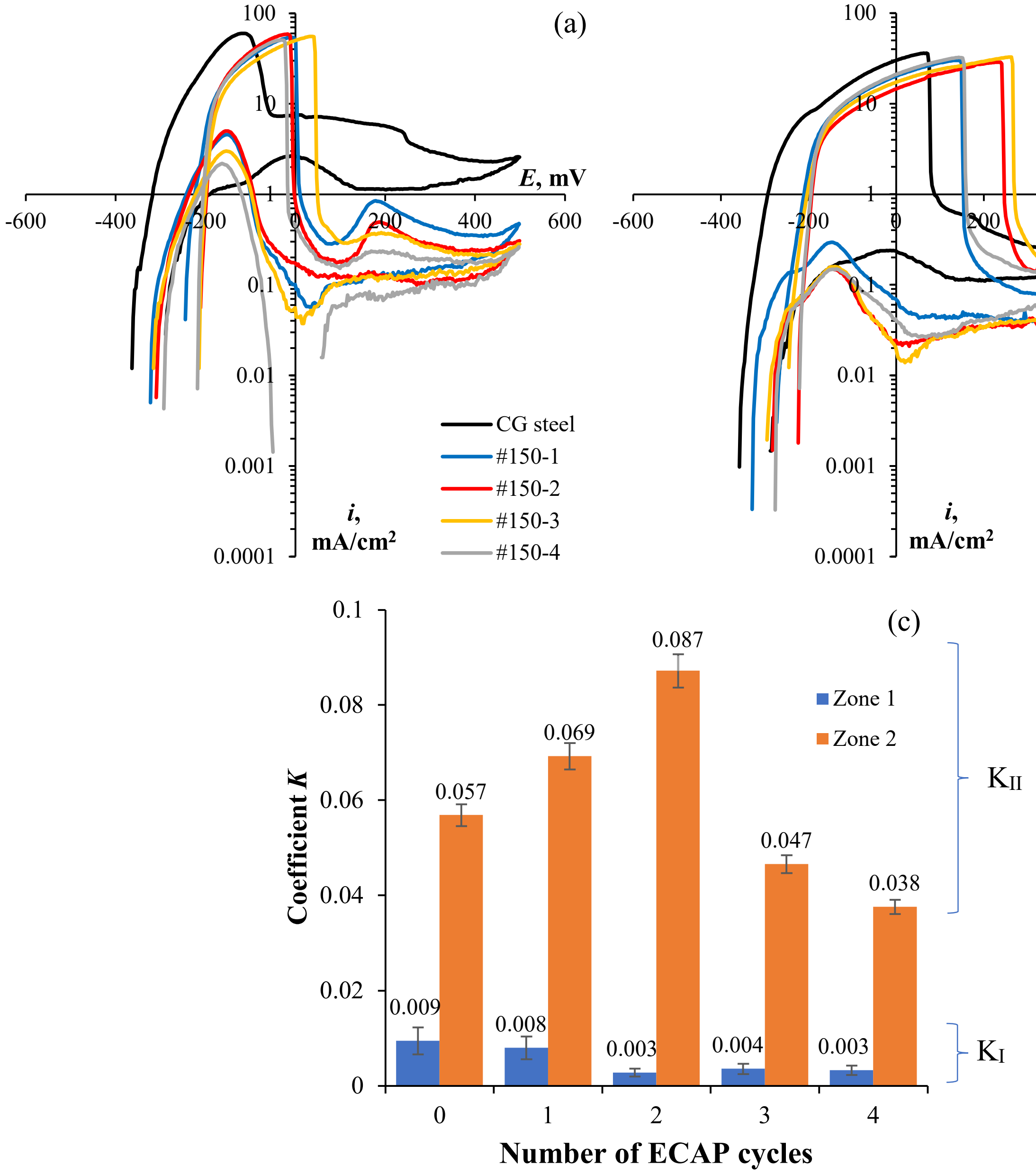

**Fig. 22** Analysis of the effect of the number of ECAP cycles on the IGC resistance: (a, b): DLEPR curves $i(E)$ for Zone II (a) and Zone I (b) of samples after tensile at RT; (c) dependence of the $K$ coefficient on the number of ECAP cycles. Test results for UFG steel produced by ECAP at 150°C

Next, the effect of ECAP on the ICG resistance of unannealed steel samples will be analyzed.

From Table 2 and Fig. 22c, it is evident that an increase in the number of ECAP cycles leads to a monotonic decrease in the IGC rate of Zone I. The dependence of the $K_{II}$ of the UFG steel samples on the number of ECAP cycles is non-monotonic: after two ECAP cycles at 150°C, the $K_{150\text{-}2}$ ~0.09, which is close to the critical value $K_{IGC} = 0.11$. With a further increase in the number of ECAP cycles, a monotonic decrease in the $K_{II}$ value is observed for the material of Zone II of the UFG samples. It is noteworthy that after four ECAP cycles at 150°C, the $K_I$ and $K_{II}$ for UFG steel is lower than that for CG steel. A similar dependence of the *K* coefficient on the SPD degree is observed for the UFG steel produced by ECAP at 450°C (Table 2). This is a rather unexpected effect, the reasons for which will be discussed in the Discussion section.

Next, the effect of test temperature on the IGC resistance of UFG steel samples will be analyzed. Fig. 23 shows DLEPR curves for UFG samples tested for SP at temperatures above 600°C.

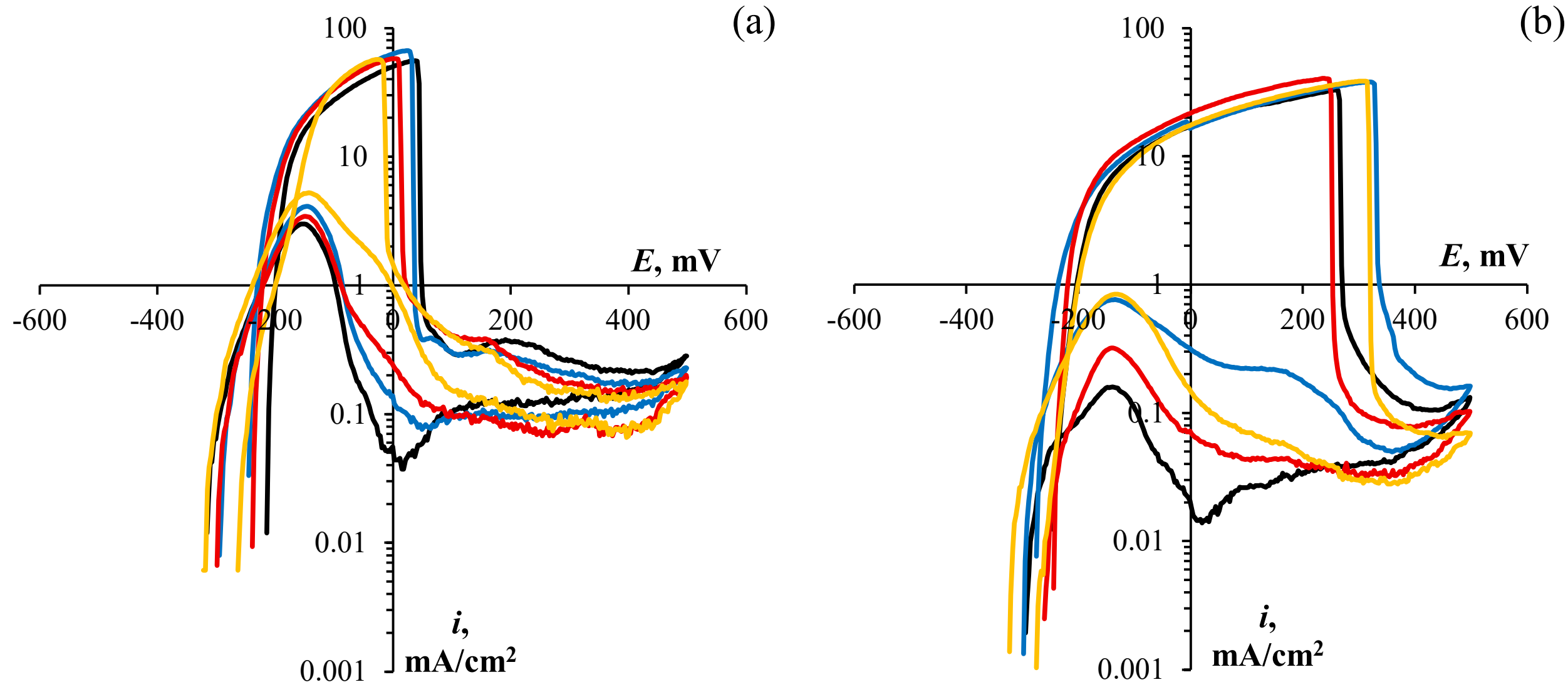

**Fig. 23** DLEPR curves *i*(*E*) for Zone II (a) and Zone I (b) of samples #450-4 after SP testing: black curve – RT, blue line – 750°C, red line – 800°C, yellow line – 900°C

Comparison of the DLEPR curves shown in Fig. 23a and Fig. 23b indicates that passivation of Zone II of the UFG samples occurs in the range from -200 mV to +50 mV ($\Delta E_{PII}$ = 250 mV). Passivation of Zone I occurs over a broader potential range, from -200 mV to +250-350 mV ($\Delta E_{PI}$ = 450-550 mV). The maximum corrosion current density after the SP-test of the UFG steel remains essentially unchanged (Fig. 23). Therefore, superplastic deformation leads to a decrease in the area of the passivation region in UFG steel ($\Delta E_{PI} > \Delta E_{PII}$).

Fig. 23 shows that Zone II of the superplastically deformed UFG samples exhibits higher activation current values than Zone I. Fig. 23a illustrates that increasing the test temperature broadens the potential range over which activation of Zone II of the UFG steel takes place.

Fig. 24 demonstrates the *K* value dependence on test temperature for the UFG steels. Analysis of the graphs presented in Fig. 24 indicates that the dependence nature of $K_I(T)$ is generally similar to the nature of the *K* dependence on annealing temperature of UFG steel [35]. The $K_I(T)$ dependence exhibits a maximum at a temperature of 700-750°C; SP at 800°C leads to a slight decrease in $K_I$, while a further increase in the SP temperature to 900°C leads to a renewed increase in the IGC rate.

The $K_{II}$ value remains virtually unchanged as the SP temperature increases to 750°C. At the 800°C, a slight decrease in $K_{II}$ is observed. However, with a further increase in the test temperature to 900°C, the IGC rate rises again. It should be noted that after superplasticity at 900°C, the $K_{II}$ value exceeds the critical value of $K_{IGC}$ = 0.11 and, consequently, the steel becomes susceptible to IGC.

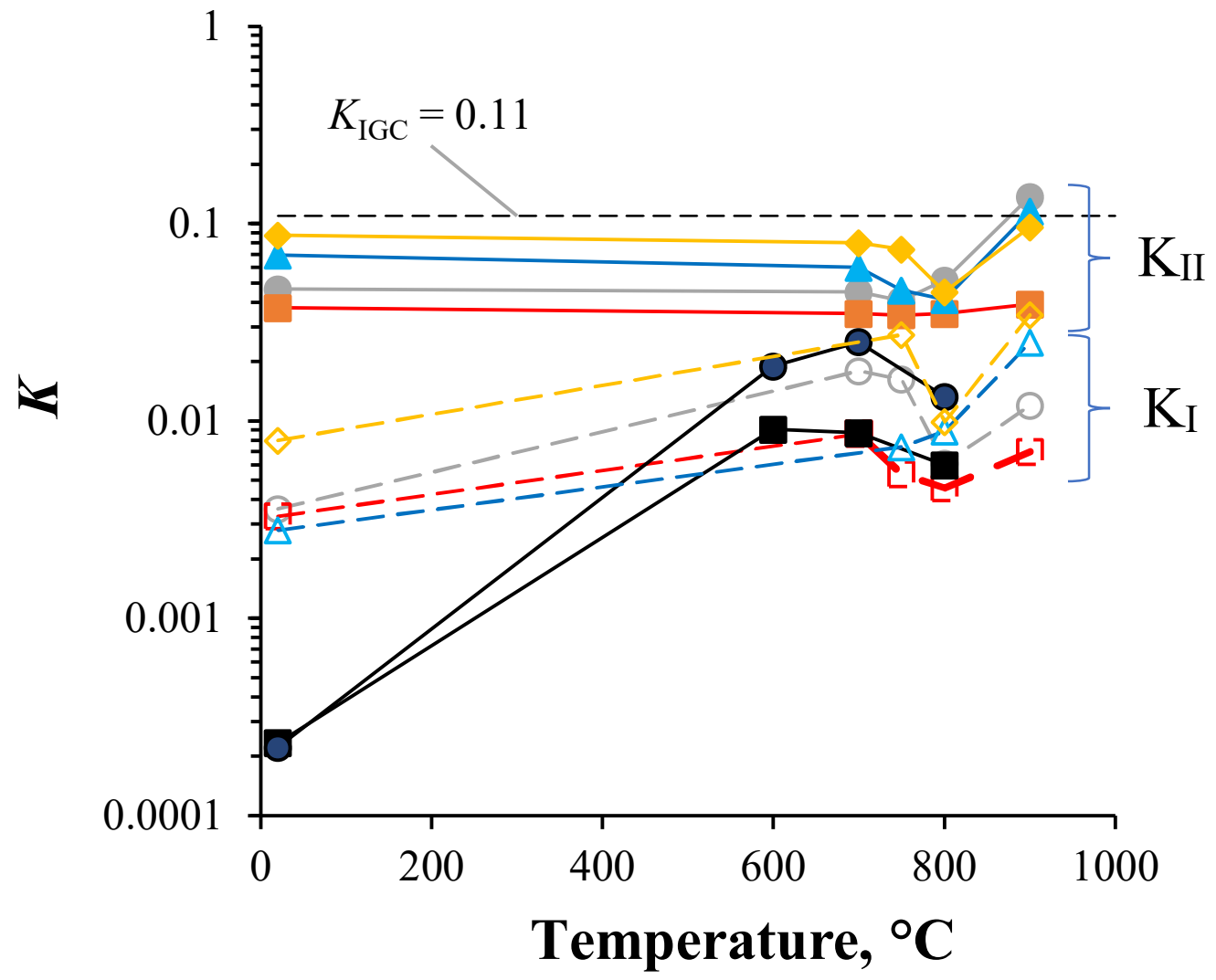

**Fig. 24** Dependence of the $K$ value on the SP temperature for samples #450-1 (blue lines), #450-2 (yellow lines), #450-3 (gray lines), and #450-4 (red lines). Solid lines indicate the Zone II, and dashed lines indicate the Zone I. For comparison, the black solid lines show the $K$ coefficient dependences on the 30-minute annealing temperature of UFG steels #150-4 and #450-4, taken from [35]

## 3.6 Microhardness vs. IGC resistance of UFG steel

Analysis of the data presented in Table 2 reveals a linear correlation between the UTS and microhardness of unannealed steel samples (Fig. 25a). It should be noted that this correlation exists for both the deformed and undeformed zones of the samples. However, lower values of the proportionality coefficient $A = HV$/UTS are observed for the metal in Zone II. The average linear correlation coefficient is quite high ($R^2 \sim 0.92$), allowing these dependencies to be used to predict the tensile strength of UFG steel based on microhardness measurement results.

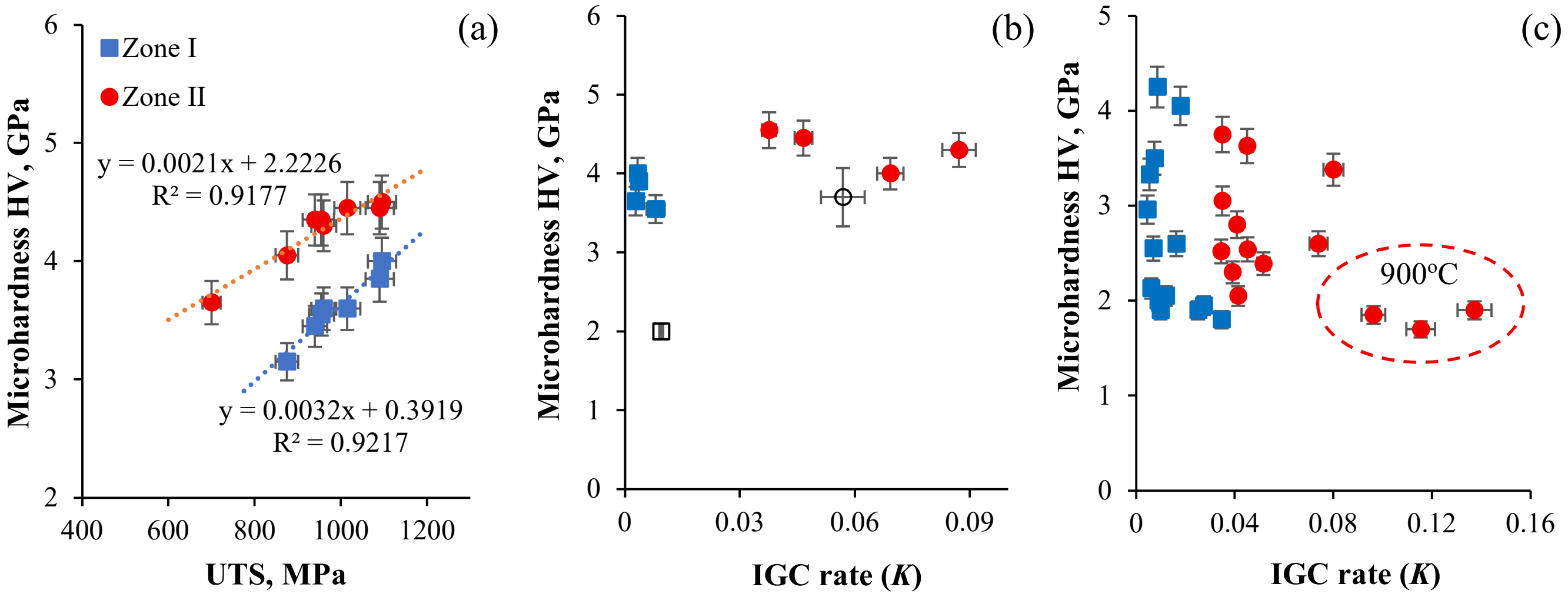

**Fig. 25** Microhardness vs. UTS (a) and *HV* vs. *K* coefficient (b, c) diagrams for tensile-tested samples at RT (a, b) and at elevated temperatures (b). The light markers in Fig. 25b indicate data for CG steel

Figs. 25b and c present diagrams of *HV* vs. *K* for samples tested at RT (Fig. 25b) and at elevated temperatures (Fig. 25c). Fig. 25b shows that, when testing UFG steels at RT, there is no consistent correlation between *HV* and the IGC rate – a significant change in the *K* is observed, while the microhardness of UFG steel remains virtually unchanged. A better correlation is observed between the *HV* and corrosion resistance of CG steel: the deformed part of the CG sample simultaneously exhibits higher *HV* and lower IGC resistance. This is likely due to the intense formation of strain-induced martensite during tensile testing of CG steel at RT.

Fig. 25c shows that, for SP deformed UFG samples, a nonlinear correlation is observed between *HV* and the *K* coefficient. UFG samples exhibiting reduced *HV* after SP deformation also demonstrate reduced corrosion resistance. UFG steel samples after SP deformation at 900°C have a low *HV* (less than 2 GPa), with the *K* exceeding the critical value $K_{IGC}$ = 0.11. Thus, despite their high values of the elongation to failure, UFG steels after SP at 900°C will have a high susceptibility to IGC (Fig. 26). The most optimal combination of *HV* and IGC resistance is possessed by UFG steels after SP at 700-750°C. Their *HV* ranges from 2.6 to 3.7 GPa, while the *K* coefficient is 0.035-0.08.

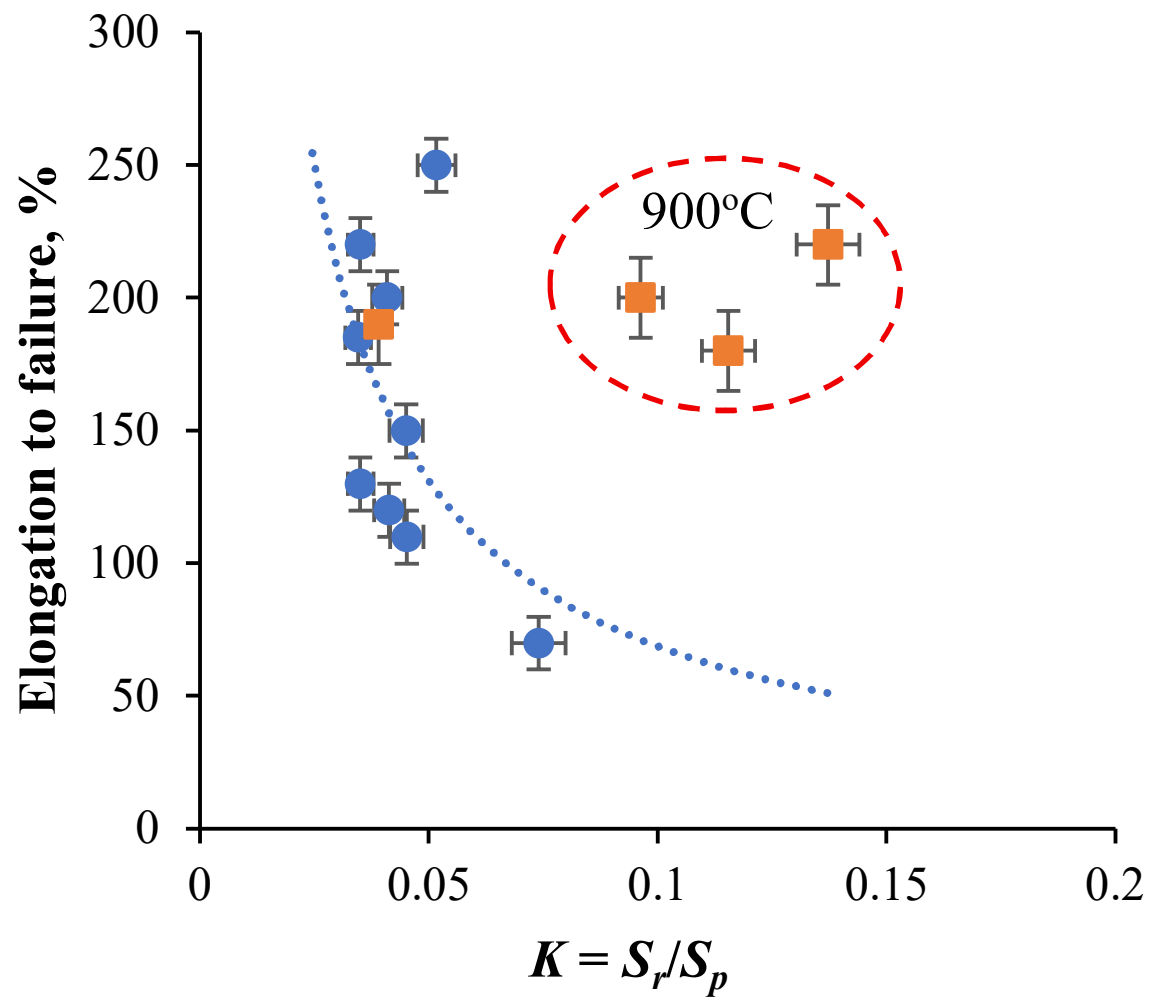

**Fig. 26** Diagram of elongation to failure (El, %) vs. *K* coefficient for UFG steel. The red dotted line indicates data for samples tested for SP at 900°C

It should be noted that the hypothesis proposed in the Introduction, suggesting a correlation between SP characteristics (elongation to failure) and the IGC rate, was not fully confirmed. As can be seen from Fig. 26, a relatively small change in the *K* coefficient corresponds to a significant change in the elongation to failure. In the long term, this will make it possible to produce samples of 321 austenitic steel that simultaneously possess enhanced hardness, ductility, and corrosion resistance. In our opinion, one of the reasons for the lack of correlation between El and the $K$ coefficient is the presence of non-metallic inclusions in the steel, which cause premature failure of the samples and limit the potential to achieve higher El values.

## 4. Discussion

4.1 General analysis of the results. The superplastic deformation mechanism in UFG steel

The main rheological equation of the SP flow is usually presented in the form [52, 53]:

$$\dot{\varepsilon} = A(\sigma^*/G)^{1/m}(b/d)^p(D_{sp}/b^2)(G\Omega/kT), \quad (1)$$

where $A$ and $p$ is a numerical parameters; $D_{SP} = D_0\exp(Q_{SP}/kT)$ is the effective diffusion coefficient in the SP conditions, $Q_{SP}$ is the activation energy of the SP, $D_0$ – the pre-exponential factor of the diffusion coefficient, $\sigma^* = \sigma - \sigma_p$ is the flow stress of the SP, $\sigma_p$ is the threshold stress of the SP, $G$ is the shear modulus, $\Omega$ is the atomic volume, $k$ is the Boltzmann constant, and $b$ is the Burgers vector.

To describe the grain boundary strain rate $\dot{\varepsilon}_b$, the equation for GBS [52] is employed usually:

$$\dot{\varepsilon}_b = A_b(\sigma/G)^2(b/d)^2(G\Omega/kT)(\delta D_b/b^3), \quad (2)$$

where $A_b$ ~100, $\delta = 2b$ is the grain boundary width, $D_b = D_{0b}\exp(Q_b/kT)$ is the grain boundary diffusion coefficient, $Q_b$ is the activation energy of the grain boundary diffusion. The possibility of realization of the GBS mechanism in the fine-grained austenite steel at elevated deformation temperature was demonstrated in [27, 30].

To calculate the intragranular strain rate $\dot{\varepsilon}_v$, the equation for the power law creep [54] is used often:

$$\dot{\varepsilon}_v = A_v(\sigma/G)^{n_1}(D_{cr}/b^2)(G\Omega/kT), \quad (3)$$

where $A_v$ is the Dorn constant, $n_1$ – the numerical coefficient depending on the creep mechanism, $D_{cr} = D_{0cr}\exp(Q_{cr}/kT)$ is the diffusion coefficient in the creep conditions, $Q_{cr}$ is the creep activation energy.

During the deformation of the UFG alloys in non-optimal for SP conditions, the $\dot{\varepsilon}_v$ gives the major contribution into total strain rate ($\dot{\varepsilon} = \dot{\varepsilon}_v + \dot{\varepsilon}_b$) where $\dot{\varepsilon}_b$ is the GBS rate: $\dot{\varepsilon}_v >> \dot{\varepsilon}_b$. If the primary SP mechanism is GBS, then $\dot{\varepsilon} \sim \dot{\varepsilon}_b$ ($\dot{\varepsilon}_b >> \dot{\varepsilon}_v$).

In accordance with Eq. (1), the strain hardening coefficient ($n$) value, which indicates the material's tendency toward uniform deformation, can be determined by the slope of the σ(ε) curve in the ln(σ)–ln(ε) coordinates. To accomplish this, the σ(ε) curves in the engineering stress – engineering strain coordinates must be plotted in the true stress – true strain coordinates.

The σ(ε) curves in the ln(σ)–ln(ε) coordinates for the CG steel exhibit a two-stage character (Fig. 27a). The slope angle at the first stage is close to $n \sim 1$ and is practically independent of the deformation conditions, which corresponds to the case of elastic strain. An increase in the deformation temperature leads to a decrease in deformation corresponding to the end of the stage with $n \sim 1$. The slope angle during the second stage of strain hardening decreases from 0.67 to 0.19–0.20 as the strain temperature increases from RT to 800 °C (Fig. 27a).

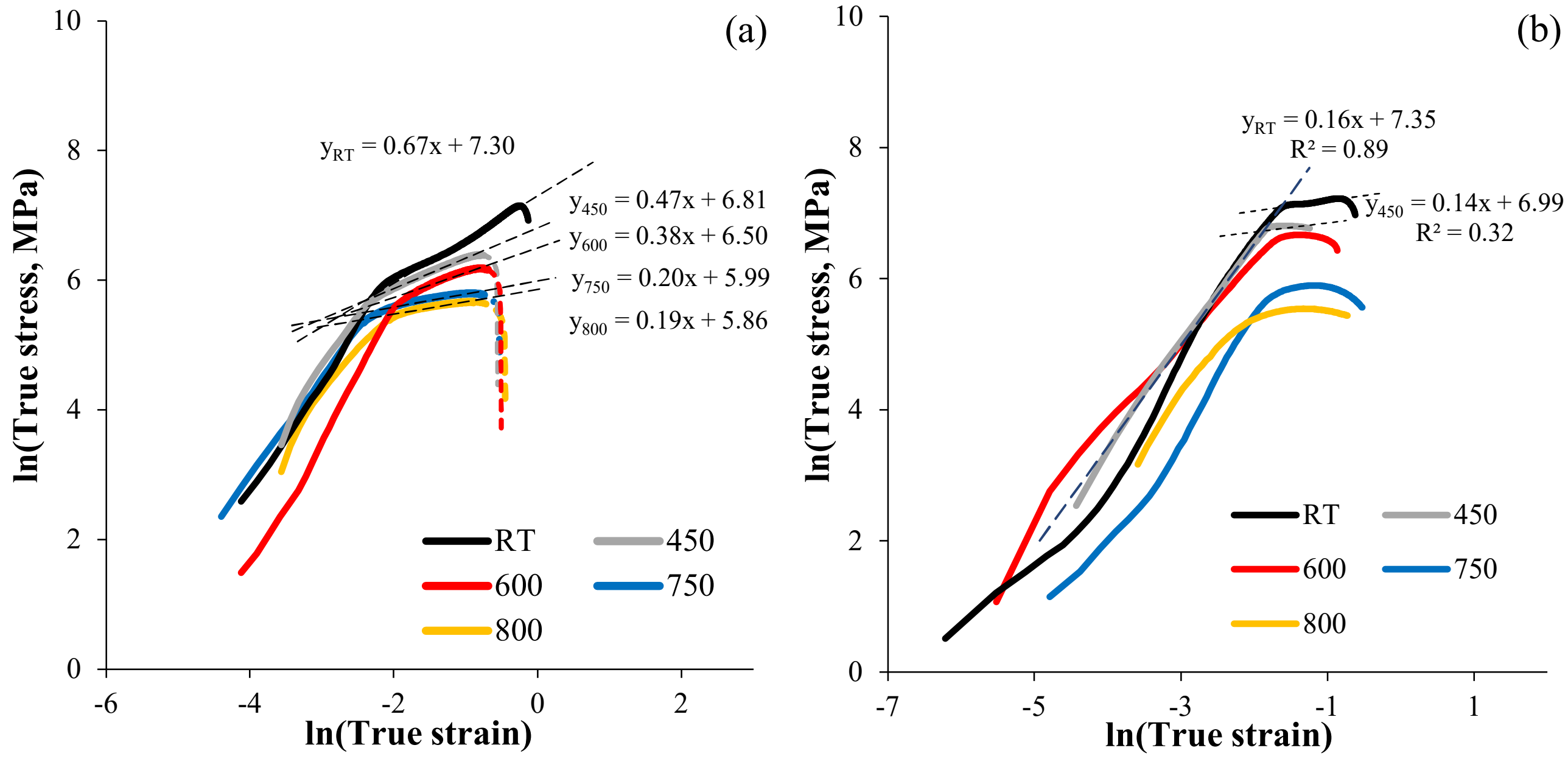

**Fig. 27** The σ(ε) curves in the ln(σ)–ln(ε) coordinates for the CG steel (a) and UFG steel #450-4 (b). Analysis of the strain hardening stage. Calculation of the strain hardening coefficient *n*

As noted above, the σ(ε) curves at 750–900°C for the UFG steels exhibit virtually no strain hardening stage. Therefore, in the ln(σ)–ln(ε) coordinates, the first strain hardening stage, characterized by a coefficient $n \sim 1$, transitions into a stationary flow stage, where the coefficient *n* value → 0 (Fig. 27b). A minor stage of strain hardening can be observed in σ(ε) curves (Fig. 27b) at low temperatures (450–600°C). However, the linear correlation coefficient for these limited regions is low ($R^2 < 0.8$). This does not allow for a correct analysis and reliable determination of the coefficient *n* value. The results obtained suggest that the σ(ε) curve for the UFG steel cannot be represented by a power law of the form $\sigma \sim \varepsilon^n$. This result indirectly corresponds to the findings of studies that investigated the rheology of plastic flow during the hot deformation of 321L steel [55, 56].

It is important to note that, in accordance with the deformation mechanism maps for CG austenitic steels 304 and 316 [54], the power law creep must be observed within the specified range of temperatures and strain rates (Eq. (3)). According to [54], the coefficient $n_1$ value for Eq. (3) varies from 4 to 10 for steels 304 and 316. The value of $n_1 \sim 5$ was obtained for steel 321 at 900–1200°C, corresponding to the dynamic recrystallization region [57].

The coefficient $1/n_1$ value can be determined from the slope of the dependence of flow stress on strain rate in the $\ln(\sigma_s)$–$\ln(\dot{\varepsilon})$ coordinates (see Eq. (3)). For this purpose, the σ(ε) curves obtained at various strain rates (ranging from $3.3 \cdot 10^{-4}$ to $3.3 \cdot 10^{-3}$ $s^{-1}$) and temperatures of 800–900°C were analyzed (Fig. 12d). The analysis of the results shows that the coefficient *n* for the UFG steels obtained by the ECAP at 150°C is ~7.2, and for the UFG steels obtained by the ECAP at 450°C is ~ 5.1. This result is consistent with the results of [57, 58].

Using Eq. (3), it is possible to estimate the effective activation energy of the plastic deformation ($Q_{cr}$) for the UFG steel. According to Eq. (3), the magnitude of $Q_{cr}$ [in $kT_m$] can be determined from the slope of the $\ln(\sigma_s)/n - T_m/T$ dependence (at $\dot{\varepsilon}$ = const) where $T_m$ = 1810 K is the melting point of steel. As illustrated in Fig. 28, the $\ln(\sigma_s)/n - T_m/T$ dependencies in the temperature range of 600–800°C can be fitted by straight lines with a good precision ($R^2 > 0.90$). The $Q_{cr}$ for the UFG steels obtained by ECAP at 450°C is 10.2–14.1 $kT_m$ (~ 153–212 kJ·mol$^{-1}$). This value is close to the activation energy of the grain boundary diffusion in the austenite ($Q_b$ ~ 159 kJ·mol$^{-1}$ [54]). For the UFG samples obtained at 150°C, the creep activation energy is $Q_{cr}$ ~ 23.8–24.0 $kT_m$ (~ 358–361 kJ·mol$^{-1}$). The $Q_{cr}$ value ~ 332–402 kJ·mol$^{-1}$ exceeds the activation energy of diffusion in the austenite crystal lattice ($Q_v$ ~ 270 kJ·mol$^{-1}$ [54]) but is close to the activation energy for the case of hot deformation of austenitic steel [54]. It is interesting to note that heterostructured steels exhibit lower creep activation energy values (~10.3 kTm) than austenitic steels with a homogeneous UFG microstructure (Fig. 28). This unexpected effect is due, in our opinion, to the increased dislocation density in heterostructured steels. Increasing the degree of SPD leads to a higher martensite content and accelerates the precipitation of σ-phase nanoparticles, which hinder dislocation motion.

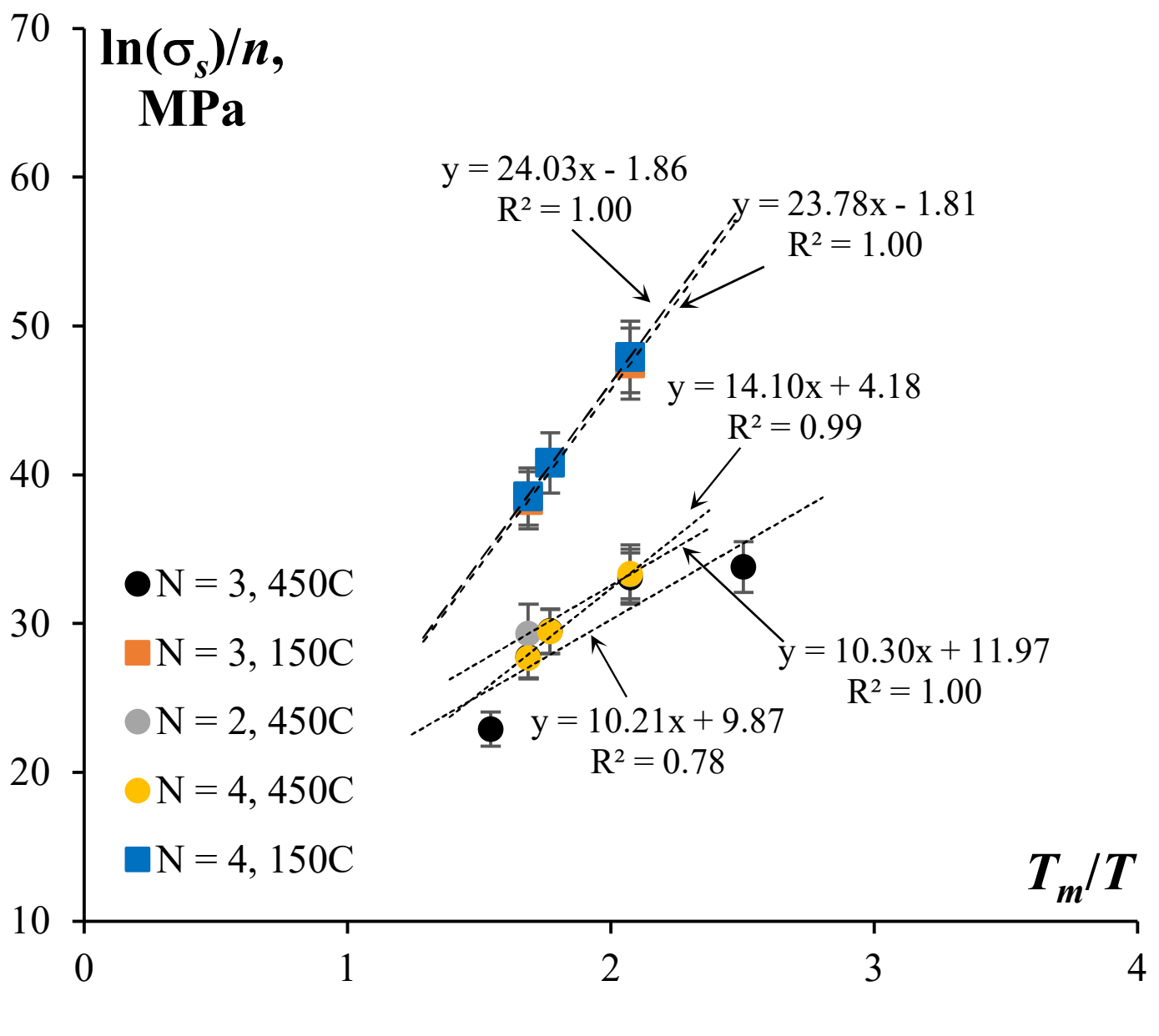

**Fig. 28** Temperature dependence of the flow stress plotted in the $\ln(\sigma_s)/n - T_m/T$ axes. Calculation of the creep activation energy in the UFG steel

Thus, the analysis of the results indicates that the primary mechanism of hot tensile deformation of UFG steel 321 in the temperature range of 600–800°C is power law creep. This conclusion is indirectly confirmed by the results of studies of the effect of the annealing temperature on the deformation behavior of steel #450-4 (Table 3, Fig. 13). As illustrated in Fig. 13 and the data presented in Table 3, an increase in the preliminary annealing temperature, along with an increase in grain size, leads to an increase in the ductility of UFG steel. This conclusion aligns well with the results of studies of the effect of grain size on the creep rate and creep life of fine-grained austenitic steels [59-61].

### 4.2 Effect of microstructure

The study results of the microstructure of 321L steel during the ECAP process and subsequent heating, which are presented in [35] and in the present work, are summarized in the following section.

(i) Steel 321L in its initial state has a homogeneous austenite microstructure, characterized by the presence of large of δ-ferrite and non-metallic inclusions.

(ii) The refinement of the austenite grains and δ-ferrite particles occurs during the ECAP process. An increase in the number of ECAP cycles results in a higher fraction of α′-martensite. An increase in the ECAP temperature from 150°C to 450°C does not significantly affect the fraction of α′-martensite. However, it does lead to a slight increase in the average grain size of austenite.

(iii) The fraction of martensite decreases to values that cannot be determined by XRD phase analysis methods when heated to temperatures 600–650°C. The formation of the σ-phase nanoparticles occurs during static heating at temperatures 650–700°C. The σ-phase particles are uniformly distributed within the volume of the grains in the UFG material. Grain growth during a 30-minute annealing process begins at a temperature of 750°C.

(iv) During SP at temperatures below 700–750°C, strain-induced grain growth occurs. The average grain size in the fracture region (Zone II) is larger than that in the undeformed region of the specimen (Zone I). During the deformation process at 800–900°C, dynamic recrystallization begins to occur, resulting in the formation of a heterogeneous microstructure.

(v) In the process of SP deformation, two types of pores are observed: large pores that form on non-metallic inclusions and on δ-ferrite, and small, micron-sized pores. Accelerated cavitation failure restricts the potential for achieving greater elongation to failure of the UFG steel.

(vi) The nature of the dependence of El($T_{ann}$) is determined by the ECAP regime. The dependence of El($T_{ann}$) exhibits a non-monotonic character, with a minimum at $T_{ann}$ = 750°C for heterostructured steels #450-1 and #450-2. An increase in the preliminary annealing temperature from 600 to 900°C leads to a monotonous increase in El of steels #450-3 and #450-4, which have a homogeneous UFG structure.

#### *a Creep vs grain boundary sliding*

Preliminary analysis of the results shows that the dominant deformation mechanism of the UFG steel at 600–900°C (~0.48–0.65 $T_m$) is the power law creep process. This result is quite unexpected, as it is generally assumed that GBS plays a significant role in the superplastic deformation process of the UFG materials within this temperature range.

It is assumed that the primary reason for this contradiction is the rapid strain-induced grain growth occurring at 600–750°C, as well as dynamic recrystallization occurring at 800–900°C. The intensive grain growth during plastic deformation of the UFG samples ($T_{ECAP}$ = 150°C) leads to a decrease in GBS in accordance with Eq. (2), and the contribution of the power law creep becomes essential. This leads to an increase in the SP activation energy so that it become close to the activation energy of the power law creep (Fig. 24). From the condition $\dot{\varepsilon}_b = \dot{\varepsilon}_v$, it is possible to find the critical grain size $d^*$, which the dominant deformation mechanism in the austenitic steel changes at:

$$d^* = \sqrt{(A_b/A_v)b(G/\sigma)\,(\delta D_b/D_v)}. \quad (4)$$

At $A_b$ = 100, $A_v$ = 4.3·10$^5$, $b$ = 2.58·10$^{-10}$ m, $G$ = 81 GPa, $\delta D_{b0}$ = 7.5·10$^{-14}$ m$^3$·s$^{-1}$, $Q_b$ = 159 kJ·mol$^{-1}$ ~ 10.6 $kT_m$, $D_{v0}$ = 1.8·10$^{-5}$ m$^2$·s$^{-1}$ [54], $T$ = 1073 K, the critical grain size $d^*$ ~ 0.16 μm is obtained. This value is an order of magnitude smaller than the average grain sizes observed experimentally (Table 4).

In our opinion, the disagreement between the calculation results and the experimental data originates mainly from the nonequilibrium state of the grain boundaries in the UFG alloys [23]. During the SP deformation, the grain boundaries in the UFG metals contain an increased density of defects: orientation misfit dislocations (OMDs) and products of delocalization of these ones. The defects trapped by the grain boundaries enhance the free (excess) volume of the grain boundaries and lead to a decrease in the activation energy of the grain boundary diffusion [23]. A good agreement between the results of calculations and the experimental data ($d^*$ ~ 2 μm, Table 4) is observed at the activation energy of diffusion in the nonequilibrium grain boundaries ~ 7.6 $kT_m$ (~ 114 kJ·mol$^{-1}$). This value of the grain boundary activation energy is typical for the UFG metals [23].

*b Cavitation failure*

Firstly, the question of the origin of the cavitation failure of the UFG steels in the SP deformation conditions will be discussed. The mechanisms of small pore formation and growth under creep conditions in CG steel have been studied in detail (see, for example, [62, 63]).

The primary cause of small pore formation is precipitation σ-particles during creep when the low strain rates. The negative effect of the σ-phase on the high-temperature plasticity of the austenite steels was also noted in [64–66].

Cavitation is one of known mechanisms of failure of materials in the SP conditions [67, 68]. A model of pore formation at the secondary particles in the SP conditions was proposed in [68]. In [68], the formation of pores at grain boundaries is attributed to the flow of lattice dislocations ($I^+$) falling into the grain boundaries at a rate of $\dot{\varepsilon}_v$: $I^+ = \varphi_1\dot{\varepsilon}_v/b$, where $\varphi_1$ is a numerical coefficient that depends on the uniformity of the lattice dislocations flow [23]. Dislocations that fall into the grain boundary begin to delocalize, resulting in the formation of both normal and tangential components of the Burgers vector [23]. According to [68], the pore nucleation in the triple grain boundary junctions is caused by the accumulation of the normal components of the delocalized dislocations. The related elastic energy and internal stresses of the defect layer formed at the particle – matrix interphase boundary increase in the course of SP deformation [68]. A detailed description of these defects is given in [23, 44, 69]. A defect nucleated at a particle with the radius $R$, in the first approximation, can be described as a disclination loop with the radius $R$ and the power $\omega(t)$. The power of the disclination loop ω(t) grows proportionally to the number of defects falling onto the grain boundary during the SP deformation: $\omega(t) = \psi_2\dot{\varepsilon}_v t$ [68], where $\psi_2$ is a geometrical factor, and $t$ is the strain time. As it has been shown in [68], at certain critical power $\omega^*$, the related elastic energy of the loop becomes so high that it becomes favorable energetically to the grain boundary to release from the source of this excess energy. At reduced deformation temperatures, it would lead to formation of microcracks near the particles. At elevated deformation temperatures, the relaxation of the accumulated material can occur due to the formation of micropores at the particle – metal interphase boundaries [68].

The disadvantage of the model [68] is the requirement of the presence of particles at grain boundaries. This condition is not typically met in this case, as the proportion of σ-particles at grain boundaries is small. In our opinion, the model [68] can effectively describe the volume cavitation failure of steel 321L. This is due to the fact that the flow of lattice dislocations can lead to the formation of disclination loops at the boundaries of incoherent particles located in the grain volume. The mechanism of formation of such defects is analogous to the Orovan mechanism for the formation of dislocation loops of incoherent particles [70]. As shown above, the nucleation of the σ-particles was observed when heating the UFG steel (Fig. 6). Cutting such incoherent σ-particles by the lattice dislocations is difficult, and the disclination loops would form around these ones in the course SP deformation.

The growth of the pores in the course of plastic deformation will be proportional to the strain and strain rate [68]. Upon achieving some critical sizes, such pores become the sources of the fracture of the UFG samples.

The primary factor contributing to the destruction at low SP temperatures (below 600°C) is the presence of large pores that form on large particles of Ti(C,N) and/or δ-ferrite. The formation of large pores on Ti(C,N) particles or δ-ferrite, which exceed the grain size in UFG steel, cannot be described using a model [68]. The primary cause of pore formation at the boundaries of these large particles is the disparity in the elastic moduli between the large particles and austenite.

Deformation at temperatures above 600°C results in the formation and growth of the σ-phase particles, which contribute to the formation of micropores. The coalescence of a large number of micropores formed on the σ-phase particles accelerates the destruction of the UFG steel specimens.

*c Effect of the test temperature*

It can be observed from Fig. 13b, Fig. 29 and Appendix B that an increase in the test temperature results in a monotonic decrease in the flow stress of the UFG steel. However, it is noteworthy that the ductility of the UFG steel exhibits a non-monotonic dependence on the test temperature (Fig. 13a). Increasing the temperature from RT to 450°C results in a reduction in El of the UFG steel by $\Delta\delta_{max} = \delta_{RT} - \delta_{450} \sim 20–35\%$, where $\delta_{RT}$ and $\delta_{450}$ represent the elongation to failure at RT and 450°C, respectively. A monotonic increase in the ductility of the UFG steel is observed with a further increase in the test temperature (Fig. 13a, 29). The dependence of $\delta_{max}(T)$ for the CG steel has a similar trend. However, the reduction in ductility at 450°C is significantly greater ($\Delta\delta_{max}$ ~ 60–65%), while the increase in ductility at elevated temperatures is considerably smaller (Fig. 13a).

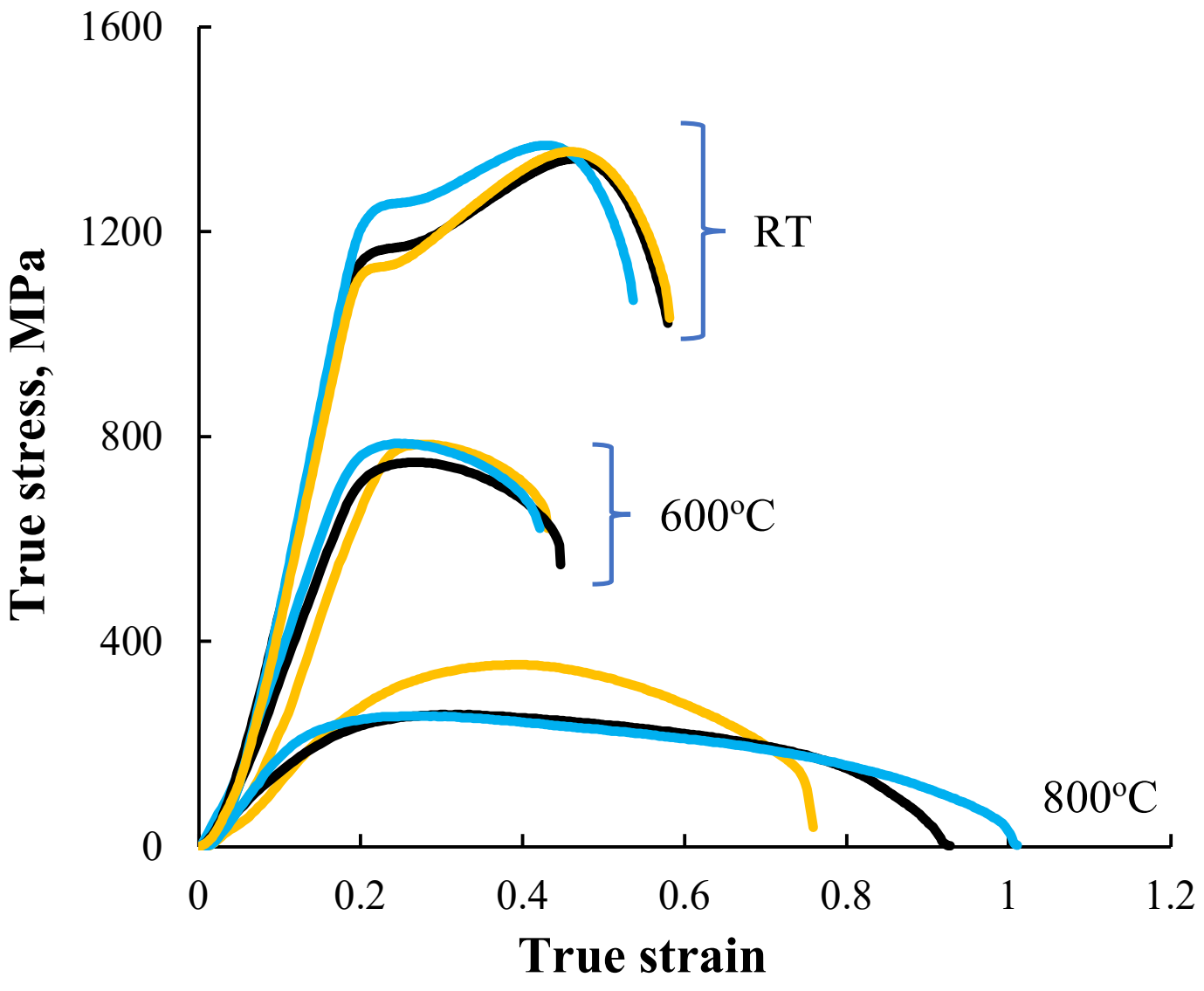

**Fig. 29** The σ(ε) curves of the steel samples #450-2 (orange line), #450-3 (black line) and #450-4 (blue line). Tensile testing at RT, 600 and 800°C. Strain rate $\dot{\varepsilon} = 3.3\cdot10^{-3}$ s$^{-1}$

The reduction in El at 450°C is due to the effect of dynamic strain aging in steel (for example, [71, 72]). The interaction between carbon atoms and dislocations in 0.12C–18Cr–9Ni austenitic steel has been previously studied in detail using the internal friction method [73]. Increasing the degree of cold deformation from 15% to 82% led to an increase in the maximum in the temperature dependence of internal friction, $Q^{-1}(T)$, occurring at 350–375°C. The addition of 0.51% wt. Ti, which forms Ti–C carbides, completely eliminated this maximum in the temperature curve $Q^{-1}(T)$.

This effect in steel 321L may be attributed to the interaction between dislocations and carbon atoms within the austenite crystal lattice. ECAP results in a high density of lattice dislocations in austenite, which interact with carbon atoms dissolved in the austenite crystal lattice. The mobility of carbon atoms within the austenite crystal lattice increases as the SP temperature rises, allowing carbon atoms to diffuse toward the nuclei of lattice dislocations. This results in a higher concentration of carbon atoms around lattice dislocations, which inhibits intragranular deformation at 450°C. Further increase in the SP temperature of UFG steel initiate strain-induced grain growth. The migration of grain boundaries facilitates a uniform distribution of carbon atoms within the crystal lattice and reduces the density of lattice dislocations. This process enhances volume strain rate at 750–900°C.

*d Effect of the number of ECAP cycles*

Table 2 and Figs. 29, 30 demonstrate that an increase in the number of ECAP cycles (*N*) leads to an increase in UTS and a slight decrease El at RT of UFG steel. This is an expected result associated with an increase in the fraction of martensite and a decrease in the average grain size.

The of ECAP cycles has a complex effect on the SP characteristics of steel. It is important to note that strongly effect of *N* shows at $T \leq 600$°C, where intensive grain growth does not occur, and the temperature range of 750–900°C, where intensive grain growth takes place. In addition, as shown in Table 3, the nature of the effect of the number of ECAP cycles depends on the preliminary annealing temperature (in the case of testing at 600°C).

It is important to emphasize once again that the creep rate and the creep life of austenitic steel depend non-monotonically on the grain size. The maximum creep life values are observed when austenite grains are ~30 μm [59-61], in some cases – at larger austenite grain size (up to 100 μm) [74, 75]. A further increase in grain size leads to a decrease in creep life and a decrease in creep rate. Consequently, it was anticipated that an increase in the preliminary annealing temperature, which increases the grain size of austenite from 0.5 μm to ~12 μm (Fig. 7), would enhance the ductility of steel. However, as illustrated in Table 3, the effect of the annealing temperature and the number of ECAP cycles is more complex than initially expected.

As is well known, at elevated temperatures, highly deformed materials with a fragmented structure exhibit lower ductility and a reduced strain rate compared to UFG materials that contain a significant proportion of high-angle grain boundaries (HAGBs). Consequently, when analyzing the results, the following factors must be taken into account:

i) An increase in the degree of SPD leads to an increase in the proportion of HAGBs in austenitic steels [40, 76-79]. Consequently, an increase in the number of ECAP cycles will lead to an increase in the proportion of HAGBs in steel 321.

ii) The recrystallization process of severely deformed materials is also accompanied by an increase in the proportion of HAGBs in metals. Consequently, an increase in the annealing temperature will lead to an increase in the average grain size (Fig. 7) and an increase in the proportion of HAGBs in fine-grained steel 321.

iii) An increase in the strain leads to a decrease in the average size of the recrystallized grain. Consequently, an increase in the number of ECAP cycles will lead to a decrease in the average size of the recrystallized grain in annealed fine-grained steel.

The SP temperature of unannealed UFG samples ($T$ = 600°C) is below the recrystallization temperature. Consequently, the average grain size or fragment of austenite in highly deformed steel is smaller than the critical grain size ($d_0 < d^* \sim 2$ μm, as indicated in Eq. (4)). This condition hinders the development of creep processes, and the strain rate of unannealed UFG steel is controlled by sliding of lattice dislocations [54]. As a result, the ductility of this steel is low, while the flow stress is high (Table 3).

Annealing at 750°C initiates recrystallization processes in steel #450-4, resulting in the formation of grains with an average size that exceeds the critical size $d^*$. This process enhances the ductility of this steel to ~ 52%. In contrast, steels #450-1, #450-2, and #450-3 retain their deformed microstructure after heating to 750°C. The fraction of recrystallized material in these steels is small and, therefore, their ductility remains low. A slight decrease in the ductility of highly deformed steels is caused by the release of σ-particles, which prevent dislocations from sliding in the austenite crystal lattice. After annealing at 800°C, the fraction of recrystallized metal increases in steels #450-2 and #450-3, while after annealing at 900°C, this increase is observed in steel #450-1. After annealing at 900°C, the differences in the microstructure of UFG steels subjected to a different number of ECAP cycles become insignificant, resulting in comparable ductility and flow stress (Table 3).

The analysis of the test results conducted at 750–900°C indicates that an increase in the number of ECAP cycles results in a simultaneous decrease in flow stress and an increase in elongation to failure (Figs. 13a, 29). It is known that an increase in the fraction of martensite results in the refinement of austenite grains during subsequent heating. Fine-grained alloys usually demonstrate higher ductility characteristics at elevated temperatures [27, 53]. However, it should be noted that the refinement of the grain microstructure in power law creep may only have an indirect effect on the ductility of the metal. In our opinion, one of the primary reasons for the increase in ductility of austenitic steel at 750–900°C is the refinement of the δ-ferrite particles. In the microstructure of CG steel, there are elongated δ-ferrite particles, with lengths reaching up to 500 μm and widths ranging from 3 to 5 μm (Fig. 3a). Such particles traverse austenite grains and can serve as effective barriers to intragranular strain. An increase in the number of ECAP cycles results in significant refinement of the δ-ferrite particles, causing them to no longer impede intragranular strain at elevated temperatures.

*e Effect of the ECAP temperature*

ECAP temperature influences the El and flow stress of the UFG steel at 600–900°C. As illustrated in Table 2 and Fig. 30, UFG steel produced at 150°C has both higher ductility and higher flow stress. Increasing the ECAP temperature to 450°C leads to a simultaneous decrease in both the El and flow stress.

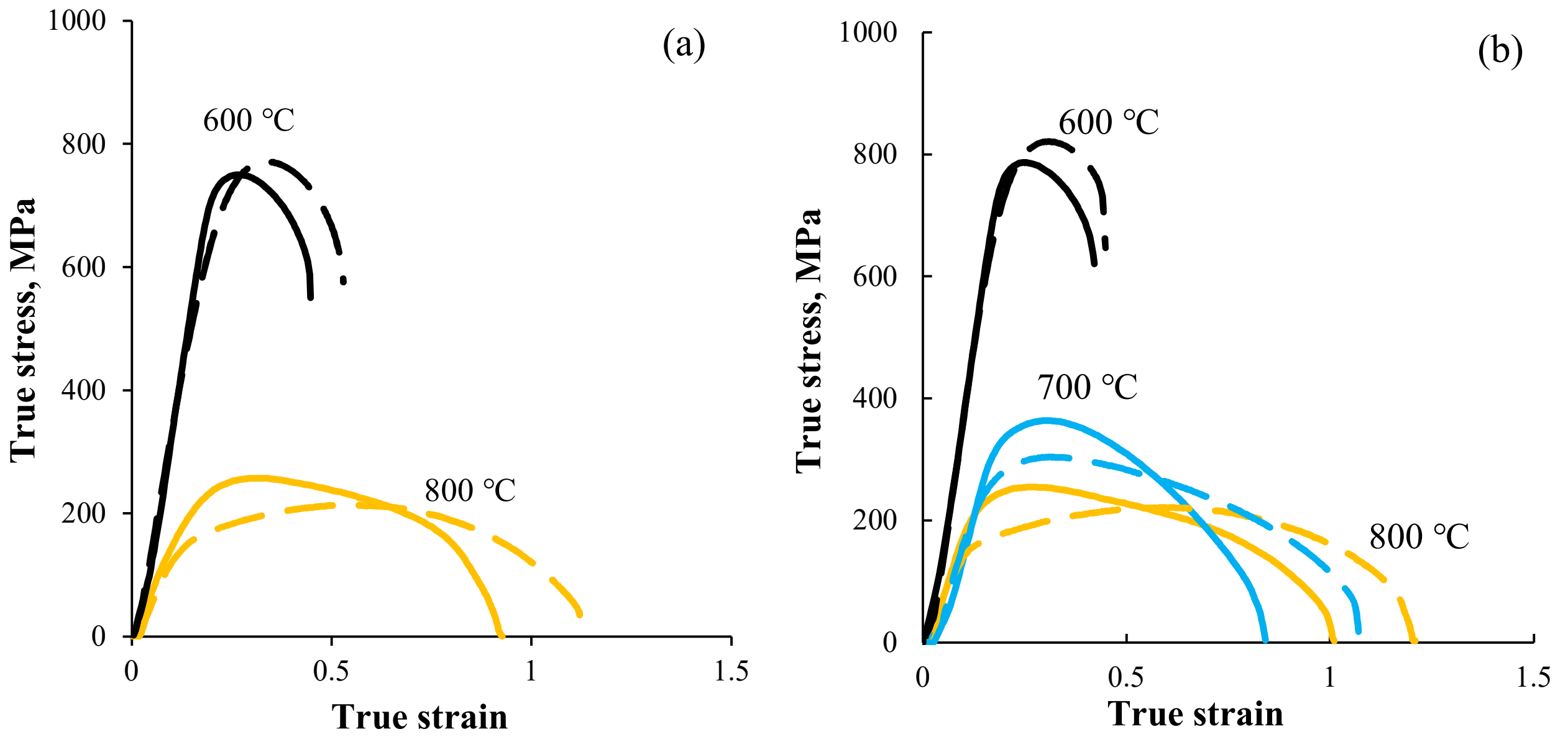

**Fig. 30** Tensile curves of the UFG samples obtained by the ECAP at 150°C (dotted lines) and 450°C (solid lines): (a) $N = 3$; (b) $N = 4$. Tests at 600°C (black lines), 750°C (blue lines) and 800°C (orange lines). Strain rate $\dot{\varepsilon} = 3.3 \cdot 10^{-3}$ s$^{-1}$

Secondly, attention should be given to the influence of the ECAP temperature on the creep activation energy of the UFG steel. Fig. 28 illustrates that the creep activation energy of the UFG steel samples obtained by ECAP at 150°C is $Q_{cr} \sim 23.8$–$24.0\ kT_m$ (~358–361 kJ/mol). The $Q_{cr}$ for the UFG steel samples obtained at 450°C is 10.2–14.1 $kT_m$ (~153–212 kJ/mol).

The ECAP temperature influences the average grain size. Samples fabricated at 150°C have smaller austenite grain size compared to those obtained at 450°C. According to Eq. (4), a decrease in grain size results in an increased contribution of GBS ($\dot{\varepsilon}_b$) to the total strain rate. Consequently, lower values of the creep activation energy are anticipated when testing samples of the UFG steel manufactured at 150°C, in comparison to the $Q_{cr}$ of the samples produced at 450°C.

This contradiction, in our opinion, is related to the influence of the δ-ferrite particles, which, when heated, can cause the formation of the σ-particles [79–82]. It is important to note that an increase in the strain accelerates the decomposition of δ-ferrite and the formation of σ-phase (see also [83]). Furthermore, it is known that σ-particles can form directly during warm and hot plastic deformation of austenitic steel [65, 66]. More efficient refinement of the δ-ferrite particles occurs during ECAP.

In this case, a more nonequilibrium state of the crystal lattice is established [84]. Consequently, it can be inferred that during creep of the UFG samples produced at 150°C, there is a more intensive decomposition of δ-ferrite, leading to a more intensive formation of the σ-particles.

The σ-phase particles prevent intragranular strain, thereby reducing the creep rate of austenitic steel. This results in an increase in the creep activation energy of UFG steel samples produced at 150°C, while simultaneously increasing the flow stress. More efficient refinement of large δ-ferrite particles will contribute to an increase in the ductility of the UFG steel produced at 150°C. The smaller austenite grain size in the steels manufactured by ECAP at 150°C additionally contributes to improved ductility at elevated SP temperatures.

### 4.3 Corrosion resistance of tension tested UFG steel samples

In this section, we will primarily focus on the study results of the IGC resistance of the deformed part of UFG samples after tension test at RT and after SP testing at elevated temperatures. In our view, two unexpected findings have emerged from the investigation of the IGC resistance of austenitic steel.

i) The number of ECAP cycles has a non-monotonic effect on the corrosion resistance of the deformed area of samples tested at RT (Fig. 22c). The value of the $K_{\mathrm{II}}$ coefficient depends non-monotonically on the number of ECAP cycles. After four ECAP cycles, the $K_{\mathrm{II}}$ coefficient of UFG samples is lower than the susceptibility of deformed samples of CG austenitic steel to IGC (Fig. 22c). The IGC rate of Zone I of UFG samples decreases slightly as the number of ECAP cycles increases.

ii) Increasing the SP test temperature leads to an increase in the IGC rate in 321 steel (Fig. 24). Increasing the SP deformation temperature to 800°C increases the IGC susceptibility of austenitic steel; after test at 900°C, the value of the $K_{\mathrm{II}}$ coefficient exceeds the critical value $K_{\mathrm{IGC}}$ = 0.11. This indicates that after SP deformation at 900°C, UFG steel becomes susceptible to IGC.

The possible causes of the observed anomalous change in the corrosion resistance of UFG steel after tensile testing will be discussed below.

It should be noted that ECAP leads to an increase in the *K* coefficient by more than two times compared to CG steel (Table 2). However, the overall magnitude of this effect is small, and the value of *K* remains low ($K \sim 10^{-4}$ [35]). UFG steel samples subjected to tensile testing at RT exhibit $K_{\mathrm{II}} \sim 10^{-2}$ (Table 2), which indicates a more pronounced negative effect of cold tensile deformation on corrosion resistance. The increase in $K_{\mathrm{II}}$ during the tensile testing of austenitic steel samples is caused by the formation of martensite during the test at RT. This is indirectly evidenced by an increase in *HV* in Zone II from 2.0–2.1 to 3.65–3.7 GPa for the CG steel. Since there is no refinement of austenite grains during tension, and the contribution of dislocations to the hardening of austenite is small [50], the observed increase in hardness by $\Delta HV \sim$ 1.55-1.7GPa is primarily associated with the formation of martensite. The hardness of Zone II in UFG steel samples subjected to tension (after warm ECAP, the initial martensite content is ~10%) increases from 3.5–3.8 GPa to 4.2–4.5 GPa ($\Delta HV$ ~ 0.7-1.0 GPa). Thus, as previously noted, the UFG steels have a lower tendency to form martensite under tension. The primary reason for this is likely the lower ductility of UFG steels and, consequently, the lower strain values observed in the tensile region of the UFG samples.

An increase in $N$ leads to a slight increase in the martensite content [35] and, simultaneously, to a refinement of the δ-ferrite particles. This is accompanied by a ~3-fold decrease in $K_I$ in the UFG steel samples (Fig. 22c). Intensive martensite formation during tension at RT leads to an increase in $K_{II}$. Thus, the tension process at RT has a more significant effect on the IGC rate than warm ECAP. The decrease in $K_{II}$ at large $N$ values (Fig. 22c) is associated with further refinement of the δ-ferrite during tension.

It is important to discuss the reasons for such a significant increase in the $K_{II}$ coefficient for the deformed metal in Zone II. As can be seen from Fig. 22c, the $K_{II}$ coefficient is an order of magnitude greater than for Zone I. Such a sharp increase in the $K$ coefficient cannot be attributed solely to the formation of martensite and, in our opinion, may be due to the formation of micropores and, as a result, an increase in the specific surface area of the samples after tensile testing (Fig. 11).

Now we should discuss the reasons for the abnormal dependence of the $K$ coefficient on the SP deformation temperature of UFG steel (Fig. 24).

It should be noted in advance that heating the metal of Zone I leads to a non-monotonic change in the $K_I$ parameter, which reaches its maximum values at 700-750°C. As shown in [35], heating UFG steel to a temperature of 700°C leads to the precipitation of σ-phase nanoparticles (Fig. 6), which negatively affect the IGC resistance of austenitic steel [1]. However, the σ-phase nanoparticles are precipitated uniformly throughout the entire volume of the material and do not form a continuous system of grain-boundary precipitates (Fig. 6).

The slight decrease in $K_I$ at 800°C is associated with the onset of intensive grain boundary migration and the disappearance of the deformed structure in the UFG steel. The temperature dependence of $K_{II}$ is similar to that of $K_I(T)$, but an important factor contributing to an increase in $K_{II}$ is the formation of large pores during SP. This, as noted above, leads to an increase in the specific surface area of the sample and, as a result, to an increase in the current.

The increase in the specific surface area of the sample due to the formation of pores is a methodological artifact that can affect the results of corrosion tests conducted using the DLEPR method. To verify the accuracy of the results obtained, tests for resistance to IGC were conducted in accordance with the Russian National Standard GOST 6032-2003 (ISO 3651-1:1998, ISO 3651-2:1998).

Samples of UFG steels, following tensile testing at room and elevated temperatures, exhibit a high susceptibility to IGC. The primary factors contributing to the corrosion fracture of the samples are δ-ferrite (Fig. 31a, b). After testing at low temperatures (RT – 450°C), in the vicinity of the fracture region, which corresponds to the maximum strain, intense corrosion fracture of the elongated α′-martensite grains is observed. The δ-ferrite bands in the fracture region are highly fragmented. At a distance from the fracture zone the amount of strain-induced martensite decreases, and the contribution of the δ-ferrite bands to the corrosion fracture of the samples becomes more significant. Corrosion dissolution of pores formed along the austenite grain boundaries is observed (Fig. 31b). Thus, the studies conducted confirm the significant contribution of pores to the corrosion-induced degradation of the UFG steels.

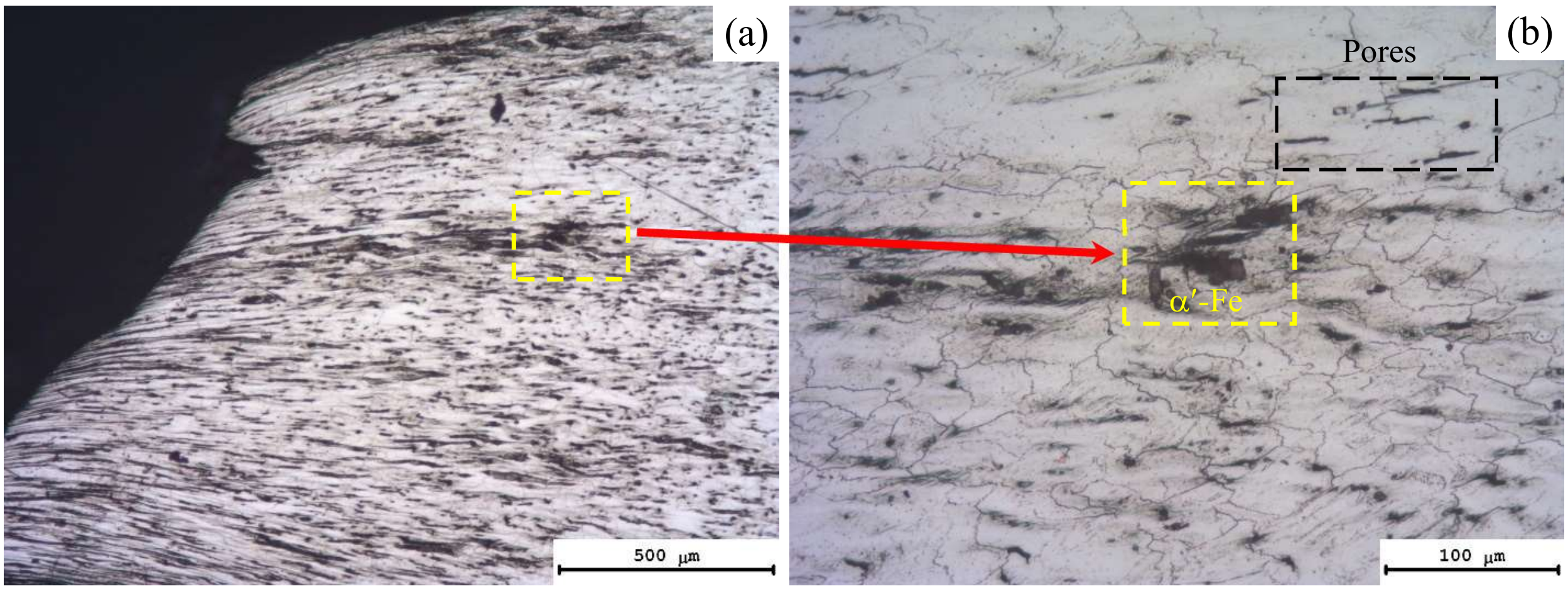

**Fig. 31** The surface of deformed UFG steel samples #2-450 after IGC tests in accordance with the GOST 6032-2003. Tensile tests at 900°C. Strain rate $\dot{\varepsilon}$ = 3.3·10$^{-3}$ s$^{-1}$

Summarizing the results obtained, it should be noted that the effective application of the SP effect for the fabrication of corrosion-resistant components from steel 321 likely requires the fulfillment of two conditions: (1) SP deformation must occur at temperatures above the temperature at which the reverse transformation of martensite to austenite is complete; and (2) SP deformation must not lead to extensive pore formation. An important factor contributing to the fulfillment of the second condition is the absence of large δ-ferrite particles and large non-metallic Ti(C,N) inclusions in the structure of 321 steel. The precipitation of σ-phase particles has a limited negative effect on the corrosion resistance of UFG steel 321, as the precipitated nanoparticles are uniformly distributed throughout the material volume and do not form large clusters along the grain boundaries.

**Conclusions**

1. The temperature dependence of the elongation to failure of the UFG steel 321L has a non-monotonic character (with a minimum point at 450-600°C). The UFG steels have good elongation to failure at 750–900°C. The maximum El ($\delta_{max}$ = 250%) in the UFG steels is limited by cavitation of the samples. The formation of large pores at the non-metallic particles Ti(C,N), as well as the micron-sized pores at the σ-phase particles, has been identified as the primary cause of cavitation failure. In specific temperature and strain rate conditions (600–900°C, $3.3\cdot10^{-3}$ – $3.3\cdot10^{-2}$ $s^{-1}$), the plastic deformation of the UFG steels is controlled by power law creep. The creep activation energy of the UFG steels produced by ECAP at 150°C is ~2 as high as the creep activation energy of the samples produced by ECAP at 450°C.

2. An increase in the number of ECAP cycles leads to an increase in ultimate strength and a slight decrease in the ductility of the UFG steel at RT. At elevated tensile temperatures (750–900°C), the opposite situation occurs. An increase in the number of ECAP cycles leads to an increase in the ductility and a decrease in the flow stress. The reason for the increased ductility is the fragmentation of δ-ferrite particles, which prevent intragranular strain. An increase in the ECAP temperature from 150°C to 450°C results in a simultaneous decrease in elongation to failure and flow stress.

3. Preliminary annealing has different effects on the ductility of heterostructured steels compared to those with a homogeneous UFG structure. Increasing the preliminary annealing temperature from 600 to 900°C leads to a monotonous increase in the ductility of UFG steel at 600°C. In contrast, the plasticity of heterosrucutured steel exhibits a non-monotonic (with a minimum) dependence on the preliminary annealing temperature.

4. Coarse-grained steel subjected to warm SP deformation (600–800°C) exhibits higher microhardness compared to CG steel samples that have been annealed at the same temperature. The increase in microhardness of CG steel is due to the intensive accumulation of defects at grain boundaries during SP deformation, as well as the refinement of austenite grains.

UFG steel subjected to SP deformation exhibits a microhardness that is 0.2–0.4 GPa lower than that of UFG steel annealed at the same temperature. An increase in the temperature during SP deformation results in a further decrease in the microhardness of the UFG steel. This decrease in microhardness is due to the martensite reduction and strain-induced grain growth. The intensive occurrence of these processes compensates for the increase in microhardness due to strain hardening during SP of UFG steel.

5. The UFG steel exhibits a low resistance to intergranular corrosion following SP tests at 600-800°C. After SP tests at 900°C, UFG steel samples have low microhardness and a high IGC susceptibility. The decrease in IGC resistance of the UFG steel is due to the accelerated release of σ-phase particles during SP deformation and the formation of pores, which lead to an increase in the specific surface area of the sample.

**Appendix A.** Finite element modeling (FEM) of deformation distribution during stretching of 321 steel samples

The modeling conducted using the finite element method in the ANSYS Workbench software package. The curves of σ(ε) presented in the Results section of Appendix B, along with the material parameters for steel 321 obtained from the ANSYS Workbench database, were used in the modeling process. The calculations were performed for the case of stretching at RT and at a temperature of 800°C. The distribution graph of the strain intensity across the working area of the sample (Fig. 1a) is shown in Fig. A1. As an example, the results of the stress intensity calculations on the surface and in the middle part of a flat sample for some CG and UFG samples are presented in Fig. A2-A5.

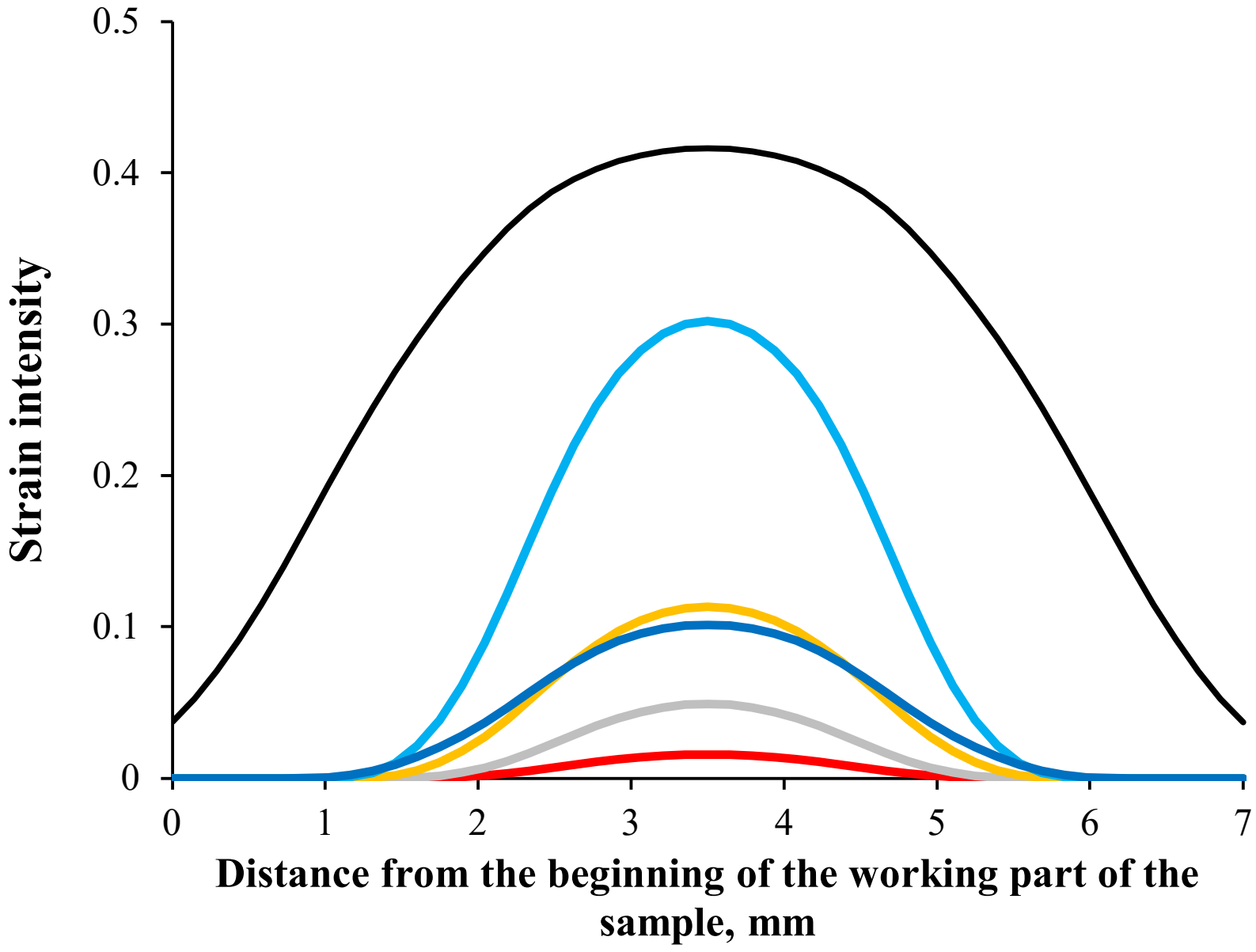

**Fig. A1** The distribution of the strain intensity along the working part of a flat sample (Fig. 1a) during tensile testing at various temperatures

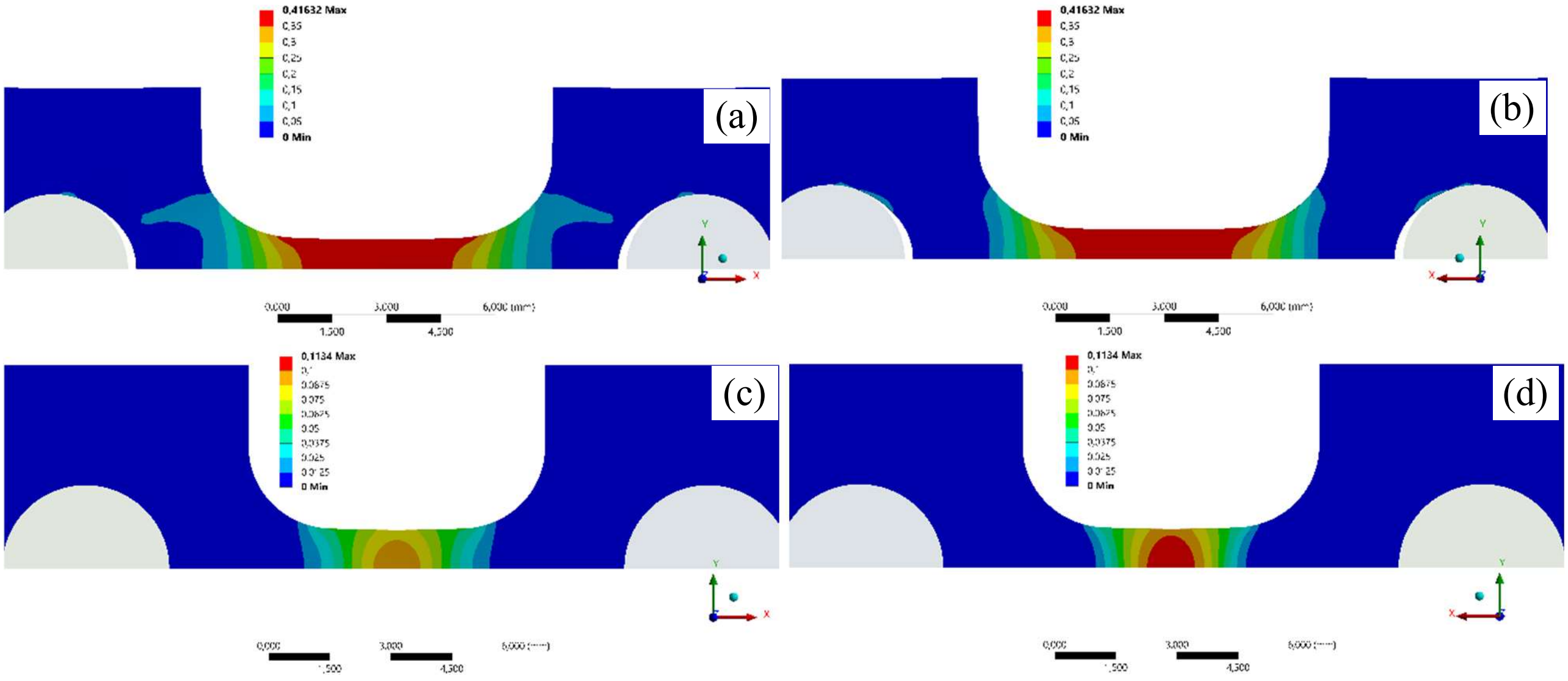

**Fig. A2** The results of modeling the strain intensity distribution during testing of CG samples at RT (a, b) and at 800°C (c, d). Fig. A2a, c shows the results of calculations of the strain intensity on the sample surface, Fig. A2b, d – in the center of the longitudinal section of the sample

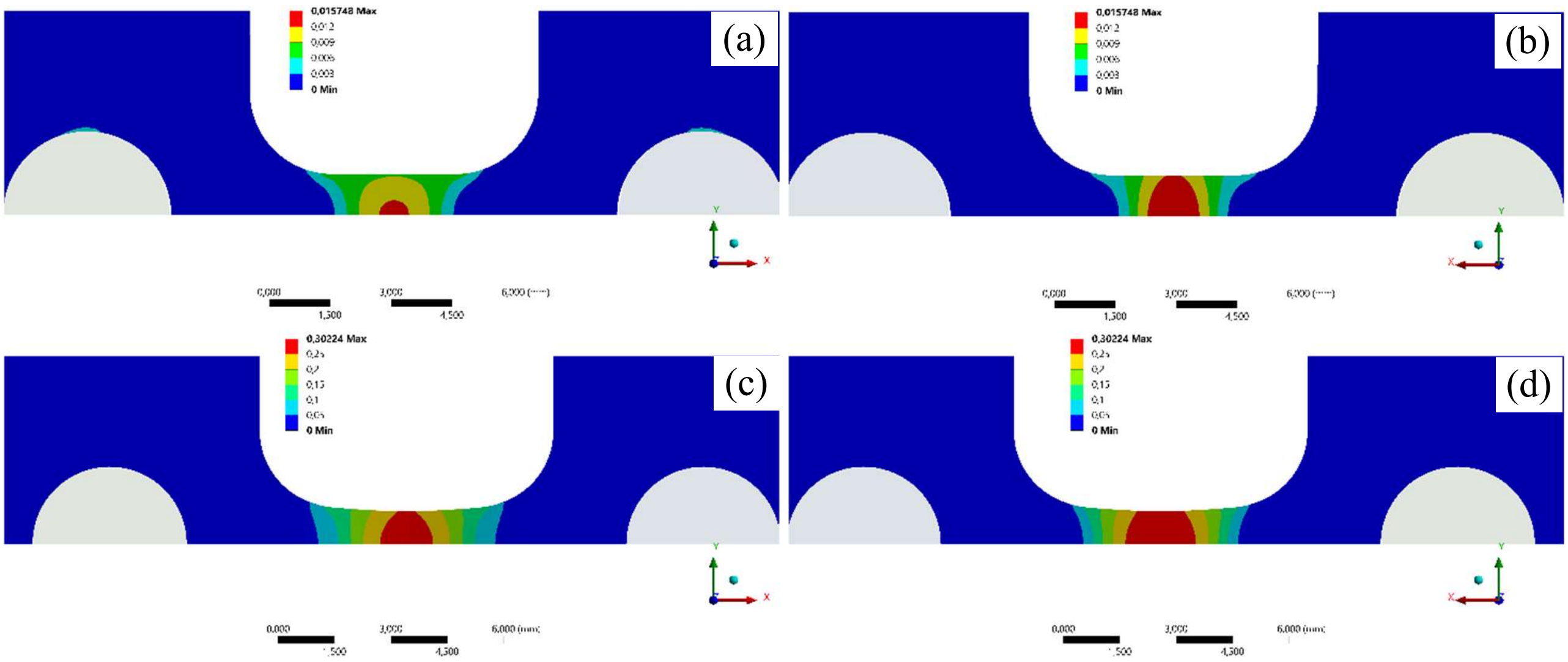

**Fig. A3** The results of FEM the strain intensity distribution during testing of samples #150-4 at RT (a, b) and at 800°C (c, d). Fig. A3a, c shows the results of calculations of the strain intensity on the sample surface, Fig. A3b, d – in the center of the longitudinal section of the sample

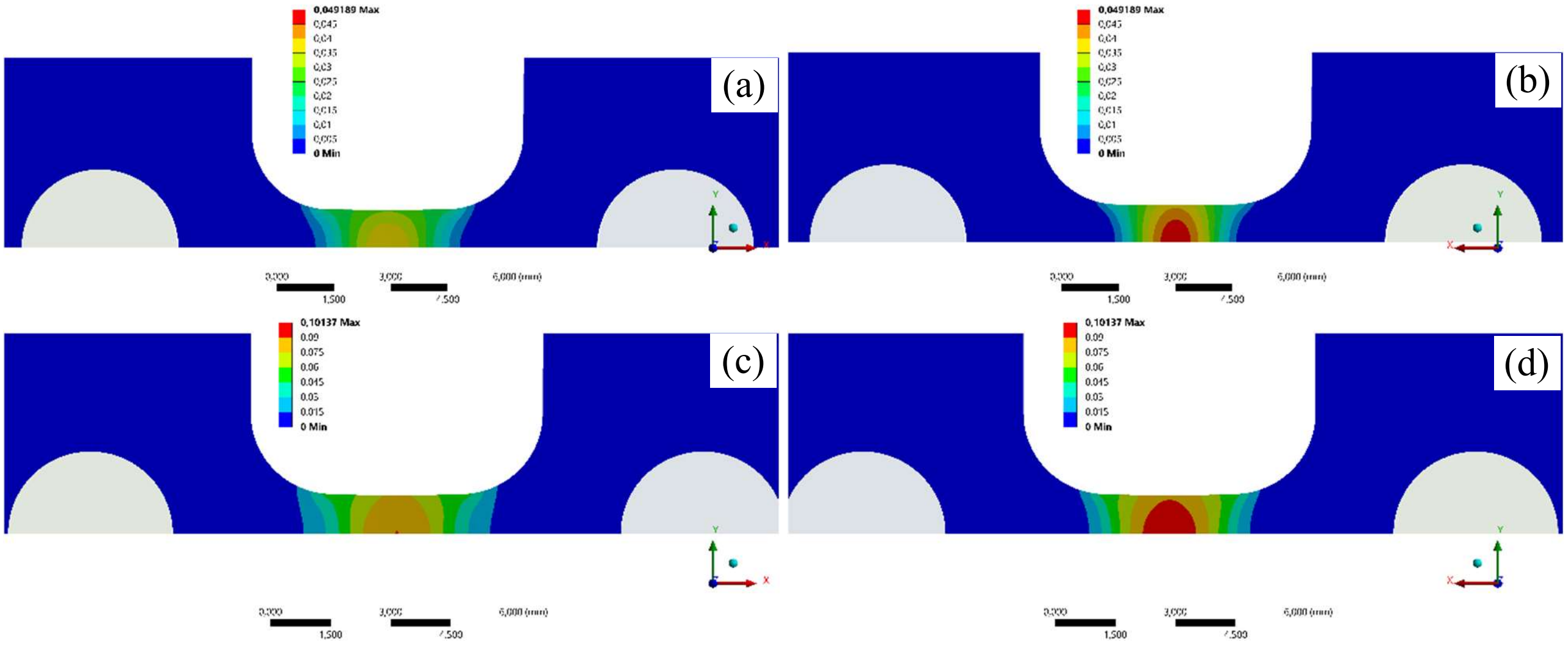

**Fig. A4** The results of FEM the strain intensity distribution during testing of samples #450-4 at RT (a, b) and at 800°C (c, d). Fig. A4a, c shows the results of calculations of the strain intensity on the sample surface, Fig. A4b, d – in the center of the longitudinal section of the sample

**Appendix B.** Results of mechanical tensile tests at various temperatures: key engineering stress - engineering strain curves for typical samples

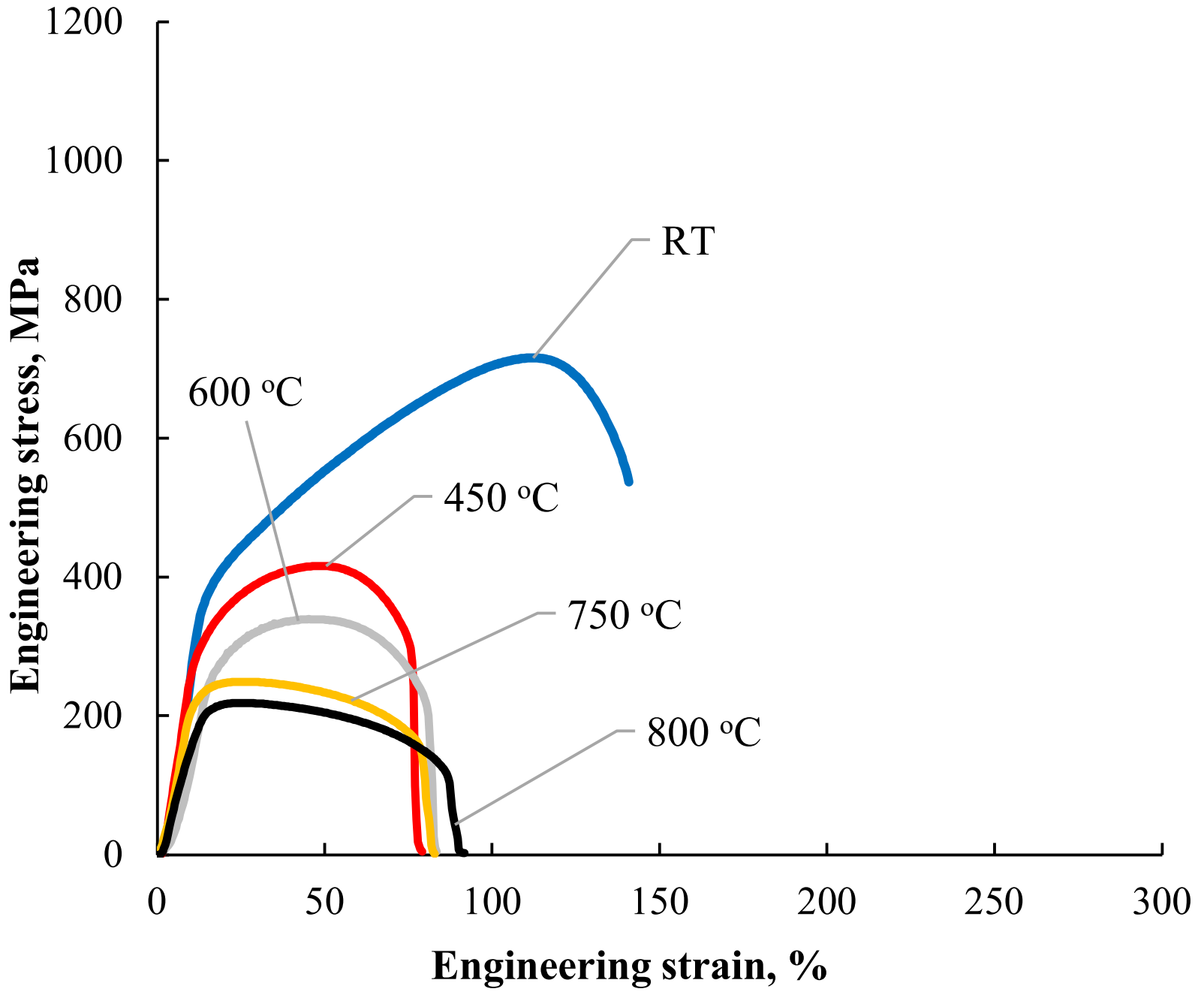

**Fig. B1** Engineering stress–engineering strain curves for the CG samples. Strain rate $\dot{\varepsilon}$ = 3.3·10$^{-3}$ s$^{-1}$

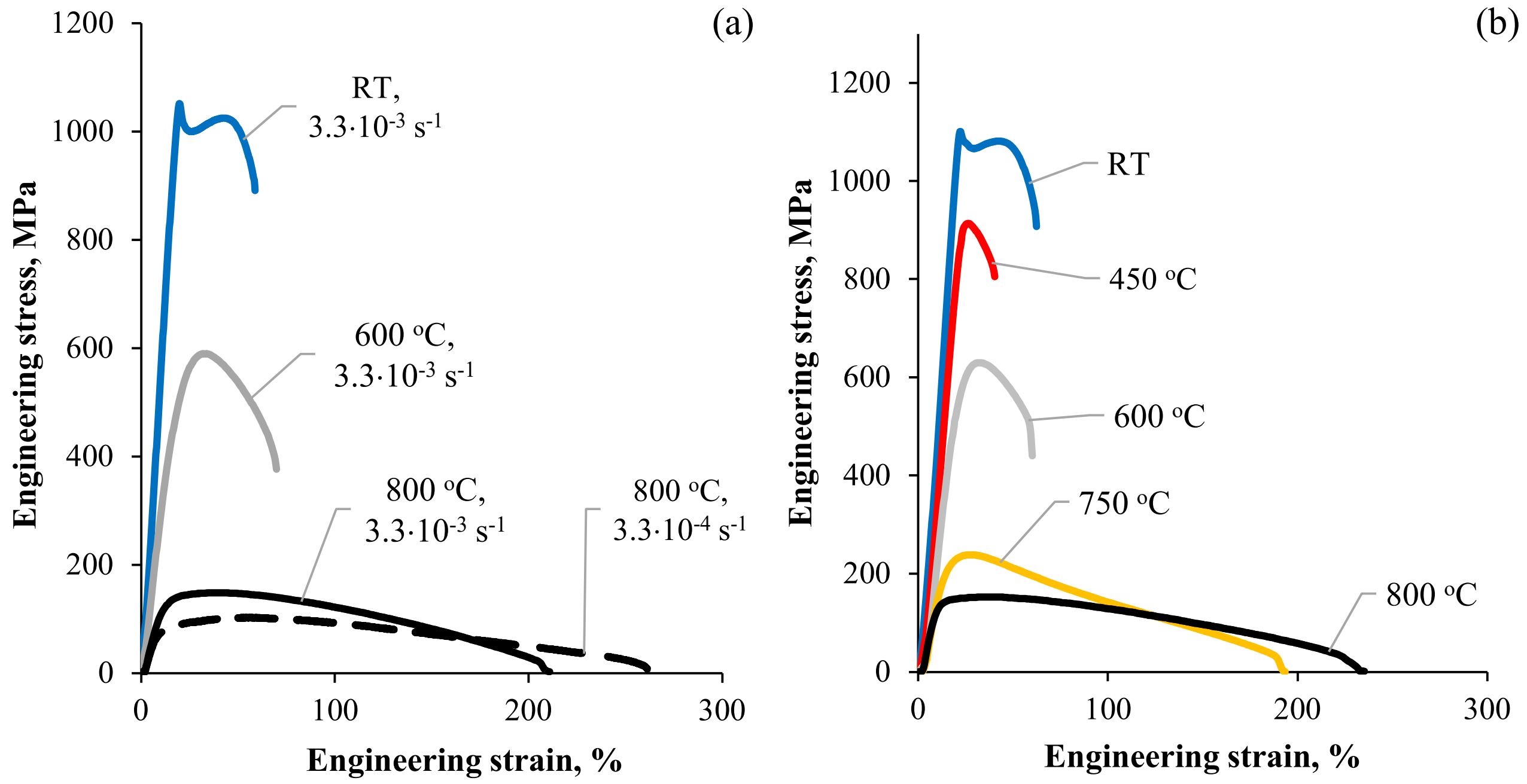

**Fig. B2** Engineering stress–engineering strain curves for the samples #150-3 (a) and #150-4 at different temperatures. Strain rate 3.3·10$^{-4}$ s$^{-1}$ and 3.3·10$^{-3}$ s$^{-1}$

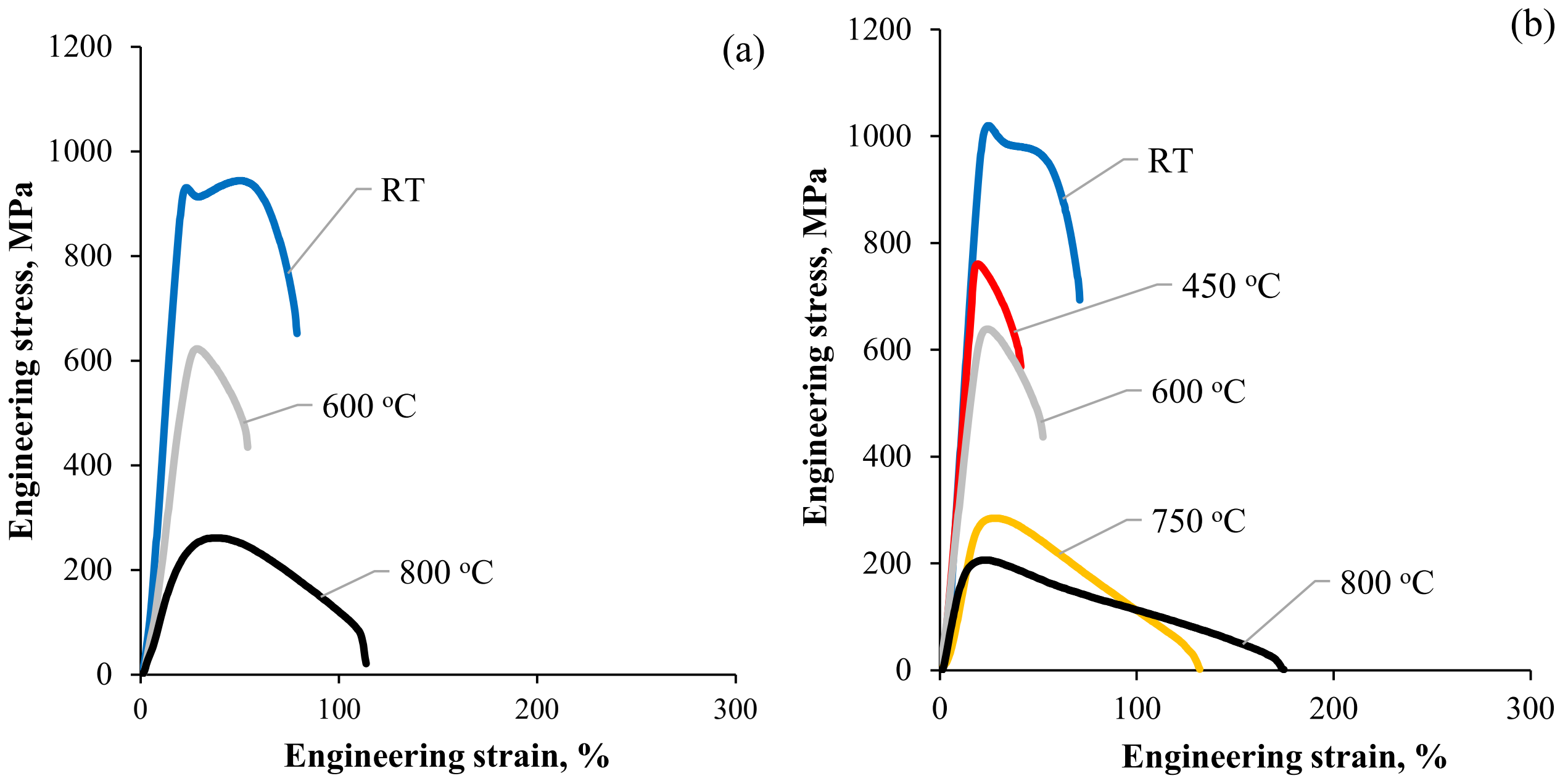

**Fig. B3** Engineering stress–engineering strain curves for the samples #450-2 (a) and #450-4 (b) at different temperatures. Strain rate $\dot{\varepsilon} = 3.3{\cdot}10^{-3}$ s$^{-1}$

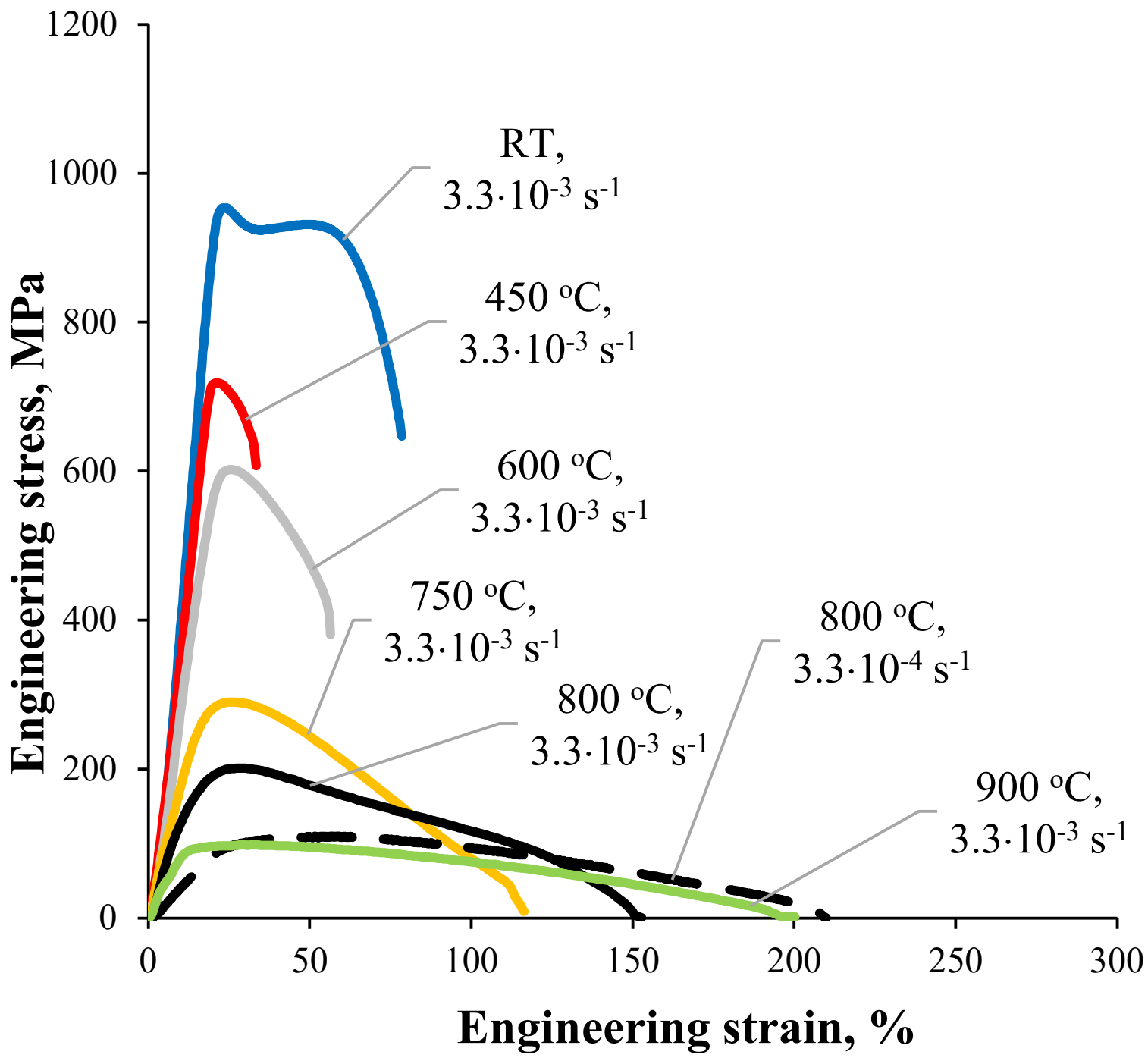

**Fig. B4** Engineering stress–engineering strain curves for the samples #450-3 at different temperatures. Strain rate $\dot{\varepsilon} = 3.3{\cdot}10^{-3}$ s$^{-1}$

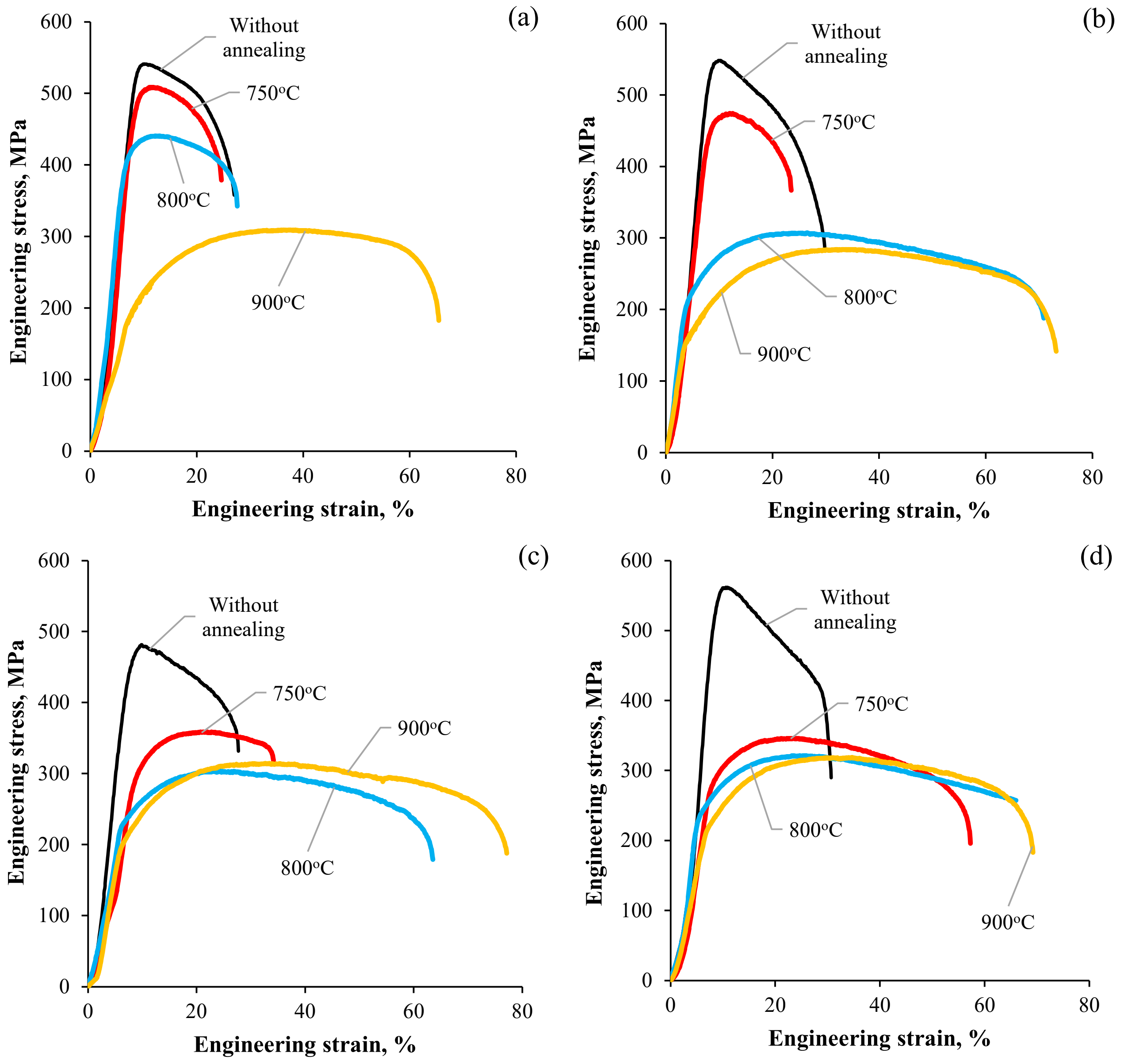

**Fig. B5** Curves of σ(ε) for UFG samples that were preliminary annealed at different temperatures: (a) #450-1; (b) #450-2; (c) #450-3; (d) #450-4. Tests at 600°C, $\dot{\varepsilon} = 3.3 \cdot 10^{-4}$ $s^{-1}$. The temperatures of 30-min annealing are indicated in the corresponding figures

**Funding:** The present study was supported by Ministry of Science and High Education of the Russian Federation (grants FSWR-2026-0039, FSWR-2026-0010).

**Acknowledgements:** The authors thank N.Yu. Tabachkova (National University of Science and Technology MISIS; A.M. Prokhorov General Physics Institute of the Russian Academy of Science, Moscow, Russia) for the results of the studies on the microstructure of steels using the TEM method. The TEM microstructure studies were carried out on the equipment of the Center Collective Use “Materials Science and Metallurgy” (National University of Science and Technology “MISIS”, Moscow, Russia).

**Author contributions**: A.V. Nokhrin: Supervision, Methodology, Writing–Review & Editing, Data curation, Visualization; V.I. Kopylov: Investigation (obtained of UFG steel by ECAP, optimization of ECAP modes, annealing), Writing–Original Draft, Formal Analysis; M.K. Chegurov: Investigation (fractography, corrosion test); S.V. Shotin: Investigation (tensile tests); C.V. Likhnitskii: Investigation (metallography, microhardness); N.V. Melekhin: Project administration, Formal Analysis, Writing–Original Draft, Recourses, Funding acquisition; D.A. Zotov: Investigation (SEM); K.A. Rubtsova (XRD); O.E. Pirozhnikova: Formal Analysis (Discussion); N.N. Berendeev & D.D. Kulagin: Formal Analysis (Appendix A).

**Conflict of interest or competing interests**: The authors declare that they have no known conflicts of financial interests or personal relationships that could have appeared to influence the work reported in this paper.

**Data and code availability**: Data will be available on the request.

**Supplementary information**: Not Applicable

**Ethical approval**: Not Applicable